\documentclass[12pt]{article}
\usepackage{latexsym, amsmath, epsfig}
\newfont{\twlvmsb}{msbm10 scaled\magstep1}   
\newfont{\ninemsb}{msbm9}                    
\newfont{\sixmsb}{msbm6}                     
\newfam\msbfam
\textfont\msbfam=\twlvmsb
\scriptfont\msbfam=\ninemsb
\scriptscriptfont\msbfam=\sixmsb
\catcode`\@=11
\def\Bbb{\ifmmode\let\next\Bbb@\else
  \def\next{\errmessage{Use \string\Bbb\space only in math mode}}\fi\next}
\def\Bbb@#1{{\Bbb@@{#1}}}
\def\Bbb@@#1{\fam\msbfam#1}

\newfont{\largeeufm}{eufm10 scaled\magstep4} 
\newfont{\twlveufm}{eufm10 scaled\magstep1}  
\newfont{\elveufm}{eufm10 at 11pt}           
\newfont{\teneufm}{eufm10}
\newfont{\nineeufm}{eufm9}
\newfam\eufam
\textfont\eufam=\twlveufm
\scriptfont\eufam=\nineeufm
\def\frak{\ifmmode\let\next\frak@\else
\def\next{\errmessage{Use \string\frak\space only in math mode}}\fi\next}
\def\frak@#1{{\fam\eufam{{#1}}}}
\catcode`@=12

\newcommand{\appsection}[1]{
\vspace{12mm}
\pagebreak[3]
\addtocounter{section}{1}
\setcounter{equation}{0}
\setcounter{subsection}{0}
\addcontentsline{toc}{section}
{\protect\numberline{\Alph{section}} {#1}}
 \begin{flushleft}
{\large\bf Appendix \thesection :~~#1}
\end{flushleft}
\nopagebreak
\medskip
\nopagebreak}
\renewcommand{\theequation}{\thesection.\arabic{equation}}
\newlength{\extraspace}
\newlength{\extraspaces}
\newcommand{\be}{\begin{equation}
\addtolength{\abovedisplayskip}{\extraspaces}
\addtolength{\belowdisplayskip}{\extraspaces}
\addtolength{\abovedisplayshortskip}{\extraspace}
\addtolength{\belowdisplayshortskip}{\extraspace}}
\newcommand{\ee}{\end{equation}}
 
\newcommand{\ba}{\begin{eqnarray}
\addtolength{\abovedisplayskip}{\extraspaces}
\addtolength{\belowdisplayskip}{\extraspaces}
\addtolength{\abovedisplayshortskip}{\extraspace}
\addtolength{\belowdisplayshortskip}{\extraspace}}
\newcommand{\ea}{\end{eqnarray}}
 
\newcommand{\bas}{\begin{eqnarray*}
\addtolength{\abovedisplayskip}{\extraspaces}
\addtolength{\belowdisplayskip}{\extraspaces}
\addtolength{\abovedisplayshortskip}{\extraspace}
\addtolength{\belowdisplayshortskip}{\extraspace}}
\newcommand{\eas}{\end{eqnarray*}}
 
\newcounter{subequation}[equation]
\makeatletter
\expandafter\let\expandafter\reset@font\csname reset@font\endcsname
\def\subeqnarray{\arraycolsep1pt
    \def\@eqnnum\stepcounter##1{\stepcounter{subequation}
        {\reset@font\rm(\theequation\alph{subequation})}}\eqnarray}

\makeatother
 
\newenvironment{theorem}[1]
{\vspace{3mm}\noindent {\bf #1 :} }{\vspace{2mm}}
\newcommand{\bt}[1]{\begin{theorem}{#1}}
\newcommand{\et}{\end{theorem}}
 
\newcommand{\newsection}[1]{
\vspace{12mm}
\pagebreak[3]
\addtocounter{section}{1}
\setcounter{equation}{0}
\setcounter{subsection}{0}
\addcontentsline{toc}{section}
{\protect\numberline{\arabic{section}}{#1}}
 \begin{flushleft}
{\large\bf \thesection. #1}
\end{flushleft}
\nopagebreak
\medskip
\nopagebreak}
 
\newcommand{\newsubsection}[1]{
\vspace{1cm}
\pagebreak[3]
 
\addtocounter{subsection}{1}
\addcontentsline{toc}{subsection}{\protect
\numberline{\arabic{section}.\arabic{subsection}}{#1}}
\begin{flushleft}
{ \bf \thesubsection. #1}
\end{flushleft}
\nopagebreak
\vspace{2mm}
\nopagebreak}

\newcommand{\newsubsubsection}[1]{
\vspace{0.8 cm}
\pagebreak[3]
 
\addtocounter{subsubsection}{1}
\addcontentsline{toc}{subsubsection}
{\protect\numberline{\arabic{section}.\arabic{subsection}.%
\arabic{subsubsection}}{#1}}
\begin{flushleft}
{ \sc \thesubsubsection. #1}
\end{flushleft}
\nopagebreak
\vspace{2mm}
\nopagebreak}

\newcommand{\NP}[1]{Nucl.\ Phys.\ {\bf #1}}
\newcommand{\PL}[1]{Phys.\ Lett.\ {\bf #1}}
\newcommand{\CMP}[1]{Comm.\ Math.\ Phys.\ {\bf #1}}
\newcommand{\PR}[1]{Phys.\ Rev.\ {\bf #1}}
\newcommand{\PRL}[1]{Phys.\ Rev.\ Lett.\ {\bf #1}}

\newcommand{\FP}[1]{Fortschr.\ Phys.\ {\bf #1}}

\newcommand{\ra}{\rightarrow}

\newcommand{\rra}{\ \longrightarrow \ }

\newcommand{\pa}{{\partial}}

\newcommand{\cL}{{\cal L}}

\newcommand{\Ds}{D\llap{/}}
\begin{document}
%
\newcommand{\Cal}{{\cal C}}
\renewcommand{\l}{\lambda}
\renewcommand{\a}{\alpha}
\renewcommand{\b}{\beta}
\renewcommand{\d}{\delta}
\renewcommand{\k}{\kappa}
\newcommand{\ld}{\buildrel \leftarrow \over \d}
\newcommand{\rd}{\buildrel \rightarrow \over \d}
\newcommand{\e}{\eta}
\renewcommand{\o}{\omega}
\newcommand{\dem}{\d_\o}
\newcommand{\p}{\partial}
\newcommand{\pmu}{\p_\mu}
\newcommand{\pmo}{\p^\mu}
\newcommand{\pnu}{\p_\nu}
\newcommand{\s}{\sigma}
\renewcommand{\r}{\rho}
\newcommand{\bpsi}{\bar\psi}
\newcommand{\dslash}{\p\llap{/}}
\newcommand{\pslash}{p\llap{/}}
\newcommand{\ve}{\varepsilon}
\newcommand{\uvi}{\underline{\varphi}}
\newcommand{\vi}{\varphi}
\newcommand{\ue}{u^{(1)}}
\newcommand{\Am}{A_\mu}
\newcommand{\An}{A_\nu}
\newcommand{\Fmnu}{F_{\mu\nu}}
\newcommand{\Fmno}{F^{\mu\nu}}
\newcommand{\Ga}{\Gamma}
\newcommand{\Gao}{\Gamma^{(o)}}
\newcommand{\Gae}{\Gamma^{(1)}}
\newcommand{\Gacl}{\Gamma_{cl}}
\newcommand{\Gagf}{\Gamma_{{\rm g.f.}}}
\newcommand{\Gainv}{\Gamma_{\rm inv}}
\newcommand{\pvi}{\partial\varphi}
\newcommand{\om}{{\bf w}}
\newcommand{\mn}{\mu\nu}
\newcommand{\Tmn}{T_{\mn}}
\newcommand{\Gmn}{\Ga_{{\mn}}}
\newcommand{\hT}{\hat T}
\newcommand{\emt}{energy-mo\-men\-tum ten\-sor}
\newcommand{\eit}{Ener\-gie-Im\-puls-Ten\-sor}
\newcommand{\Tc}{T^{(c)}}
\newcommand{\Tcr}{\Tc_{\rho\sigma}(y)}
\newcommand{\ha}{{1\over 2}}
\newcommand{\dalam}{{\hbox{\frame{6pt}{6pt}{0pt}}\,}}
\newcommand{\wtm}{{\bf W}^T_\mu}
\newcommand{\nnp}{\nu'\nu'}
\newcommand{\mnp}{\mu'\nu'}
\newcommand{\emn}{\eta_{\mn}}
\newcommand{\mpm}{\mu'\mu'}
\newcommand{\tbw}{\tilde{\bf w}}
\newcommand{\tw}{\tilde w}
\newcommand{\hw}{\hat{\bf w}}
\newcommand{\bw}{{\bf w}}
\newcommand{\bW}{{\bf W}}
\newcommand{\ubW}{\underline{\bW}}
\newcommand{\hmn}{h^{\mn}}
\newcommand{\gmn}{g^{\mn}}
\newcommand{\ga}{\gamma}
\newcommand{\Gf}{\Gamma_{\hbox{\hskip-2pt{\it eff}\hskip2pt}}}
\newcommand{\T}{\buildrel o \over T}
\newcommand{\Lf}{{\cal L}_{\hbox{\it eff}\hskip2pt}}
\newcommand{\np}{\not\!\p}
\newcommand{\lp}{\partial\llap{/}}
\newcommand{\ah}{{\hat a}}
\newcommand{\han}{{\hat a}^{(n)}}
\newcommand{\hak}{{\hat a}^{(k)}}
\newcommand{\dv}{{\d\over\d\vi}}
\newcommand{\zze}{\sqrt{{z_2\over z_1}}}
\newcommand{\zez}{\sqrt{{z_1\over z_2}}}
\newcommand{\Hmn}{H_{\(\mn\)}}
\newcommand{\hfrac}[2]{\hbox{${#1\over #2}$}}  
\newcommand{\smdm }{\underline m \p _{\underline m}}
 \newcommand{\tsmdm }{\underline m \tilde \p _{\underline m}} 
 \newcommand{ \Wh }{{\hat {\bf W}}^K}
\newcommand{ \CS }{Callan-Symanzik}
\newcommand{ \G}{\Gamma}
\newcommand{ \bl }{\b _ \l}
\newcommand{ \kdk }{\k \p _\k}
\newcommand{\mdm }{m \p _m} 
\newcommand{ \te }{\tau_{\scriptscriptstyle 1}}
\newcommand{ \mhi }{m_H}
\newcommand{ \mf }{m_f}
\newcommand{\bare}{^o}
\newcommand{\Pol}{{\mathbf P}}
\newcommand{\brs}{{\mathrm s}}
\newcommand{\cw}{\cos \theta_W}
\newcommand{\cws}{\cos^2 \theta_W}
\newcommand{\sw}{\sin \theta_W}
\newcommand{\sws}{\sin^2 \theta_W}
\newcommand{\cg}{\cos \theta_G}
\newcommand{\sg}{\sin \theta_G}
\newcommand{\cwg}{\cos (\theta_W- \theta_G)}
\newcommand{\swg}{\sin (\theta_W- \theta_G)}
\newcommand{\cv}{\cos \theta_V}
\newcommand{\sv}{\sin \theta_V}
\newcommand{\cvg}{\cos (\theta_V- \theta_G)}
\newcommand{\svg}{\sin (\theta_V- \theta_G)}
\newcommand{\fsc}{{e^2 \over 16 \pi ^2}}
\newcommand{\cvt}{\cos \Theta^V_3}
\newcommand{\svt}{\sin \Theta^V_3}
\newcommand{\cvf}{\cos \Theta^V_4}
\newcommand{\svf}{\sin \Theta^V_4}
\newcommand{\cgt}{\cos \Theta^g_3}
\newcommand{\sgt}{\sin \Theta^g_3}



\newcommand{\Identity}{\scalebox{1}[.95]{1}\hspace{-3.5pt}{\scalebox{1}[1.1]{1}}}
\newcommand{\slashed}{\hspace{-1.1ex}/}
\newcommand{\Slashed}{\hspace{-1.7ex}/}
\newcommand{\equ}[1]{\begin{gather} #1 \end{gather}}
\newcommand{\equa}[1]{\begin{align} #1 \end{align}}
\newcommand{\derivative}[2]{\ensuremath{\frac{\text{d} #1}{\text{d} #2}} }
\newcommand{\pderivative}[2]{\ensuremath{\frac{\partial #1}{\partial #2}} }
\newcommand{\fderivative}[2]{\ensuremath{\frac{\delta #1}{\delta #2}}}
\newcommand{\m}{\mu}
\newcommand{\n}{\nu}
\newcommand{\g}{\gamma}
\newcommand{\tI}{\tilde{I}}
\newcommand{\RE}{\mbox{Re }}
\newcommand{\IM}{\mbox{Im }}
\newcommand{\Gm}{\Gamma^{mass}_{Dirac}}
\newcommand{\intd}{\int \! d ^4 x}
\begin{titlepage}
%
\renewcommand{\thefootnote}{\fnsymbol{footnote}}
\begin{flushright}
BN-TH-98-18\\
NIKHEF 98-027\\
hep-th/9809069\\
August 1998
\end{flushright}
\vspace{1cm}
 
\begin{center}
{\Large {\bf Renormalization of the
Electroweak  Standard Model}}
\\[4mm]
{\bf Lectures given at the Saalburg Summer School 1997}
\\[1.5cm]
{\bf Elisabeth Kraus}
\\[3mm]
{\small\sl Physikalisches Institut, Universit\"at Bonn} \\
{\small\sl Nu\ss allee 12, D-53115 Bonn, Germany} \\
\vspace{1cm}
written by:\\[5mm]
Stefan Groot Nibbelink \\[3mm]
{\small\sl NIKHEF, Amsterdam}\\
\vspace{1.5cm}

{\bf Abstract}
\end{center}
\begin{quote}
These lecture notes give an introduction to the algebraic
renormalization of the Standard Model.  We start with the  construction
of the
tree approximation and give the classical action and
 its defining symmetries in functional form.  These are
the Slavnov-Taylor identity,
Ward identities of rigid symmetry and 
the abelian local Ward identity.  The abelian 
Ward identity   ensures coupling of the electromagnetic current
 in higher orders of perturbation theory, and is the
functional form
of the Gell-Mann--Nishijima relation.
 In the second part of the
lectures we present in simple examples the basic properties of
renormalized perturbation theory: scheme dependence of counterterms and
the quantum action principle. Together with an algebraic characterization
of the defining symmetry transformations they are the ingredients 
for a scheme independent unique construction of Green's functions 
 to all orders of perturbation theory.

\end{quote}
\vfill
\renewcommand{\thefootnote}{\arabic{footnote}}
\setcounter{footnote}{0}
\end{titlepage}


\tableofcontents
\newpage
\newsection{Introduction}

In these lectures we give an introduction to the algebraic
renormalization of the Standard Model of electroweak interaction. The
Standard Model of elementary particle physics is a renormalizable
quantum field theory and allows consistent predictions of physical
processes
in terms of a few parameters, as masses and couplings,
 order by order in perturbation theory. The Standard Model includes
electromagnetic, weak and strong interactions and the classical model is
a non-abelian gauge theory with gauge group $U(1) \times SU(2) \times
SU(3)$. The $U(1) \times SU(2)$  gauge group is spontaneously broken to the
electromagnetic subgroup providing masses for
 the charged leptons and
quarks and for the vector bosons of weak
interactions via the Higgs mechanism,
 but leaving the photon as a massless particle 
\cite{GLA61,WEI67,SAL68}. Since the
electromagnetic subgroup does not correspond to the abelian factor subgroup it
turns out that weak interactions cannot be described consistently
without the electromagnetic interactions, but we are able to split off
the unbroken $SU(3)$-colour gauge group responsible for the strong
interactions without destroying the physical
structure of the theory. In these lectures we only consider  the $SU(2)
\times U(1)$-structure of electroweak interactions.

The Standard Model of electroweak interactions has been tested to high
accuracy with the precision experiments at the Z-resonance at LEP
 \cite{datasum96}.  The degree of precision enforces to take into account also
 contributions  beyond the tree approximation in the perturbative formulation. For
this reason an extensive calculation of 
 1-loop processes and also 2-loop processes has been carried out in the
past years 
 and compared to the experimental results. 
(For reviews see \cite{hollik,JB1} and references therein; for a recent review
see
\cite{HO97}.) A careful
analysis shows that the theoretical 
predictions  and the experiments are in excellent agreement with each
other
\cite{YeRep}. 

A necessary prerequisite for carrying out precision tests of the Standard
Model
is the consistent mathematical and physical formulation of the Standard
Model in the framework of its perturbative construction.
Explicitly one has to prove the following properties 
in order to bring it into the predictive power, which the  Standard
Model
is expected to have:

\begin{itemize}
\item The Green's functions of the theory are {\it uniquely} determined as 
functions of a finite (small) number of free parameters 
to all orders of perturbation theory.
This property is called renormalizability.
\item  
The physical 
scattering matrix constructed from the Green's functions is unitary and
gauge parameter independent.
In particular these properties ensure
 a probability interpretation of S-matrix elements
 and  guarantee at the same time that unphysical particles are
cancelled in physical scattering processes. Only then the theory has indeed a
physical
interpretation.
\item  It has to be shown, that
the theory is in agreement with the experiments by calculating
 different processes as accurately as possible.
\end{itemize}

In the present lectures we only treat the first point,
 the unique construction of
the Green's functions to all orders of perturbation theory.
 We want to point out, that unitarity and gauge
parameter independence of the S-matrix  are not rigorously derived
 in the Standard Model by now, but are commonly assumed to hold.
However, its analysis includes  the  important problem of 
unstable particles, whose solution will have far 
reaching consequences in phenomenological applications (see for example
 \cite{BEBE96}).

 Renormalizability of gauge theories has been first shown  in the framework
of dimensional regularization
\cite{HOVE72a,HOVE72b}.  One has used that dimensional
regularization is an invariant scheme for gauge and BRS invariance,
 respectively, as long as parity is conserved. In this scheme
it has been proven that all the
divergencies can be absorbed into invariant counter terms to the coupling, the
field redefinitions and the masses 
of the classical action. This method implies the unique construction
of the Green's functions. These proofs are  not applicable to the Standard Model,
since there parity is broken. It is also well-known, that the group structure
of the Standard Model
allows the presence of anomalies. For this reason an invariant scheme
is very likely not to exist.
 The algebraic method of renormalization provides a 
proof of renormalizability also in such cases where an invariant scheme
does not exist. It gives in a scheme-independent way the symmetry
relations of   finite Green's functions to all orders.

The  algebraic method has been applied to gauge theories with semi-simple
gauge groups \cite{BRS75,BRS76}.
 Necessary prerequisite for the algebraic method to work was
the discovery of the BRS symmetry \cite{BRS75,TYU75}
named after Becchi, Rouet and
Stora. In its functional form BRS symmetry  is called 
the Slavnov-Taylor identity. This identity
 is the defining symmetry of gauge theories
 in renormalizable and Lorentz
invariant gauges and includes the gauge-fixing action and the
action of the Faddeev-Popov ghosts. 

To gauge theories with non-semisimle groups the algebraic method has
been
applied in \cite{BABE78}. In particular this paper includes an 
 investigation of
the anomaly structure and an investigation of the instability of 
abelian factor groups, but the authors do not consider massless particles and
do not care about physical normalization conditions. 
The Green's functions of  the electroweak Standard Model  and its 
defining symmetry 
transformations  are constructed   in \cite{KR98} by algebraic
renormalization to all orders. 
In this paper we have given also special attention  to on-shell
 normalization conditions and to a careful analysis of free parameters.
In the present lecture we will give an introduction to this 
construction: In the first lectures, section 2,
we construct the classical action as an $SU(2) \times U(1)$ gauge theory.
Special attention is paid to the uniqueness of the action and transformations
and their algebraic  characterization. In section 3 we introduce the 
renormalizable gauges, BRS symmetry and Faddeev-Popov ghosts. Finally we 
summarize the defining symmetry transformations of the tree approximation
in functional form, the Slavnov Taylor identity, the Ward identities of
rigid symmetry and the local abelian Ward identity. In the last lecture,
section 4, the construction to all orders by the algebraic method
is outlined. In particular we present the basic ingredients of
the algebraic method, namely scheme dependence of counterterms and
the action principle.
In Appendix A we collect the important formulae of the tree approximation:
the classical action and the symmetry transformations of the Standard Model.
The exercises that were given during the lectures can be found in  Appendix B.

Since we  assume in these lectures, that the reader has a basic
knowledge about quantum field theory and renormalization, we 
give  a few books and reviews separated from the usual references,
 which introduce the foundations of
perturbative  quantization and renormalization. 
The books and reviews that we have  selected  are mostly close to our
presentation and these lectures continue the methods presented therein
to the Standard Model of electroweak interactions.


\newpage

\newsection{The classical limit of the Standard Model}
\newsubsection{Particle content of the Standard Model}

The particles of the Standard Model are divided into groups according
to their particle properties, as spin and electric charge. 
The first group consists of particles with spin $\frac 12 $, the fermions.
The group of fermions has two  subgroups, the  leptons and quarks. 
Whereas leptons only participate in weak interactions,
the quarks   interact by weak and  strong interactions. Accordingly all quarks
are colour vectors. Strong interaction is described by  $SU(3)$ colour 
gauge theory, weak and electromagnetic by a $SU(2) \times U(1)$ gauge theory,
so that the complete Standard Model is a $SU(3) \times SU(2) \times U(1)$
gauge  theory. In these lectures strong interaction will
not be taken into account, so we restrict ourselves in treating
the $SU(2) \times U(1)$ gauge theory of electroweak interactions and
consider colour $SU(3)$ as a global symmetry. (We 
come back to this point at the end of this subsection.)
Quarks and leptons are also distinguished by their electric 
charge: There exist two types 
leptons, charged leptons $e, \m, \tau$ with electric charge $Q_e = -1 $
and the neutral neutrinos $\nu_e, \nu_\mu , \nu_\tau$.
 Up-type quarks, the $u$p, $c$harm, and $t$op,
have charge $Q_u=\frac 23$,
down type quarks, the $d$own, $s$trange and $b$ottom have
electric charge $Q_d = - \frac 13$.
fermions in the Standard Model are furthermore arranged into three families
according to the following scheme:
\begin{equation}
\begin{array}{ccc}
\qquad \nu_e \qquad& \qquad \nu_\mu \qquad & \qquad \nu_\tau \qquad\\
\qquad e \qquad& \qquad \mu \qquad & \qquad \tau \qquad\\
\qquad u \qquad& \qquad c \qquad & \qquad t \qquad\\
\qquad d \qquad& \qquad s \qquad & \qquad b \qquad
\end{array}
\end{equation}
In the following we only consider the
first generation of fermions ($e, \nu_e, u, d$). In particular we disregard
any mixing effects between different generations. 
In generality  mixing between three families leads to
CP violation via the Cabibbo-Kobayashi-Maskawa matrix \cite{CKM72}, 
which makes proving the renormalizability
 more difficult.
\par
The second group of particles consists of  the vector  bosons, which are 
particles with spin 1.
The 
gauge bosons of electroweak interactions are the photon ($A_\mu$), 
the $Z$-boson ($Z_\mu$) and the
$W^\pm$-bosons ($W^\pm_\mu$). The photon and $Z$-boson are neutral, 
$W^\pm$-bosons have electric charge $+1$ and $-1$, respectively.
The full Standard Model in addition contains
eight gluons of strong interactions, which are not considered in the course of
these lectures.
 The photon is massless and  couples to all the
electric charged particles, in particular it couples also to the charged
bosons of electroweak interactions. The other three bosons are massive, which
makes the weak interactions important only at small distance scales.
The weak force is responsible for the decay of the muon and the
$\beta$-decay of the neutron:
\equ{
\begin{array}{ccc}
\epsfig{file = 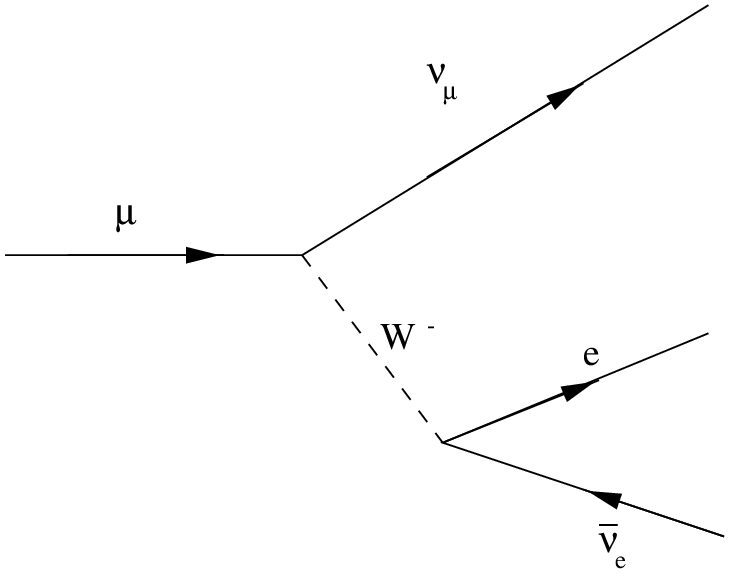, width = 4cm, angle = 270}
& &
\epsfig{file = 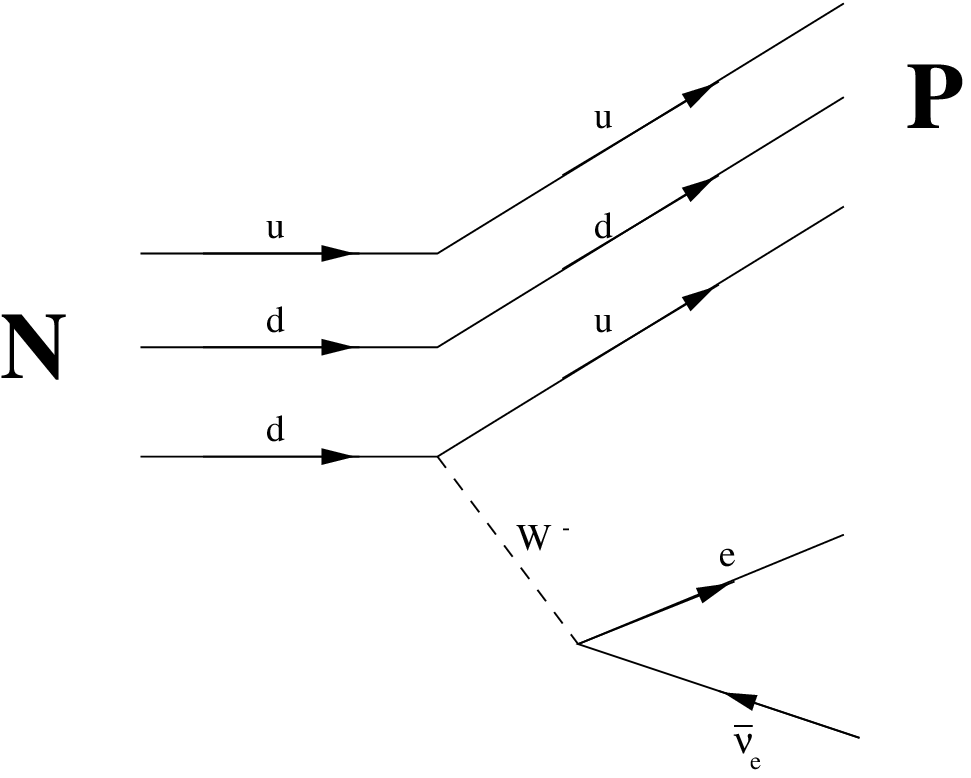, width = 4cm, angle = 270}
\nonumber \\
\mu \longrightarrow e + \bar{\nu}_e + \nu_\mu,
& &
n \longrightarrow p + e + \bar{\nu}_e. 
\end{array}
}
\par
In the  Standard Model there is one scalar particle, which has spin 0 and
is neutral with respect to electric charge. It is called the Higgs boson.
In the theoretical prescription of electroweak interactions
it is needed to give masses to the vector
bosons  and to the fermions  in agreement with $SU(2) \times U(1) $ gauge
symmetry \cite{HIG64,BRO64}.
 However, the Higgs particle
 has not been observed until now. 
All the particles of the Standard Model 
 with  their  properties are listed in
table \ref{tab: Standard Model}. 
\par
The following remarks should be made about the exclusion of the strong interactions in these notes. 
The QCD coupling constant is by far the largest coupling in the full Standard Model for energy scales that are reached in experiments now and in the near future. 
This means that the QCD corrections are more important in phenomenological applications than the electroweak corrections:
for the precision tests of the physics at the $Z$ resonance at LEP1 the
following calculations were needed: 3 loop QCD corrections, 
1 loop electroweak corrections with 1 loop QCD corrections on 
top of that; 2 loop electroweak was of minor importance. In the lectures we disregard
 QCD corrections, but this does not mean that QCD corrections factor out of
the calculations of scattering matrix elements. Indeed 
 this is not at all the case, as can been seen from the following two diagrams:
\begin{center}
\epsfig{file = 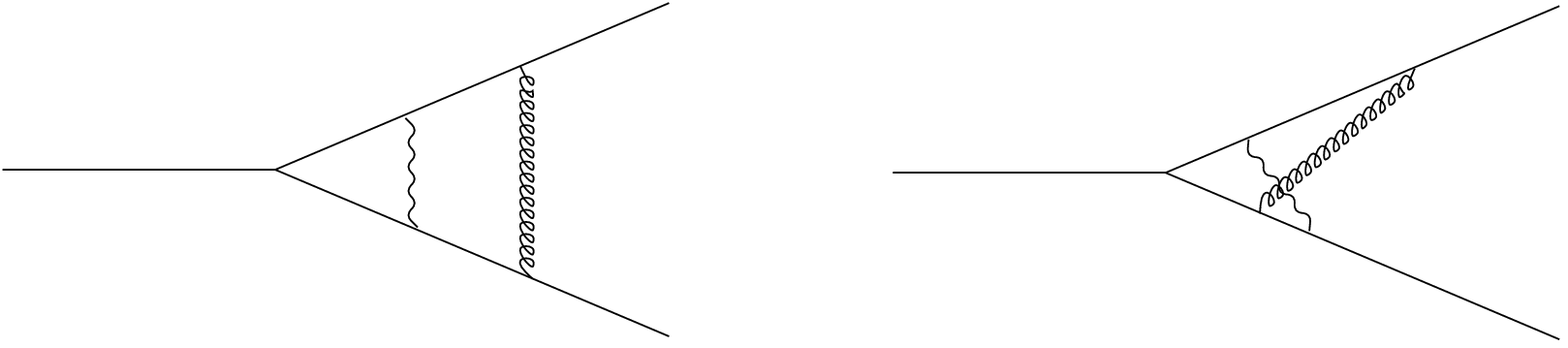, width = 3cm, angle = 270}
\end{center}
The diagram on the left can be understood as first an electroweak correction is applied and then a QCD correction, so this diagram is factorable.  
But this analysis can't clearly be done to the diagram on the right.
\par
Disregarding QCD corrections in the proof of renormalizability is
justified, since the colour group $SU(3)$ is unbroken in the
Standard Model and its generators of global symmetry do
not mix with the one of 
$SU(2) \times U(1)$ symmetry. In contrast to this renormalizing only
$SU(2)$ instead of the $SU(2) \times U(1)$ symmetry means to treat
a different theory, since the symmetry of the electroweak model is
spontaneously broken in such a way
 that the abelian subgroup cannot be factorized out anymore. 
(The electromagnetic charge operator is a linear combination of a genuine
abelian operator $Y$ and the third component of weak isospin.)  
So if we  understand the renormalization there, then the inclusion of QCD 
requires just the addition of an unbroken local symmetry, whose
global symmetry is conserved by construction.
\begin{table}
\begin{center}
\begin{tabular}{|cc|c|c|c|c|c|}
\hline
Type & Spin &Particle  & Charge & Mass$(MeV)$ & \\
\hline
& & & & & \\
& &           $g  
    $&$    0 $&$ 0          $&$               $\\
& &           $\gamma $&$    0 $&$ 0          $&$               $\\
Vector& $1$ & $Z      $&$    0 $&$ 91.1884 GeV$&$ \pm 0.022 GeV $\\
& &           $W^{\pm}$&$\pm 1 $&$ 80.26 GeV   $&$ \pm 0.16       $\\
& & & & & \\
\hline
& & & & & \\
& & $\nu_{e}   $&$   $&$ < 7.3 eV   $&$  CL = 90\%    $\\
& & $\nu_{\mu} $&$  0$&$ < 0.17     $&$  CL = 90\%    $\\
& & $\nu_{\tau}$&$   $&$ < 24       $&$  CL = 95\%    $\\
Leptons & $\frac{1}{2}$ & & & & \\
& & $e         $&$    $&$ 0.51099907 $&$ \pm 0.0000015 $\\
& & $\mu       $&$ -1 $&$ 105.658389 $&$ \pm 0.000034  $\\
& & $\tau      $&$    $&$ 1777.0     $&$ +3.0, -2.7    $\\
&  & & & & \\
\hline
& & & & & \\
& & $u$&$            $&$ 5.6        $&$ \pm 1.1       $\\
& & $c$&$ \frac{2}{3}$&$ 1350       $&$ \pm 50        $\\
& & $t$&$            $&$ 180 GeV    $&$ \pm 12 GeV    $\\
Quarks& $\frac{1}{2}$ & & & & \\
& & $d$&$            $&$ 9.9        $&$ \pm 1.1       $\\
& & $s$&$-\frac{1}{3}$&$ 199        $&$ \pm 33        $\\
& & $b$&$            $&$ 5 GeV      $&$ \pm 1 GeV     $\\
& & & & & \\
\hline
& & & & & \\
Higgs& $0$ &  $H      $&$    0 $&$>58.4 GeV   $&$ CL = 95\%     $\\
& & & & & \\
\hline
\end{tabular}
\end{center}
\caption[]{Properties of the particles which make up the electroweak
Standard Model \cite{PDG96}.}
\label{tab: Standard Model}
\end{table}


\newsubsection{The construction of gauge theories}
\newsubsubsection{The free Dirac equation}

We start our discussion with the Dirac equation of  free fermions:
\equa{
\label{eq: Direqu}
(i \gamma^\mu \pa_\mu -m_f) f &= 0, \\
\bar{f} (i\gamma^\mu {\overset{\leftarrow}{\pa}}_\mu + m_f) &= 0.
}
Here $f$ is a four component Dirac spinor and $\bar{f} = f^\dag \gamma^0$  the
adjoint spinor. 
 $\gamma^\mu$ are the Dirac matrices which form a Clifford algebra,
\equ{
\lbrace \gamma^\mu, \gamma^\nu\rbrace = 2 g^{\mu\nu} \Identity,
}
with the metric $g^{\mu\nu} = (1, -1, -1, -1)$.
For a set of fermions $\{f\} $ the
 equations of motion can be derived from the action 
\equ{
\Gamma_{Dirac}^{bil} = \sum_f \intd \bar{f} (i  \pa\slashed -m_f)f
\label{eq: Free Dirac}
}
by the classical principle of least action,
i.e.
\begin{equation}
\label{leastaction}
\delta \Ga_{Dirac}^{bil} = 0 .
\end{equation}
Here we have defined $\pa\slashed = \gamma^\mu \pa_\mu$ and
 The summation
is understood over  all fermions in question, as for example
$ f = \nu, e, u, d$, if we include
the first fermion generation of the Standard Model.
\par
To evaluate the variation of a functional we  introduce -- for later use --
the functional derivative. 
 A functional $F$ assigns to functions $u^i$ of some function
space $B$ a complex number: $F: u^i\in B \rra F[u]\in \Bbb{C}$. In
generalization of ordinary variations of functions the
variation $\delta F$ of the functional is given by:
\equ{
\delta F = \sum_j \int {d}^4x \fderivative{F}{u^j(x)} \delta u^j(x).
}
The $\fderivative{}{u^j(x)}$ denotes the functional derivative with
respect to $u^j$ at $x$, which is defined by the usual properties of a derivative together with:
\equ{
\fderivative{u^i(x)}{u^j(y)} = \delta_{ij} \delta^4(x-y).
}
If we apply functional variation for determining the
variation of the Dirac action $\Ga_{Dirac} ^{bil}$ (\ref{eq: Free Dirac}) 
we get:
\equ{\label{varDir}
\delta \Gamma_{Dirac}^{bil} = \sum_f \int {d}^4x 
\left\lbrace
\delta \bar{f}(x)
\fderivative{\Gamma_{Dirac}^{bil}}{\bar{f}(x)} +
\fderivative{\Gamma_{Dirac}^{bil}}{f(x)} 
\delta f(x)
\right\rbrace \\
= \sum_f \int {d}^4x
\left\lbrace
\delta \bar{f}(x) (i \pa\slashed-m_f) f(x) + \bar{f}(x)(i\overset{\leftarrow}{\pa\slashed} +m_f)
\delta f(x) 
\right\rbrace  \nonumber
}
Since $\delta \bar{f}$ and $\delta f$ are independent variations,
the
Dirac equation of the  fermions and adjoint fermions follow 
from the  principle of least action (\ref{leastaction}).
Note that in (\ref{varDir})
spinor variation is applied from the right and
variation with respect to the adjoint spinor from the left for consistency.

\newsubsubsection{The electromagnetic interaction}

Noether's theorem tells that current conservation and charge conservation
is connected with the symmetries of the action.
For this reason
we now want to look for symmetries of the Dirac action. 

Of course the Dirac action is invariant,
\begin{equation}
\label{inv}
\Ga_{Dirac}^{bil} (\bar f , f) = \Ga_{Dirac}^{bil} (\bar f ' , f'),
\end{equation} 
if we
redefine  all fermions by 
a single phase factor:
\equ{\label{eq: ab group}
f \rra f ' = e^{-i\ve q_f} f \mbox{ and } \bar{f} \rra 
\bar f ' = e^{i\ve q_f}\bar{f} .
}
Here $\ve$ denotes a real parameter, $q_f $ are numbers associated to the
different fermions.
These transformations form an abelian group for arbitrary $q_f$
as long as no further symmetries
are considered.
Assigning to $q_f$ the electric charge $Q_f$ of the respective fermion 
the transformation (\ref{eq: ab group}) is related to electric current
and finally charge
conservation. We could also assign 
\begin{equation}
q_f = \left\{\begin{array}{ccl} 1,& \quad \mbox{if} \quad &f = \nu, e \\
                                0 ,& \quad \mbox{if} \quad &f = u, d 
             \end{array} \right.
\qquad \mbox{or} \qquad
q_f = \left\{\begin{array}{ccl} 0,& \quad \mbox{if} \quad &f = \nu, e \\
                                1 ,& \quad \mbox{if} \quad &f = u, d 
             \end{array} \right.
\end{equation}
Then the transformation corresponds to lepton or baryon family number
conservation. In the following we restrict ourselves to
electromagnetic transformations ($q_f \equiv Q_f$),
 since -- 
in
contrast to lepton and baryon number symmetry --
electromagnetic symmetry  is gauged in the electroweak
Standard Model. 
            
From now on we do not consider the group transformations,
but expand the exponential function for small $\ve$
and only  consider the corresponding infinitesimal transformations.
(So we restrict ourselves to the Lie algebra of the Lie group.). 
The infinitesimal transformations of (\ref{eq: ab group}) have the form:
\equ{
\d^{em}(\ve)f = \ve \d^{em} f= -i \ve Q_f f \qquad\mbox{and} \qquad 
\d^{em}(\ve)\bar{f} = \ve \d^{em}\bar f= i \ve  Q_f \bar{f}.
\label{eq: Global Gauge}
}
If we apply these  transformations to the bilinear action 
\eqref{eq: Free Dirac}, we find the infinitesimal version of \eqref{inv}
\equ{
\d^{em}(\ve) \G_{Dirac}^{bil} = \ve {\cal W}_{em} \G_{Dirac}^{bil} = \ve\int
{d}^4x 
 {\bf w}_{em}(x) \G_{Dirac}^{bil}
= 0.
}
In this equation we have introduced the functional operators
which correspond to electromagnetic transformations:
 ${\bf w}_{em}$ is the functional operator of the infinitesimal
{\it local} electromagnetic transformations,
\equa{
\label{eq: wem}
\ve {\bf w}_{em}(x)& = \sum_f\left(\d^{em}\bar f(x) \fderivative{}{\bar f(x)} +
\fderivative{}{{f}(x)}
\d^{em} {f}(x)\right)\\
& = \sum_f \left( i \varepsilon Q_f \bar f(x) \fderivative{}{\bar f(x)} - i
\varepsilon Q_f
\fderivative{}{{f}(x)}
 {f}(x)\right), \nonumber
}
and ${\cal W}^{em}$ is the one of {\it global} or {\it rigid} 
electromagnetic transformations:
\equ{
{\cal W}_{em} = \int d^4x \;{\bf w}_{em}(x).
}
Now we are able to derive immediately Noether's first theorem:
Since the Dirac action is invariant under global transformations, the
local transformations can be only broken by a total derivative:
\equ{
\label{eq: currcons}
 {\bf w}_{em}(x) \G_{Dirac}^{bil} =- \pa^\m j^{em}_\m(x) .
}
The electromagnetic current, 
\equ{\label{jem}
j^{em}_\m(x) = \sum_f Q_f \bar{f}(x) \g_\m f(x),
}
 is seen to be conserved by applying the
equations of motions
\begin{equation}
{\delta \Ga_{Dirac}^{bil} \over \delta f} = 0
 \qquad \mbox{and} \qquad
{\delta \Ga_{Dirac}^{bil} \over \delta \bar f } = 0
\end{equation}
on the left-hand-side of eq.\ (\ref{eq: currcons}).
According to Noether's second theorem we are able to gauge the symmetry by
coupling the electromagnetic current to a vector field $A_\mu$:
\equ{ \label{eq: curr coup}
\G_{matter} =\G_{Dirac}^{bil} -\intd\, e\, j^{em}_\m A^\m .
}
Here  $e$ is the electromagnetic coupling constant
and we may interpret $A_\m$ as the electromagnetic vector potential.
The action \eqref{eq: curr coup} is indeed invariant 
under local gauge transformation as it stands if we assign
the transformation
\equ{
\d^{em}(\ve(x)) A_\m = \frac{1}{e} \pa_\m \ve(x)
}
as abelian gauge transformation to  $A_\m$. 
The local gauge invariance of the new action $\G_{matter}$ can be
expressed in functional form:
\equ{\label{eq: gaugeward}
\left(
{\bf w}_{em} - \frac{1}{e} \pa^\m \fderivative{}{A_\m}
\right) \G_{matter} = 0.
}
To interpret $A_\mu$ as a dynamical physical field, namely the photon,
 it  needs to have a kinetic action as well:
The free field action
\equ{\label{eq: gauge act}
\G_{gauge} =  - \frac{1}{4}\intd F^{\m\n} F_{\m\n}
}
with the antisymmetric field strength tensor 
\begin{equation}
F_{\m\n} = \pa_\m A_\n - \pa_\n A_\m
\end{equation}
is invariant under the local transformation \eqref{eq: gaugeward}.

 If we put the invariant actions \eqref{eq: curr coup} and
\eqref{eq: gauge act} together, we arrive at the
action of classical electrodynamics:
\equ{
\G_{em} = \intd \left( \sum_f \bar{f}(i\pa\slashed-m_f)f - e j^{em}_\m A^\m -
\frac{1}{4}  F^{\m\n} F_{\m\n} \right).
\label{eq: Classical EM}
}
The field equations that follow from the classical 
 action \eqref{eq: Classical EM} are:
\equ{
\pa^\m F_{\m\n} = e j^{em}_\n, \\
(i\pa\slashed - m_f)f = e Q_f \g^\m A_\m f. \nonumber
}
 Gauge invariance of the electromagnetic
 action can be expressed by the functional
identity:
\equ{
\left(
{\bf w}_{em} - \frac{1}{e} \pa^\m \fderivative{}{A^\m}
\right)
\G_{em} = 0.
\label{eq: functional gauge invariance}
}
In perturbation theory this equation will be continued  to
the electromagnetic Ward identity, which plays an important role for
the definition of  Green's functions in higher orders (see $[$Q4,R5$]$).
For this reason one has to note that the most general solution of
the Ward identity for local actions with dimension less than or equal
four is given by \eqref{eq: Classical EM} up to the field and coupling
redefinitions:
\begin{eqnarray}
f \to z_f f &\qquad & \bar f \to z_f \bar f \\
A \to z_A A & \qquad& e \to z_A^{-1} e \nonumber
\end{eqnarray}
Note that  these redefinitions  leave the operator in
 \eqref{eq: functional gauge invariance} 
invariant.

One final remark about the dimensions of the fields $[$Q3$]$: 
If one scales the coordinates by: $ x^\m \rra e^{-\l} x^\m$, then a field
$B$ may scale as $B \rra e^{\a\l} B$.
The number $\a$ is called the (naive scale or mass) dimension of a field $B$.
This leads to the following table.
\begin{center}
\begin{tabular}{|c|c|c|c|c|c|}
\hline
Field & $x^\m$& $\pa_\m$ & $f$ & $\bar{f}$& $A_\m$ \\
\hline
Dim  & -1      & 1       & $\frac{3}{2}$ & $\frac{3}{2}$ & 1 \\
\hline
\end{tabular}
\end{center}

\newsubsubsection{Beyond the Fermi model}

In the previous subsection the electromagnetic interaction was
discussed, we now turn to the weak interactions. 
In this discussion we take all fermions to be massless to start with.
Low energy experiments, like the decay of neutrons or muons, suggested
the existence of charged currents:
\equa{
\label{eq: CC}
J^+_\m &= \frac{1}{2\sqrt{2}} \Bigl(\bar{e} (\Identity - \g_5) \nu +
\bar{d} (\Identity -\g_5) u \Bigr), \\
J^-_\m= (J^+_\m)^\dag &= \frac{1}{2 \sqrt{2}} \Bigl(
\bar{\n} (\Identity - \g_5) e +
 \bar{u} (\Identity -\g_5) d\Bigr). \nonumber 
}
 Their interaction could be described by the effective Lagrangian:
\equ{
\cL _{eff} = -4 \sqrt 2  G_\mu J^{+\m} J^-_\m,
}
where the coupling constant $G_\mu$ is called the Fermi constant. 
This is  the Fermi model of weak interactions, which worked
phenomenologically quite well for describing the low energy processes
of weak interactions.
In  the charged currents \eqref{eq: CC} we have introduced
the $\g_5$ matrix, which  is defined by  
\equ{
\g_5 = i\g^0\g^1\g^2\g^3,
}
and has the following properties:
\equ{
\lbrace \g^\m, \g_5 \rbrace = 0, \qquad 
\mbox{and} \qquad
(\g_5)^2 = \Identity .
}
Out of the $\g_5$-matrix two projection operators can be constructed,
\equ{
P^L = \frac{\Identity -\g_5}{2}, \qquad
P^R = \frac{\Identity +\g_5}{2},
}
with the properties:
\begin{eqnarray}
P^L+P^R = \Identity , & \qquad  & 
P^R P^L = P^L P^R = 0 
\\ P^L P^L = P^L , &\qquad  &
P^R P^R = P^R. \nonumber
\end{eqnarray}
Next we introduce the notation of left- and right-handed fermions:
\equ{\label{leftright}
f^L = P^L f = \frac{\Identity -\g_5}{2} f, \qquad
{f^R}  = P^Rf =  \frac{\Identity +\g_5}{2} f,
}
with the Dirac conjugates $\overline{{f}^R} = \Bigl(f^L\Bigr)^\dag \gamma_0$
 and 
$\overline{{f}^L} = \Bigl(f^R\Bigr)^\dag \gamma_0$. 
 To the charged currents only the left-handed fermions
contribute. 
The left-handed fermions can be combined into doublets, one doublet for the
leptons and one doublet for the quarks:
\equ{
F^L_l = 
\begin{pmatrix}
\n^L \\ e^L
\end{pmatrix}
\mbox{ and }
F^L_q = 
\begin{pmatrix}
u^L \\ d^L
\end{pmatrix}
.
}
The charged currents 
(\ref{eq: CC}) can then be cast in the explicit $SU(2)$ form:
\equa{
J^+_\m  &=  \overline{F^L_l} \g_\m \frac{\tau_-}{2} F^L_l +
\overline{F^L_q} \g_\m \frac{\tau_-}{2} F^L_q, \\ 
J^-_\m  &=  \overline{F^L_l} \g_\m \frac{\tau_+}{2} F^L_l +
\overline{F^L_q} \g_\m \frac{\tau_+}{2} F^L_q, \nonumber
}
where 
\equ{
\tau_+ = 
\begin{pmatrix}
0 & \sqrt{2} \\
0 & 0
\end{pmatrix} 
, \qquad
\tau_- = 
\begin{pmatrix}
0 & 0 \\
\sqrt{2} & 0
\end{pmatrix}
\qquad \mbox{and} \qquad
\tau_3 = 
\begin{pmatrix}
1 & 0 \\
0 & -1
\end{pmatrix}.
}
These matrices form a representation of the $SU(2)$ algebra with the
commutation relation:
\begin{equation}
\label{su2alg}
\bigl[ \tau_\alpha , \tau _\beta \bigr]  =  2 i \epsilon _{\alpha
\beta \gamma} \tau_\gamma^T 
\end{equation}
 The structure constants $\epsilon _{\alpha\beta \gamma}$ are
completely antisymmetric in all three indices and  $\epsilon _{+-3} = -i $. 

So we see a $SU(2)$ representation structure emerging for the charged
currents of weak
interactions. 
From current algebra  one also expects the existence of  a
neutral current $J^3_\mu$ which corresponds to the generator $\tau_3$:
\equ{
J^3_\m  = \bar{F^L_l} \g_\m \frac{\tau_3}{2} F^L_l +
\bar{F^L_q} \g_\m \frac{\tau_3}{2} F^L_q.
}
(Since in $J_3^\m$ only left-handed fermions occur, it is not possible to
identify this current with the electromagnetic current.)
As in the case of electromagnetic interactions, also the weak currents
can be identified as conserved currents when acting with
the following functional $SU(2)$-generators 
\equ{\label{eq: w^fermion}
{\bf w}_\a(x) = i \sum_{\d = l, q} \left( \overline{F^L_\d}(x)
 \frac{\tau^T_\a}{2}
\fderivative{}{\overline{F^L_\d}(x)} -
 \fderivative{}{F^L_\d(x)} \frac{\tau^T_\a}{2} F^L_\d (x)\right)
}
on the massless Dirac action $\Ga_{Dirac}^{bil}$ 
 \eqref{eq: Free Dirac} with $f = e, \nu , u, d $:
\equ{
{\bf w}_\a(x) \left.
\Ga_{Dirac}^{bil}\right|_{m_f = 0}  = - \partial^\mu J_\mu^\a(x).
}
Indeed 
we see that
the bilinear
action \eqref{eq: Free Dirac}  is invariant under  rigid
$SU(2)$ transformations as long as all fermion masses vanish:
\equ{
{\cal W}_\a \left. \G_{Dirac}^{bil} \right|_{m_f = 0} = 0, 
\qquad {\cal W}_\a = \intd  {\bf w}_\a(x) . 
}
The functional generators satisfy the local and global $SU(2)$ algebra:
\begin{equation}
\big[ {\bf w}_\a(x) , {\bf w}_\b(y) \big] = \delta^4 (x-y) \ve_{\a \b \g}
\tilde I _{\g \g' } {\bf w}_{\g'} (x)  
\end{equation}
\begin{equation}
\label{algferm}
\big[ {\cal W}_\a , {\cal W}_\b \big] = \ve_{\a \b \g}
\tilde I _{\g \g' } {\cal W}_{\g'}  
\end{equation}
The charge conjugation matrix $\tI_{\a \a'}$ 
\equ{\label{tI}
\tI = \left(
\begin{array}{cccc}
0 & 1 &0 &0 \\
1 & 0 &0 &0 \\
0&0 & 1 &0 \\
0& 0&0 & 1
\end{array}
\right) 
 }
removes various transpositions from the
formulae.

\newsubsubsection{$SU(2)\times U(1)$ gauge theory}

Now we note the following remarkable fact: If one subtracts the
electromagnetic charge operator $ {\bf w }_{em}$  (\ref{eq: wem})
from the third component of the
weak isospin ${\bf w} _3$, one finds a generator, denoted by ${\bf w}_4^Q$,
 which
commutes with all $SU(2)$ operators and consequently with the charge
operator:
\equ{
[{\bf w}_\a, {\bf w}_4^Q] = 0 \quad
\mbox{ with } \quad
{\bf w}_4^Q = {\bf w}_{em} - {\bf w}_3
\label{eq: w_4^Q}
}
Therefore the symmetry operators ${\bf w}_\a$ and 
${\bf w}_4^Q$ build a closed $SU(2) \times U(1)$ algebra and imply
current conservation of weak and electromagnetic currents, when applied
to the massless Dirac action:
\equ{
\bigl({\bf w}_{em} - {\bf w}_3\bigr) \left. \Ga_{Dirac}^{bil} \right|_{m_f = 0} =
- \partial^\mu ( j_\mu^{em} - J^3_\mu).
}
(The electromagnetic current $j_\mu^ {em} $ is defined in \eqref{jem}.)
 For the procedure of quantization
it is important to note that ${\bf w}_4^Q $ is not uniquely determined by  the 
characterization that it commutes with the $SU(2)$ operators:
any generator ${\bf w}_4$ is abelian with respect to  ${\bf w}_{\a}$
when it has the form:
\equ{
{\bf w}_4(x) =   i\frac{Y_W^l} 2 \biggl(\overline{{F}_l^L} (x)
\fderivative{}{\overline{{F}^L_l}(x)} - \fderivative{}{F_l^L(x)} 
F_l^L (x)\biggr) + 
i\frac{Y_W^q}2
 \biggl(\overline{{F}_q^L}(x)  \fderivative{}{\overline{{F}^L_q}(x)}
- \fderivative{}{F_q^L(x)}  F_q^L(x) \biggr)  \nonumber \\
+ \sum_f iQ_f \biggl(\overline{f^R} (x) \fderivative{}{\overline{f^R}(x)}
- \fderivative{}{f^R(x)} 
f^R (x)\biggr) 
\label{w4gen}
}
with arbitrary values of $Y_W^l$, $Y_W^q$ and $Q_f$. 
(This means there are $5$ linearly independent abelian operators in
 ${\bf w}_4$.)
Applying ${\bf w}_4(x)$ to
the massless Dirac equation all these symmetry operators are connected with
classically conserved currents. Since only the 
 electromagnetic symmetry is gauged, the parameters $Y_W^l$, $Y_W^q$ and $Q_f$
  are determined by the relation
\eqref{eq: w_4^Q}, which is the  functional form equivalent to the 
well-known Gell-Mann--Nishijima relation:
\equ{
Q = \frac{Y_W} {2} + T_3.
}
From (\ref{eq: w_4^Q}) one derives the following values for the
weak hypercharge of leptons and quarks 
\equ{
Y_W^l = -1 \qquad Y_W^q = \frac{1}{3}
}
and identifies
$Q_f$ with the electric charge of the respective fermions:
\equ{ Q_e = -1 \qquad Q_u = \frac 23 \qquad Q_d = - \frac 13 . }
\par
Having constructed the relevant symmetry transformations we are able
to proceed as in the case of abelian gauge theories, when we want to
construct
the gauge theory belonging to the conserved currents. We couple the currents
 $J_\mu^\pm$, $J_\mu^3$ and $J_\mu^4 \equiv j_\mu^ {em} - J_\mu ^3$ 
to the vector fields
 $W ^\mu_\pm$, $W^\mu_3$ and $W^\mu_4$ and enlarge the free field action
by these terms:
\equ{\label{Gamatter}
\G_{matter} = \left. \G_{Dirac}^{bil} \right|_{m_f = 0} - \intd \left( g_1
  J^4_\m  W_4^{\m} 
- g_2 \left( J^3_\m W_3^{\m} + J^+_\m W_-^{\m} + J^-_\m W_+^{\m}\right)\right).
}
Since the gauge group of the Standard Model is a direct product of two
 groups, the couplings ($g_2$ and $g_1$) of  $SU(2)$ and $U(1)$ are 
independent from each other. 
From the $SU (2)$ algebra of the
functional operators (\ref{algferm}) as well as from global invariance
\begin{equation}
{\cal W}_\a \Ga_{matter} \stackrel != 0 
\end{equation}
it is derived that the
 ${\bf w}_\a$'s have to be extended to  include the vector 
bosons.
If we now indicate the ${\bf w}$ we had on the fermions explicitly by ${\bf
w}^{fermion}$, we have now:
\equ{\label{wfermvec}
 {\bf w}_\a \to  {\bf w}_\a \equiv {\bf w}_\a^{fermion} + {\bf w}_\a ^{vector}
}
with
\equ{\label{eq: w^vector}
{\bf w}_\a^{vector}(x) = I_{\a\a'} W_\b^\m(x) \ve_{\b\g\a'}
\tI_{\g\g'} \frac{\d}{\d W^\m_{\g'}(x)} \qquad \hbox{and}
\qquad {\bf w}_4 ^{vector}(x) = 0,
}
where $\a = +,-,3,4$. The structure constants $\ve_{\a\b\g}$ are defined as 
in (\ref{su2alg})
but with $\ve_{a\b4} = 0$. The matrix $\tI $ is defined in (\ref{tI}).

Since the $SU(2)\times U(1)$ algebra
 uniquely determines the abelian transformation of
vectors to vanish, it is possible 
to  determine the charge of the vector bosons by looking at
\equ{\label{wemvector}
{\cal W}_{em}^{vector}  =
 {\cal W}_3^{vector} + {\cal W}_4^{vector} = -i \intd
\left( W_+^\m \frac{\d}{\d
W_+^\m} - W_-^\m \frac{\d}{\d W_-^\m}\right),
}
thus $W_\pm$ has got charge $\pm 1$.

With the functional operators ${\bf w}_\a$ (\ref{wfermvec})
gauge invariance  of $\G_{matter}$ (\ref{Gamatter})
is expressed in functional form by
the identities:
\begin{eqnarray}
\label{nagauge}
\Bigl(
{\bf w}_\a + \frac 1{g_2} \partial^\mu \tI_{\a\a'} \frac{\d}{\d W^\m_{\a'}}
\Bigr)
\G_{matter} & = & 0  \qquad \a = +,-,3\\
\Bigl(
{\bf w}_4^Q - \frac 1{g_1} \partial^\mu  \frac{\d}{\d W^\m_{4}}
\Bigr)
\G_{matter} & = & 0 . \nonumber
\end{eqnarray}
By introducing the covariant derivatives
\equa{
D_\m F_\d ^L &= \left( \p_\m - ig_2 \frac{\tau_\a}{2} W_{\a\m} +ig_1 \frac{Y^\d
_W}{2}
W_{4\m}\right) F^L_\delta, \qquad \d = l,q , \\
D_\m f^R & = \left( \p_\m +ig_1 Q_f W_{4\m}\right) f^R, \nonumber
}
the matter action of fermions can also be rewritten into the form:
\equ{
\label{eq: Gamatter}
\G_{matter} = \intd \left( \sum_{\d=l,q}\overline{{F}_\d^L} i\g^\m D_\m F_\d^L 
+ \sum_f\overline{{f}^R} i\g^\m D_\m f^R\right).
}
\par
Finally we have to add
 kinetic terms   for the gauge fields to the action 
in such a way that the gauge
invariance remains. The Yang-Mills action
\equ{
\label{eq: GaYM}
  \Gamma_{YM}  = 
  -\frac 1 4 \intd \Bigl( G_\alpha^{\mu\nu}\tilde{I}_{\a \a'}G_{\mu\nu \a'}
                          + F^{\mu\nu} F_{\mu \nu} \Bigr)
}
with  the abelian and  non-abelian field strength tensors
\equa{
  F^{\mu\nu} & =  \partial ^\mu W_4^\nu - \partial^{\nu} W_4^\mu \\
\nonumber
  G^{\mu\nu}_\a & =  \partial ^\mu W_\a^{\nu} - \partial^{\nu} W_\a^\mu
  +  g_2 \tilde{I}_{\a\a'} 
  {\epsilon}_{\a'\b \gamma} W^{\mu}_\beta W^{\nu}_\gamma ,\qquad
\a,\beta, \gamma = +,-,3
}
is the  properly normalized  solution of the functional
identities \eqref{nagauge} with dimension 4.

The complete action containing massless vector bosons and massless fermions is 
the sum of the matter and Yang-Mills action:
\equ{
\label{eq: Gasym}
\G_{sym} = \G_{YM} + \G_{matter}.
}
In the same way as the electromagnetic action (\ref{eq: Classical EM})
is characterized by 
electromagnetic gauge invariance (\ref{eq: functional gauge invariance}),
 $\Ga_{sym}$ is characterized up to field
and coupling redefinitions by the functional identities of $SU(2)\times
U(1)$ gauge symmetry 
\begin{eqnarray}
\label{nagaugeid}
\Bigl(
{\bf w}_\a + \frac 1{g_2}  \tI_{\a\a'} \partial^\mu\frac{\d}{\d W^\m_{\a'}}
\Bigr)
\G_{sym} & = & 0 \\
\Bigl(
{\bf w}_4^Q
 - \frac 1{g_1} \partial^\mu  \frac{\d}{\d W^\m_{4}}
\Bigr)
\G_{sym} & = & 0 .
\end{eqnarray}
Here the operators ${\bf w}_\a$ 
are the sum of fermion and boson functional operators (\ref{wfermvec})
defined
in (\ref{eq: w^fermion}) and (\ref{eq: w^vector}).
The abelian operator ${\bf w }_4 ^Q$
is defined by the relation (\ref{eq: w_4^Q}) and
the electromagnetic charge operator includes fermions (\ref{eq: wem})
and vector bosons (\ref{wemvector}):
\equ{
{\bf w}_4^Q = {\bf w}_{em} - {\bf w}_3 \qquad {\bf w}_{em} =
{\bf w}^{fermion}_{em}+{\bf w}^{vector}_{em}
}

\newpage

\newsubsubsection{Higgs mechanism and masses}

For deriving
 the $SU(2)\times U(1)$ gauge invariant action in the previous section
we have assumed that all fermions are massless.
In reality the charged leptons as well as the up-type and down-type
quarks are massive, i.e.~$m_e, m_u, m_d \neq 0$. 
The Dirac action (\ref{eq: Free Dirac})
in terms of left- and right-handed fields (\ref{leftright}) has
the following form:
\equ{\label{GaDiracmass}
\Ga^{bil}_{Dirac} = \intd \left( \sum_{f}(\bar{f^L} i \pa \slashed f^L
 + \bar{f^R} i \pa \slashed f^R ) +
\sum_{f = e, u, d}  m_f (\bar{f^R} f^L + \bar{f^L} f^R)  \right)
}       
 Applying the $SU(2)$ transformations (\ref{eq: w^fermion}) on the
free field action for massive fermions it is seen that the
mass terms break the global $SU(2)$ symmetry:
\begin{equation}
{\cal W}_\a \Ga_{Dirac}^{bil} =  i  \Delta_{\a} \equiv i \intd Q_{\a}(x)
\end{equation}
where
\begin{eqnarray}
\label{massbreak}
\Delta_+ & = & \intd \frac 1 {\sqrt 2} \biggl(m_u \bar {d^L } u^R - 
                                       m_d  \bar {d^R} u ^L 
                                       - m_e \bar {e^R} \nu ^L \biggr)    \\ 
\Delta_- & = &  \intd \frac 1 {\sqrt 2} \biggl(m_d \bar {u^L } d^R - 
                                       m_u  \bar {u^R} d ^L 
                                       + m_e \bar {\nu^L} e ^R   \biggr)  
\nonumber\\       
\Delta_3 & = &  \intd \frac 1 { 2} \biggl(m_u  (\bar {u^L } u^R - \bar {u^R }
                                       u^L)  
                                 - m_d  (\bar {d^L } d^R - \bar {d^R } d^L)
                                 - m_e  (\bar {e^L } e^R - \bar {e^R } e^L)
\biggr)
\nonumber
 \end{eqnarray}
Electric charge invariance, of course, is not broken by the mass terms:
\begin{equation}
{\cal W}_{em} \Ga^{bil}_{Dirac} = 0 \qquad \Longrightarrow
\qquad {\cal W}_4^Q \Ga^{bil}_{Dirac} = -i \Delta_3
\end{equation}
For including fermion masses and  vector boson masses in agreement
with $SU(2) \times U(1)$  gauge symmetry into the Standard Model,
 the  symmetry is spontaneously broken to
the electromagnetic subgroup.

In these lectures we present a construction of spontaneous symmetry
breaking which is purely algebraic and can be compared to the Noether
construction of gauge theories, which we have carried out in the last
sections. In contrast to the usual construction, which is presented in
the books on quantum field theory (see for example $[$Q3$]$), it does not
start from the symmetric theory, but from the bilinear massive Dirac action
of free fermions. Eventually, if one carries out the algebraic
characterization of the classical action in the course of algebraic
renormalization (see section~4.3), the computation is equivalent
 to the analysis
and construction presented here.

First we couple the breaking of $SU(2)$ transformations (\ref{massbreak}) to
scalars $\phi^\pm$
and $\chi$ in such a way that the breaking terms can be expressed as field
differentiations with respect to these scalars:
\equ{
\Gm \to \hat \Gamma^{mass}_{Dirac} \equiv
\Gm - \frac 2 v \intd \bigl( \phi^+ Q_- - \phi^- Q_+ -
                                i \chi Q_3 \bigr).
\label{eq: SU(2) invariant mass terms}
} 
and
\equ{
\label{shifttrans}
\Bigl({\cal W}_\pm  \mp i \frac v 2 \intd\frac {\d}{\d \phi^\mp } \Bigr)
\hat \Gamma^{mass}_{Dirac} 
\bigg|_{{{\phi^\pm = 0}\atop {\chi= 0}}} =   0   \\
\Bigl({\cal W}_3  -  \frac v 2 \intd\frac {\d}{\d \chi } \Bigr)
\hat \Gamma^{mass}_{Dirac} 
\bigg|_{{{\phi^\pm = 0}\atop{\chi= 0}}} =   0  \nonumber
}
Here $\Gm$ denotes the mass term of the Dirac action:
\equ{
\Gm =  \sum_{f = e, u, d} \intd \, m_f (\bar{f^R} f^L + \bar{f^L} f^R)  .
}
We assign quantum numbers to the scalar fields in such a way that the
enlarged action $\hat \Gamma^{mass}_{Dirac} $
(\ref{eq: SU(2) invariant mass terms}) is
CP invariant,
neutral with respect to electric charge  and hermitian, 
i.e.~the fields $\phi^\pm$ carry charge $\pm 1$  and transform under CP
according to $\phi^+ \stackrel {\mathrm CP} \rightarrow - \phi^-$  
and $\phi^- \stackrel {\mathrm CP} \rightarrow - \phi^+,$  and the field
$\chi$ 
is a neutral field which is CP-odd. 
(Global signs and normalization in
eq.~(\ref{eq: SU(2) invariant mass terms})
are chosen according to usual conventions.)
It is seen that  the transformation operators appearing in
(\ref{shifttrans}) are not yet algebraically closed, the commutation relations
yield e.g.
\begin{equation}
\Bigl[{\cal W}_\pm  \mp i \frac v 2 \intd\frac {\d}{\d \phi^\mp }, 
{\cal W}_3  -  \frac v 2 \intd\frac {\d}{\d \chi } \Bigr] 
= \pm i {\cal W}_\pm ,
\end{equation}
and on the right-hand-side the inhomogeneous contributions of the shift
are missing.
It is also seen that
  $\hat \Gamma^ {mass}_{Dirac} $ is not invariant under these
transformations at
$\phi^\pm, \chi \neq 0$. For this reason we have to enlarge the action as well
as the transformation operators in such a way that the action is invariant
under the enlarged   transformations and that the algebra closes in presence of
the inhomogeneous shifts. 

For proceeding we note that the breaking terms $Q_\pm(x)$
and $Q_3(x)$ together with the mass term
$Q_m(x)$
\equ{
Q_m = \frac 12 \sum_{f = e, u, d}  m_f (\bar{f^R} f^L + \bar{f^L} f^R)  
}
 can
be arranged into a $SU(2)$ doublet and its complex conjugate:
\equa{
\label{masstrans}
{\cal W}_\a {  - Q _+ \choose \frac 1 {\sqrt 2}(Q_m + Q _3) }
& = - i \frac {\tau^T _\a}2 { - Q _+ \choose \frac 1 {\sqrt 2}( Q_m +
 Q _3) } \\
{\cal W}_\a {   Q _- \choose \frac 1 {\sqrt 2} (Q_m - Q _3) }
& = + i \frac {\tau _\a} 2 {  Q _- \choose \frac 1 {\sqrt 2} (Q_m - Q _3) }
 \nonumber }
\equ{
{ - Q _+ \choose \frac 1 {\sqrt 2} (Q_m + Q _3) } ^*
={   Q _- \choose \frac 1 {\sqrt 2} (Q_m - Q _3) }
 }
From (\ref{masstrans}) one
reads off that one has  to introduce a further CP even scalar
field $H$,
when one wants to complete the enlarged mass action 
$\hat \Gamma^ {mass}_{Dirac} $ in such a way, that it is invariant under
$SU(2)$ transformations.
The Yukawa action 
\equa{
\label{Yuk1}
\Gamma_{Yuk}& \equiv 
\Gm - \frac 2 v \intd \bigl( \phi^+ Q_- - \phi^- Q_+ + H Q_m -
                                i \chi Q_3 \bigr) \\
            & =  \Gm - \frac 2 v
 \intd \left( {  Q _- \choose \frac 1 {\sqrt 2} (Q_m - Q _3) }^ T 
\cdot
{\phi^+ \choose \frac 1 {\sqrt 2} (H + i\chi) }  \right. \nonumber \\
& \phantom{\Gm + \frac 2 v \intd }\left.\qquad + 
{\phi^- \choose \frac 1 {\sqrt 2} (H - i\chi) }^ T
\cdot
{ - Q _+ \choose \frac 1 {\sqrt 2} (Q_m + Q _3) } \right) \nonumber
}
continues $\hat \Gamma^ {mass}_{Dirac} $ 
(\ref{eq: SU(2) invariant mass terms})  in a minimal way to a
$SU(2)$ invariant action:
\equ{
{\cal W}_\a \Ga_{Yuk} = 0 \qquad \Bigl[{\cal W}_\alpha , {\cal W}_\b \Bigr]
= \epsilon_{\a \b \ga} \tilde I_{\ga \ga'}{\cal W}_{\ga'} .
}
The transformation operators ${\cal W}_\a $ consist of the fermion, the
vector and the scalar transformation operators.
The latter ones are defined
to include  the shift which we have introduced
 for absorbing the  breaking terms of the masses  (\ref{shifttrans})
and in this way they are  uniquely determined by the construction:
\begin{equation} 
\label{eq: global functional operators}
{\cal W}_\a = {\cal W}^ {fermion}_\a+ {\cal W}^ {vector}_\a + {\cal W}^ 
{scalar}_\a
\end{equation}
with
\equa{\label{eq: W^scalar}
{\cal W}_\pm^{scalar} & = \intd \Bigl( \Phi  ^\dag (x)
\frac{i\tau_\mp} {2}
\frac{\d}{\d \Phi^\dag(x)} - \frac{\d}{\d \Phi(x)} \frac{i\tau_\mp}{2} 
\Phi(x) \mp i \frac v 2 \frac {\d}{\d \phi^\mp } \Bigr), \\
{\cal W}_3^{scalar} & = \intd \Bigl( \Phi  ^\dag (x)
\frac{i\tau_3} {2}
\frac{\d}{\d \Phi^\dag(x)} - \frac{\d}{\d \Phi(x)} \frac{i\tau_3}{2} 
\Phi(x)   - \frac v 2 \frac {\d}{\d \chi } \Bigr). \nonumber
 }
Here we have arranged the scalars into $SU(2)$ doublets and have introduced
the notation:
\equ{ 
 \Phi \equiv\left(
    \begin{array}{c}
      \phi^+(x)\\
    \frac  1{\sqrt 2}(H(x) + i\chi(x))
    \end{array}
  \right) 
\qquad  \Phi^ *  = \left(
    \begin{array}{c}
      \phi^-(x)\\
    \frac  1{\sqrt 2}(H(x) - i\chi(x))
    \end{array}
  \right).
\nonumber
} 
With 
$
\tilde \Phi = i \tau_2 \Phi^ *
$
it is  straightforward to calculate that the Yukawa action
(\ref{Yuk1}) can be written in the conventional form
\equa{
\G_{Yuk} = &   - \intd \sum
_{f=e,u,d} m_f (\bar{f}^R f^L + \bar{f}^L f^R) \label{eq: Yukawa}\\
&   - \, 
\frac {\sqrt 2} v\intd \left(
 m_e \bar{F}^L_l \Phi e^R + m_d \bar{F}^L_q \Phi d^R
+ m_u \bar{F}^L_q \tilde{\Phi} u^R  + \mbox{h.c.} \right)
\nonumber \\
 = & - \frac {\sqrt 2} v\intd \left(
 m_e \bar{F}^L_l (\Phi + \hbox{v}) e^R + m_d \bar{F}^L_q (\Phi +\hbox{v})  d^R
+ m_u 
\bar{F}^L_q (\tilde{\Phi} + i \tau_2\hbox{v})  u^R  + \mbox{h.c.} \right) .
\nonumber
}
Here $\hbox{v}$ denotes the shift in vector notation:
\equ{
\hbox{v} = \left( \begin{array}{c}
0 \\ v/\sqrt{2} \end{array}\right)
}

The Yukawa interaction is invariant not only  under spontaneously broken 
global
 $SU(2) \times U(1)$ transformations, but even under the local ones: 
\begin{eqnarray}
\label{invyuk}
\Bigl({\bf w}_\a + \frac 1 {g_2}\partial^\nu \tI_{\a \a '}
{\d\over \d W^\nu _{\a'}} \Bigr) \Ga_{Yuk} & =&  0, \qquad \a = +,-,3, 
 \\
\Bigl({\bf w}^Q_4 - \frac 1 {g_1} \partial^\nu 
{\d\over \d W^\nu_4} \Bigr) \Ga_{Yuk} & =&  0 .
\end{eqnarray}
 The operators
$ {\bf w}_\a $ are the non-integrated version 
of (\ref{eq: global functional operators}): 
\begin{equation}
\label{eq: full functional operators}
{\cal W}_\a \equiv \intd {\bf w}_\a (x)=
\intd \Bigl(
{\bf w}^ {fermion}_\a(x) +  {\bf w}^ {vector}_\a  (x) + {\bf w}^ {scalar}_\a(x)
\Bigr)
\end{equation}
with 
\equa{\label{eq: w^scalar}
{\bf w}_\a^{scalar}(x) & = \Bigl( \Phi(x) + \hbox{v} \Bigr)^\dag
\frac{i\tau^T_\a} {2}
\frac{\d}{\d \Phi^\dag(x)} - \frac{\d}{\d \Phi(x)} \frac{i\tau^T_\a}{2} \Bigl(
\Phi(x)  +
\hbox{v} \Bigr).
}
 The abelian operator ${\bf w}_4^Q$ is defined by eq.~\eqref{eq: w_4^Q}
${\bf w}_4^Q = {\bf w}_{em} - {\bf w}_3$ and the electromagnetic operator
includes also the charged scalars (see (\ref{chargecons})). Explicitly we find
\equ{
{\bf w}_4^{Q\,scalar} (x) = \Bigl( \Phi(x) + \hbox{v} \Bigr)^\dag 
\frac{Y_W^s i }{2}
\frac{\d}{\d \Phi^\dag(x)} - \frac{\d}{\d \Phi(x)} \frac{Y_W^s i }{2} \Bigl(
\Phi(x) 
+ \hbox{v} \Bigr)}
with
\equ{ Y_W^s =  1.
}
Since the symmetric action (\ref{eq: Gasym}) does not depend on scalars
it is trivially invariant also with respect to the 
spontaneously broken $SU(2) \times U(1) $ gauge transformations
 (\ref{nagaugeid}) with the enlarged local operators 
(\ref{eq: full functional operators}).


\par
Looking for the most general, local  action invariant under the
local spontaneously broken gauge transformations  \eqref{invyuk}
with mass dimension less or equal 4  we 
find in addition the  kinetic and potential terms of the scalars:
\equ{\label{eq: Gascalar}
\G_{scalar} = \G_{kin\; scalar} + \G_{pot\; scalar}.
}
They read in the conventional normalization:
\equa{\label{Gascalarpot}
\G_{pot\; scalar} & = -\intd  \l
 (\Phi^\dagger\Phi +  
 {\hbox{v}}^\dagger \Phi + \Phi^\dagger   {\hbox{v}})  ^2 , \\
\label{Gascalarkin}
\G_{kin \; scalar} & = 
-\intd \Bigl(D^\mu (\Phi + \hbox{v})
 \Bigr)^\dagger D_{\mu} (\Phi + \hbox{v})
}
with the covariant derivative
\equ{ 
D_\mu \Phi =
\bigl(\partial_\mu - i  ( g_2
 \frac{{\tau}_\a}{2} W_{\mu \a}  - g_1 \frac {1}2 
 W_{\mu 4} )\bigr) \Phi.
} 
(We have already omitted the invariant, which is  linear in the Higgs field,
$ \Phi^\dagger\Phi +  
 {\hbox{v}}^\dagger \Phi + \Phi^\dagger   {\hbox{v}}, $ from the action.)

The bilinear term of the potential gives the mass term for the scalars, i.e. 
\equ{
-\l (\Phi^\dag \hbox{v} + \hbox{v}^\dag \Phi)^2 = - \frac \l 2 v^2 (H -i\chi
+ H + i\chi)^2 = -\frac 12\,  4\l v^2 H^2.
}
So only the real, CP-even scalar $H$ gets a mass $m_H^2 = 4 \l v^2 $, the
scalars
$\chi$ and $ \phi^\pm$ are massless.
\par
We have seen that the fermion masses are generated by Yukawa
couplings to the Higgs doublets and the Higgs mass arises from 
the scalar potential. Eventually  also gauge bosons get mass via the covariant
derivative of the scalars.
Evaluating (\ref{Gascalarkin}) one gets for the gauge boson masses:
\equ{
\Bigl(D^\mu ( \hbox{v}) \Bigr)^\dagger D_{\mu} (\hbox{v})
= 
{g_2 ^2 v^2 \over 4} 
W^\mu_+ W_{\mu -}
+ 
\frac 12 {v^2 \over 4}
(g_2 W^\mu_3 + g_1 W^\mu_4 )^2.
}
Since the mass terms are non-diagonal in the vector fields, 
 the fields $W^\mu_\a$  are not the physical fields. Physical on-shell fields
are constructed,
 if one rotates the fields $ W^\mu_3$, $W^\mu_4$ by an orthogonal
matrix $O_{\a a } (\theta_W)$ over an
angle $\tan \theta_W = \frac{g_1}{g_2}$:
\begin{equation}
W^\mu_\alpha = O_{\a a} (\theta_W) V^\mu_a
 \quad \mbox{with} \quad 
O_{\a a} (\theta_W) = 
\left(\begin{array}{cccc}
1&0&0&0\\ 0&1&0&0\\ 0&0&\cw&-\sw\\ 0&0&\sw & \cw \end{array}\right) .
\label{eq: thetaW rot}
\end{equation}
 Then the mass term is diagonalized,
\equ{
{g_2 ^2 v^2 \over 4} 
W^\mu_+ W_{\mu -}
+ 
\frac 12 {v^2 \over 4}
(g_2 W^\mu_3 + g_1 W^\mu_4 )^2 =  M_W ^2 W^\mu_+ W_{\mu -} + \frac 12
M_Z^2 Z^\mu
Z_\mu,
}
and the masses   are determined to 
\equ{
M_W = \frac{g_2 v}{2}, \quad
M_Z = \frac{g_2 v}{2\cos \theta_W} \quad \mbox{and} \quad
M_A = 0.
}
It is important to note that
the kinetic terms of the vector bosons remain diagonal after an orthogonal
rotation. From now on we denote  
with $V_a = ( W_+, W_-, Z, A)$ the physical on-shell vector fields 
of the Standard Model.
\par
At this stage a few remarks on notation should be made. 
With the indices $\a, \b,\ldots$ we denote the $SU(2)\times U(1)$-indices:
 $+, -, 3, 4$.
On the other hand the indices $a, b, \ldots$ refer to the indices of the 
physical fields in the theory: $+, -, Z, A$. 
\par
Having given masses to the vectors the massless Goldstone bosons
$\phi^\pm , \chi$
become unphysical fields. This can be understood qualitatively as follows:
A massless vector boson has only 2 transverse polarizations.
A massive vector boson has one more, the longitudinal polarization. 
Before the symmetry breaking there were 4 vector bosons, each with 2
degrees of freedom and one Higgs doublet with 4 degrees of freedom.
After the symmetry breaking there is  left  one  scalar,
the  Higgs boson.
The other three degrees of freedom of the scalar doublet make the longitudinal
polarization of the vector bosons physical. 
So the total number of physical degrees of freedom has not changed.
Figuratively  one says that the Goldstone bosons ($\phi^\pm, \chi$) are 
eaten up by the vector
bosons for giving them masses. These results are obvious in the unitary
gauge, whose lowest component we construct in exercise 5.
\par
The full classical action is combined from the single invariant
4-dimensional actions (\ref{eq: Gamatter})(\ref{eq: GaYM})
(\ref{eq: Yukawa}) and (\ref{eq: Gascalar})
and is called the Glashow-Salam-Weinberg model:
\equ{
\G_{GSW} =  \G_{YM}+ \G_{matter} + \G_{scalar} + \G_{Yuk}.
\label{eq: tree SM}
}
This action is uniquely determined up to field and parameter redefinitions 
by spontaneously broken $SU(2) \times U(1) $ gauge transformations:
\begin{eqnarray}
\Bigl(
\label{nagaugeid2}
{\bf w}_\a + \frac 1{g_2} \partial^\mu \tI_{\a\a'} \frac{\d}{\d W^\m_{\a'}}
\Bigr)
\G_{GSW} & = & 0 \\
\Bigl(
{\bf w}_4^Q - \frac 1{g_1} \partial^\mu  \frac{\d}{\d W^\m_{4}}
\Bigr)
\G_{GSW} & = & 0 .
\end{eqnarray}
The local operators are defined in \eqref{eq: full functional operators}
as the sum of fermion, vector and (shifted) scalar operators.
\par
By now a lot of parameters are introduced. 
But not all of these are 
independent for there were a couple of relations between them. 
So one question which one should ask, is which of these are taken to be
fundamental. 
This fundamental set should be applicable in any order of perturbation
theory and should also characterize the particle properties of the model. 
It is therefore natural to take physical on-shell parameters as
fundamental. 
\par
The  free parameters we choose are
\begin{equation}
\label{onshellpar}
  M_W ,  M_Z ,  m_f ,  m_H  \quad \mbox{and} \quad  g_2 
\end{equation}
and the vectors are expressed in physical on-shell fields $V^\mu_a$.
The weak mixing angle $\theta_W$ is not taken to be fundamental, 
but is defined
 by the relation  \cite{SIMA80}:
\equ{\label{wma}
\cw = \frac{M_W}{M_Z}.
}

As an illustration let us calculate the interaction of the photon to
the electromagnetic current:
If we apply the orthogonal rotation $O_{\a a}(\theta_W)$
to the interaction of the gauge fields with the currents $\Ga_{matter}$
\eqref{eq: Gamatter},
 we get:
\equ{
- g_2 ( J_\mu^+ W_-^{\mu} + J_\mu^- W_+^{\mu}
 + J_\mu^3 W_3^{\mu}) -g_1
J_\mu^4 W_4^{\mu} = \\
- g_2 \Bigl( J_\mu^- W_+^{\mu} + J_\mu^+ W_-^{\mu}
 + \frac{1}{\cos
\theta_W} (J_\mu^3 + \sws j_\m^{em})Z ^\m \Bigr) - g_2 \sw j^\m_{em} A^\m.
\nonumber}
It is seen in an explicit form that the electromagnetic current couples
to the massless vector boson $A^\mu$ which is identified by this
property as  the
photon field. 
The same conclusion is derived by transforming the unphysical fields
$W^\mu_3$ and $W^\mu_4$ into the physical
on-shell fields  $Z^\mu$ and $ A^\mu$ 
in the functional operators of gauge transformations.
 There one reads off as well, that
the photon couples to the electromagnetic current and is the massless
field corresponding to the unbroken subgroup: 
\equ{\label{wemid}
\Bigl(g_2 \sw
 {\bf w}_{em} - \partial^\m \frac{\d}{\d A^\m} \Bigr) \G_{GSW} = 0.
}
For this reason we introduce the electromagnetic coupling constant
 $e = g_2 \sw$ as fundamental coupling of the electroweak Standard Model.
The QED-like on-shell parameters are then given by
\begin{equation}
\label{QEDonshellpar}
  M_W ,  M_Z ,  m_f ,  m_H  \quad \mbox{and} \quad  e 
\end{equation}
(For fixing the coupling  $e$ to its experimental value
a physical process has to be chosen, as it is 
for example Compton scattering at low energies or Bhabha scattering at
LEP energies.)

The on-shell parameters have been used by several groups as fundamental
parameters for calculating higher order processes in perturbation theory
\cite{PAVE79,CO79,SIMA80,FLJE81,AOHI82,BACH80,COLO83,BOHO86,HO90}.
We want to mention already here, that
in higher orders   it is crucial  for infrared
existence of Green's functions  to choose a parameterization,
which ensures that the photon propagator has
a pole at $p^2 = 0$.  Unfortunately it turns out that the QED-Ward identity
corresponding to the
functional
identity \eqref{wemid} cannot be proven in  perturbation theory. 
So the
photon will be  characterized   by the property of being
the massless vector boson, and not by its property of coupling to
the electromagnetic current. 

\newsubsubsection{Other (global) symmetries}

In the previous section we have looked at the consequence of the local
$SU(2)\times U(1)$ gauge invariance. 
We have built a phenomenologically acceptable model around this symmetry. 
It turns out that this model also has some extra global symmetries. 
\par
The un-quantized Standard Model action \eqref{eq: tree SM} is invariant
under the combined transformation CP and under  T. Parity is broken
in the fermion sector, since only left-handed fermions  contribute to
the charged currents.
(C denotes charge conjugation, P the parity reflection and T time
reversal.) 
We should stress here that this not true in the Glashow-Salam-Weinberg
model with three generations of fermions. 
In its most general form mixing between three families leads to
CP violation via the Cabibbo-Kobayashi-Maskawa matrix.
In that case the model is only invariant under the combined transformation
CPT.
\par
Two other symmetries of $\Ga_{GSW}$
are conservation of lepton and baryon numbers. For one generation the
corresponding symmetry operators are
\equa{
{\cal W}_l &= i \intd \Bigl(\bar{{e}} \frac{\d}{\d \bar{{e}}} 
+ \bar{\n^L} \frac{\d}{\d \bar{\n^L}} 
-  \frac{\d}{\d  {e}} {e} -  
    \frac{\d}{\d  {\n}^L} {\n}^L 
\Bigr), \\
{\cal W}_q &= i \intd \Bigl(\bar{u} \frac{\d}{\d \bar{u}}
+ \bar{d} \frac{\d}{\d \bar{d}}
-  \frac{\d}{\d {u}} {u}
-  \frac{\d}{\d {d}} {d} 
\Bigr).
}
These operators are abelian operators and
are included in the abelian operators we have found in generality in 
\eqref{w4gen}. These symmetries are not gauged in the Standard Model, but are
global symmetries in the classical theory,
\equ{\label{wlqGSW}
{\cal W}_l \G_{GSW} = 0 \quad \mbox{and} \quad
{\cal W}_q \G_{GSW} = 0,
}
and 
in higher orders of perturbation theory.
(Of course in principle these symmetries can also be made local
in the classical theory, but then one needs extra $U(1)$ gauge fields as
we demonstrate in 
 exercise 7.
In nature they are not observed, thus in the Standard
Model the lepton and baryon numbers are globally conserved quantum numbers.)

\newpage


\newsection{Gauge fixing and BRS transformations}
\newsubsection{Free field propagators and gauge fixing}

In the previous section we presented the $SU(2) \times U(1)$ gauge
 structure of the Glashow-Salam-Weinberg model.
This section is devoted to the quantization of the theory
in perturbation theory and in particular to
the definition of the action in the tree approximation.
First we want to review 
how 
the perturbative expansion of time ordered Green's functions is constructed
(see e.g.\ $[$Q1$]$ -- $[$ Q5$]$).

The basic formula for the perturbative construction is the 
 Gell-Mann--Low formula, which relates time ordered expectation values of
interacting fields $\varphi_k$ to time ordered expectation values of free field
 $\varphi_k^{(0)}$:
\equ{ 
\langle T \varphi_{i_1}(x_1) \cdots \varphi_{i_n}(x_n) \rangle = {\cal R}
\langle T \varphi^{(0)}_{i_1}(x_1) \cdots \varphi^{(0)}_{i_n}(x_n) e^{i 
\G_{int}(\varphi^{(0)}_k)} \rangle.
\label{eq: Gell-Mann-Low}
}
$\Ga_{int}$ includes all the interaction polynomials appearing in the model,
and is obtained by splitting off from the classical action the bilinear
part:
\begin{equation}
\label{Gabil}
\Ga_{cl} = \Ga_{bil} + \Ga_{int}
\end{equation}
After having expanded the exponential function
 in its Taylor series, the vacuum
expectation values of free fields are decomposed into a sum of products of
free field propagators and certain vertex factors according to Wick's theorem.
The combinatoric and vertex factors are summarized in the Feynman rules.
However, due to the well-known ultraviolet divergencies of the formal
perturbative expansion the Gell-Mann--Low formula is not meaningful in
higher loop orders of perturbation theory and has to be rendered meaningful
in the course of renormalization. (This is the sense of ${\cal R}$
 in eq.\ \eqref{eq: Gell-Mann-Low}.) Let us now 
have a closer look to the free field propagators
of the various particles.

The free scalar field obeys the Klein-Gordon equation:
\equ{
( \Box + m^2 ) \varphi = 0.
}
The time ordered expectation value of  the free scalar field is
given by the
solution of the inhomogeneous equation:
\equ{
( \Box + m^2 ) \Delta_{\varphi\varphi}(x-y) = i \d^ 4(x-y).
}
The solutions of such equations are the causal Green's functions, which  are
called the free field propagators 
\equ{
\Delta_{\varphi\varphi}(x) = 
\raisebox{1mm}{\epsfig{file = 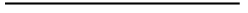, height = 15mm, angle = 270}}
=\langle T \varphi^{(0)}(x) \varphi^{(0)}(0) \rangle = \int \frac{\mbox{d}^4
k}{(2\pi)^4} e^{-ikx} \frac{i}{k^2 - m^2 + i\epsilon}.
}
(Here we have also given the Feynman diagram corresponding to the propagator.)
For fermions the free field propagators are
calculated  similarly by solving the inhomogeneous Dirac equation:
\equ{\label{fermprop}
(i \g^\m \partial_\m - m) \Delta_{\psi\bar{\psi}}(x-y) = i \d^ 4(x-y).
}
From (\ref{fermprop}) the fermion propagator is determined:
\equ{
\Delta_{\psi\bar{\psi}}(x) = 
\raisebox{1mm}{\epsfig{file = 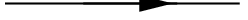, height = 15mm, angle = 270}}
= \langle T \psi^{(0)}(x) \bar \psi^{(0)}(0) \rangle =
\int \frac{\mbox{d}^4 k}{(2\pi)^4} e^{-ikx} \frac{i(\g^\m k_\m +m)}{k^2 -
m^2 + i\epsilon}.
}
In general, also in the case of several particles with non-diagonal
bilinear parts, one can calculate the free field propagators in the following
way:
\equ{
\label{defprop}
\sum_\g \int \mbox{d}^4 z\; \Ga^{(0)}_{\varphi_\a\varphi_\g}(x,z)
\Delta_{\varphi_\g \varphi_\b}(z,y) = i \d_{\a\b} \d^ 4(x-y).
}
Here $\Delta_{\varphi_\a \varphi_\b}(x,y) = \langle T \varphi^{(0)}_\a(x)
\varphi^{(0)}_\b(y) \rangle$ and $\G^{(0)}_{\varphi_\a\varphi_b}(x,y) $
is derived from the bilinear part of the classical action  (\ref{Gabil}):
\equ{
\G^{(0)}_{\varphi_\a\varphi_b}(x,y) = 
\frac{\d^2 \G_{bil} }{\d \varphi_\a(x) \d
\varphi_\b(y)}.
}
\par 
 If we try to apply the formula \eqref{defprop}
 for determining  the photon propagator of the electromagnetic action
\eqref{eq: Classical EM},
 we get
into trouble. 
The equations of motion for a free photon are given by
\equ{
\partial^\m F_{\m\n} = ( \eta^{\m\n} \Box - \partial^\m\partial^\n )
A_\n = 0 
}
and the respective inhomogeneous equations by
\begin{equation}
( \eta^{\m\n} \Box - \partial^\m\partial^\n )
\Delta_{\nu\rho} = i \d^ 4(x-y) \delta^{\mu}_\rho.
\end{equation} 
Since the operator which acts on $A_\mu$ is not invertible,
 the naive
way of calculating the propagator does not work. The reason can be found
in gauge invariance of the theory, which brings about, that the vector field
$A^\mu$
is determined up to a gauge freedom by the classical equations of motion.
In perturbation theory  one usually adds
a gauge-fixing action to the gauge invariant action:
\equ{
\label{eq: Ga QED}
 \G^{QED}_{cl} = \G_{em} - \frac{1}{2\xi} \intd (\partial^\m A_\m)^2,
}
with $\xi$, the gauge  parameter. In this way it is possible to fix the
gauge and  to maintain at the same time
Lorentz invariance and locality of the action.
 The propagator of the vector field is determined  from (\ref{eq: Ga QED}) as
 solution of the inhomogeneous equations
\equ{
\Bigl( \eta^{\m\n} \partial^2 - \partial^\m\partial^\n + \frac{1}{\xi}
\partial^\m \partial^\n \Bigr) \Delta_{\n\rho}(x-y) = i\d^ 4(x-y)\d ^\mu_\rho.
}
It is given by
\equ{
\Delta_{\m\n}(x-y) = 
\int \frac{d^4 k}{(2\pi)^4}
e^{-ikx}\frac{-i}{k^2 +i\epsilon} \Bigl( P^T_{\m\n}  + \xi P^L_{\m\n}\Bigr),
}
$P^L$ and $P^T$ are the projectors for longitudinal and transverse
polarization:
\equ{
P^L_{\m\n} = \frac{k_\m k_\n}{k^2} \quad \mbox{and} \quad
P^T_{\m\n} = \eta_{\m\n} - \frac{k_\m k_\n}{k^2}.
}
The complete action of QED \eqref{eq: Ga QED}
 is not invariant under gauge transformations, but  gauge invariance is
broken by the gauge-fixing action
linearly 
\equ{\label{QEDwi}
\Bigl(e {\bf w}_{em} - \partial^\m \frac{\d}{\d A^\m} \Bigr) \G^{QED}_{cl} =
- \frac 1 \xi \Box  \partial A.
}
In fact one has  introduced an unphysical scalar $\partial A$
with spin $0$ and negative norm into the theory.
 For making QED meaningful
one has to prove  that the S-matrix constructed from the action
(\ref{eq: Ga QED}) indeed  describes a physical 
theory with  a  spin $1$ particle and that the resulting 
theory has a probability
interpretation in the sense of quantum theory. 
In QED  one is finally able  to show  that $\partial A$ 
does not contribute to  physical scattering processes and that
the physical S-matrix indeed has positivity properties (see $[$Q4,R5$]$).
The proof is based on the QED Ward identity 
\equ{
\label{QEDWI}
\Bigl(e {\bf w}_{em} - \partial^\m \frac{\d}{\d A^\m} \Bigr) \G =
-\frac 1 \xi (\Box + \xi m_{ph}^2 )  \partial A.
}
This identity has to be proven 
for the Green's function of QED to all orders of perturbation  in the
course of renormalization.
In our notations $\Ga $ denotes the generating functional of 
one-particle-irreducible (1PI) Green's functions. Its  lowest order coincides 
with the classical action 
\equ{
\Ga^{(0)} = \Ga_{cl}^{QED}.
}
(For infrared definiteness we have introduced a photon mass term 
$m_{ph}$ in addition,
which breaks abelian gauge invariance not worse than the gauge fixing.)
The final proof is then carried out by Legendre transforming the
1PI Green's functions to connected Green's functions and finally
by applying the LSZ reduction formula (see $[$Q2,Q3$]$) on the Ward identity.
Then the operator identity
\equ{
\frac 1 \xi (\Box + \xi m_{ph}^2 ) \partial A ^{op} = 0
}
is deduced, i.e.\  $\partial A^{op}$ satisfies the Klein-Gordon equation and
does not interact. For the purpose of  these lectures we only want to indicate
how this result appears for the classical theory: 
Therefore we consider  the Ward identity of QED \eqref{QEDWI} 
for the classical action $\Gacl$. When we use the equations of
motion for fermions and the vector bosons,
 the left-hand-side vanishes and 
 we are left with the free field equation for the scalar
part of the vector field
\equ{
\label{pAcl}
\frac 1 \xi (\Box + \xi m_{ph}^2 ) 
 \partial A = 0 .
}
This equation proves that $\partial A$  does not interact in the classical
theory.

In non-abelian gauge theories one fixes the gauge for  the vectors
 as we have done
it in QED and one gets  the same expression for the free field propagators.
 But in contrast to QED a Ward identity as  (\ref{QEDWI}) does not exist,
which would allow to draw conclusions for the physical interpretation.
 This role is taken over
by BRS symmetry and by the Slavnov-Taylor identity. For this reason
these symmetries  are
the basis for the definition of the non-abelian gauge theories in renormalized
perturbation theory.

\newsubsection{Gauge fixing in the Standard Model}

For the massive vector bosons it is possible to determine the
propagators without the difficulty described above. 
So in the unitary gauge the $W$ propagator is given by:
\equ{\label{wunit}
\langle T W_+^\m (x) W_-^\n (0) \rangle
 = \int \frac{\mbox{d}^4k}{(2\pi)^4} e^{-
ikx}
\frac{-i}{k^2 - M_W^2} 
\Bigl( \eta^{\m\n} - \frac{k^\m k^\n}{M_W^2} \Bigr).
}
(That (\ref{wunit})
 is a gauge choice will become clear below, as well as why it is 
called unitary.) 
However this propagator does not allow for naive power counting
arguments of renormalizability to go through, 
since it behaves as a constant  for asymptotically large
momenta, i.e.\ when $k^2 \ra -\infty$. If we want to apply the arguments of
power counting renormalizability, the boson
propagators have to behave as $1/k^2$ for asymptotic  $k^2$. 
One-loop  calculations within the
Standard Model in the unitary gauge \cite{BACH80} have been carried out,
 but it is 
hard to see how these
calculations are  extended to higher orders. 
In order to have renormalizability by  power counting one has to fix the
gauge similarly as in QED by adding the gauge fixing part. For the purpose of
algebraic renormalization we choose a (linearized) generalization of the 
usual $R_\xi $-gauges \cite{HO71} and couple the gauge-fixing functions to the
auxiliary field $B_a ,a  = +,-,Z,A$
\equ{\label{gford}
\G_{g.f.}^{(B,\xi)} = \intd \left(
 \xi_W B_+ B_- + \frac 12 \xi_Z B_Z^2 + \frac 12
\xi_A B_A^2 + B_a \tI_{aa'} F_{a'} \right),
}
The gauge-fixing functions of the $R_{\xi}$-gauges
 fix the scalar part of vectors and introduce
mass terms for the would-be Goldstone fields  $\phi^{\pm}$ and $\chi$
\begin{eqnarray}
\label{F_a}
F_{\pm}&\equiv&\partial_{\mu}W^{\mu}_{\pm} \mp
iM_{W}\zeta_{W}\phi_{\pm},
\nonumber{}\\
F_Z&\equiv& \partial_{\mu}Z^{\mu}-M_{Z}\zeta_{Z}{{\chi}},\\
F_A&\equiv & \partial_{\mu}A^{\mu}.  \nonumber 
\end{eqnarray}
Various choices of the parameters $\xi_a, \zeta_a$ have special names:
the Landau gauge has $\xi_W = \xi_Z = \xi_A = 0$ and $\zeta_W = \zeta_Z = 0$,
 the 't Hooft gauges
have $\xi_Z = \zeta_Z, \quad \xi_W = \zeta_W$ and the 't Hooft-Feynman
gauge has in addition $\xi_W = \xi_Z = \xi_A = 1$. 
(The unitary gauge is retrieved in the limit $\zeta_W = \xi_W^2 \ra \infty$
and $\zeta_Z = \xi^2_Z \ra \infty$.)
The $B_a$-fields can be  
eliminated from the action  fields by their equations of motion and in this
way
one comes back to the usual $R_\xi $-gauges:
\equ{
\label{Rxi}
{\delta \Ga \over \delta B_a} = 0 \Longrightarrow
\G_{g.f.}^{\xi} = \intd \left(
 - \frac{1}{\xi_W} F_+ F_- - \frac{1}{2\xi_Z} F_Z^2 -
\frac{1}{2\xi_A} F_A^2 \right).
}
On a first sight the gauge-fixing with $B_a$-fields
 seems to be less practical than the $R_\xi$-gauges,
since one introduces  extra non-diagonal propagators, like 
$\langle T B_\pm (x)W_\mp (y) \rangle$, into  the theory. 
But,
 as we discuss in section 3.3, in this formulation
 BRS transformations are nilpotent on all fields and the algebraic method is
applied much easier  as it is in the naive approach. One has
 already
to note at this stage, that in the linear $(B,\xi)$ gauges the
gauge fixing part of the action does not get loop corrections and remains a
local field polynomial as in the tree approximation. This observation
is simply deduced from the observation that
 there are no interaction vertices of the $B_a$-fields with
other propagating fields.

All the propagators now behave such that naive power counting is
possible. In the 't~Hooft gauges ($\xi_W = \zeta_W$) one finds for example:
\equa{
\langle T B_+ (x) W_-^\m (0)\rangle & = 
\int \frac {d ^4 p}{(2\pi)^4}e^{-ipx}
\frac{-p^\m}{p^2 - \xi_W M_W^2},
\nonumber \\
\langle T B_+ (x) \phi^- (0)\rangle & = 
\int \frac {d ^4 p}{(2\pi)^4}e^{-ipx}
\frac{-M_W}{p^2 - \xi_W M_W^2},
\nonumber \\
\langle T W_+^\m (x)W_-^\n (0)\rangle & = 
\int \frac {d ^4 p}{(2\pi)^4}e^{-ipx}\left(P_T^{\m\n} \frac{-i}{p^2 - M_W^2} +
P_L^{\m\n}  \frac{-i\xi_W}{p^2 - \xi_W M_W^2}\right), \nonumber \\
\langle T \phi^+  (x) \phi^- (0) \rangle & = 
\int \frac {d ^4 p}{(2\pi)^4}e^{-ipx} \frac{i}{p^2 - \xi_W M_W^2}.
\nonumber 
}
(A complete list of the free field propagators of the Standard Model
 in a general linear gauge 
can be found in \cite{KRWE98}.)
\par 
In section 2 we have constructed the $SU(2) \times U(1)$ 
gauge invariant part of
the action of the electroweak Standard Model.
 We  have to look how the gauge symmetries (\ref{nagaugeid2}) act on
the gauge-fixing part of the action (\ref{gford}).
  In the $B_a$-gauges
we have to extend the symmetry transformations by 
 the contributions of the auxiliary fields in a way that $\intd B_a
\tilde I_{ab}
\partial V_b$
is  invariant under {\it rigid} transformations:
\begin{equation}
{\cal W}_\a \intd B_a \tilde I_{ab}\partial V_b = 0 
\qquad  {\cal W}_{em} \intd  B_a
\tilde I_{ab} \partial V_b= 0.
\end{equation}
By this requirement the transformation behaviour of $B_a$-fields is
uniquely determined:
\equa{\label{eq: w^B}
{\cal W}_\a^{B} &=  \tilde I_{\a \a'} 
\intd B_b O^ T_{b\b}(\theta_W)\epsilon_{\b\g \a'} O_{\g c}(\theta_W)
\tilde I_{c c'}
\fderivative{}{B_{c'}}, \qquad \a = +,-,3 ;\\
{\cal W}_{em}^{B} &= -i \intd
\left( B_+ \fderivative{}{B_+} - B_- \fderivative{}{B_-}\right).
}
The matrix $O_{\a a}(\theta_W) $ is defined in (\ref{eq: thetaW rot}) and
$\epsilon_{\a\b\g} $ as in (\ref{su2alg}).
The abelian (hypercharge)  operator is defined according to \eqref{eq: w_4^Q}
 by  ${\bf w}_4^{Q} = {\bf w}_{em} - {\bf w_3}$.
It is seen that the rigid as well as the gauge symmetries are broken by
the gauge fixing.
To be precise, the rigid $SU(2)$ symmetries   are broken,
\equ{
{\cal W}_\a \G_{g.f.}^{(B, \xi)} = \Delta_\a^{g.f.},
}
but electric charge is conserved,
\equ{
\quad 
{\cal W}_{em} \G_{g.f.}^{(B, \xi)} = 0.
}
However,  electromagnetic gauge symmetry is broken by a non-linear expression,
\equ{
\Bigl(e{\bf w}_{em} -
\partial \frac {\d}{\d A}
\Bigr) 
\G_{g.f.}^{(B, \xi)} = \Box B_A - i\partial_\m( B_+ W_-^\m -
B_- W_+^\m),
}
which reads in the $R_\xi$-gauges (\ref{Rxi})
\equ{
\label{emwibr}
\Bigl(e{\bf w}_{em} -
\partial \frac {\d}{\d A}
\Bigr) 
 \G_{g.f.}^{ \xi} = 
- \frac 1 {\xi_A}\Box \partial A +  \frac i {\xi _W}
\partial_\mu \Bigl( ( \partial
W_+ -  i M_W\zeta_W \phi^+) W_-^\m - \mbox{h.c.}\Bigr).
}
From the last  expression it is immediately clear that the situation is
dramatically changed compared to pure QED (cf.~(\ref{QEDwi})).
In the case of QED we have derived from the QED Ward identity 
 that $\partial A $  is a free field in the classical theory (cf.~the
 derivation of eq.~(\ref{pAcl})). 
The same arguments applied to eq.~(\ref{emwibr}) show, 
that $\partial  A $
 interacts with $W_+$ and
$W_-$ and will therefore indeed contribute to physical scattering processes.
To cancel these  contributions in the physical scattering matrix
additional fields, the Faddeev-Popov ghost fields \cite{FAD67}, 
have to be introduced into
 the theory and gauge symmetry has to be  replaced by 
BRS symmetry \cite{BRS75,TYU75}.
This is the topic of the following subsection 3.3.
\par
Another complication of the  gauge fixing in spontaneously broken
theories and in particular in the Standard Model is that it 
does not even  maintain rigid $SU(2)\times U(1)$ symmetry. 
Instead the gauge-fixing action and  the gauge parameters
(\ref{gford}) have been chosen as
though they have been built around several $U(1)$ factors. 
In order not to spoil the group structure of global
$SU(2) \times U(1)$ symmetry, the following choices for the
gauge-fixing parameters are made:
\equ{\label{xispecial}
\xi_A = \xi_W = \xi_Z =\xi \quad \mbox{and} \quad \zeta_W = \zeta_Z =
\zeta.
}
Then the 4-dimensional terms of the gauge fixing
 are invariants
\begin{equation}
{\cal W}_\a \intd \left(\frac \xi 2 B_a \tilde I_{ab} B_b + B_a \tilde I_{ab}
\partial V_b \right) = 0,
\end{equation}
 whereas the 3-dimensional
ones are seen to transform in the same covariant way as the fermion
mass terms under $SU(2) \times U(1)$ transformations.
The mass breakings  of the gauge fixing cannot be coupled to the scalar 
doublet, since the corresponding expression vanishes identically, but
we are able to couple it to an {\it external} scalar doublet
$\hat \Phi $ and
 its hermitian conjugate:
\begin{equation}
\hat \Phi = {\hat \phi^ + \choose \frac 1 {\sqrt 2 }
( \hat H + i \hat \chi )}, \qquad
\hat \Phi^\dagger = {\hat \phi^ - \choose \frac 1 {\sqrt 2 }
( \hat H - i \hat \chi )}  .
\end{equation} 
It is transformed in the same way  as
 the scalar doublet $\Phi$ under rigid $SU(2) \times U(1)$ (see 
(\ref{eq: w^scalar})),
 only the shift can be
chosen differently in including the gauge parameter $\zeta$ of 3-dimensional
breakings:
\equa{
\label{gtextscal}
{\cal W}_\a^{\hat \Phi} & = \intd\Bigl(\bigl( 
\hat \Phi + \zeta \hbox{v} \bigr)^\dag \frac{i\tau^ T_\a}{2}
\frac{\d}{\d \hat \Phi^\dag} - \frac{\d}{\d \hat \Phi} 
\frac{i\tau^ T_\a}{2} \bigl(\hat \Phi + \zeta
\hbox{v} \bigr)\Bigr) . 
}
algebraically this construction is similar to the one that was
 applied when we did introduce the scalar doublet and spontaneous breaking of
the symmetric gauge theory in section 2.2.5. 
However, since the construction here is done  for a 
non-propagating scalar doublet, $\hat \Phi$ does not have 
 a physical interpretation.

The gauge-fixing functions including the external fields read now:
\begin{equation}
\label{gfrig}
F_a \to {\cal F}_a  = \partial V_a - 
i \frac e{\sin \theta_W}
\Bigl((\hat \Phi + \zeta \mbox{v} )^\dagger \frac {\tau^T_{a}(\hat G)} 2
( \Phi +  \mbox{v} ) -
( \Phi + \mbox{v} )^\dagger \frac {\tau^T_{a}(\hat G)} 2
( \hat \Phi + \zeta \mbox{v} ) \Bigr) . \nonumber 
\end{equation}
Here we have introduced the following notations:
 \begin{eqnarray}
\label{tauphys}
\tau_Z(G)  & = & \cw \tau_3 + G \sw {\mathbf 1},  \nonumber \\
\tau_A (G) & = & - \sw \tau_3 +G \cw {\mathbf 1}. 
\end{eqnarray}
When we choose the parameter $\hat G$
\equ{
\label{Gspecial}
\hat G =  -\frac{\sw}{\cw}, 
}
 we recover the gauge-fixing  functions
 \eqref{F_a} with $\zeta_W = \zeta_Z = \zeta $.
Explicitly,
with this choice the  gauge-fixing action at $\hat \Phi = 0$ reads:
\equ{
\label{gaugesp}
 \intd \Bigl.
\Bigl( \frac{1}{2} \xi B_a \tI_{ab} B_b + B_a \tI_{ab} {\cal F}_b 
\Bigr) \Bigr|_{\hat \Phi = 0 \atop
 \hat G = - \tan \theta_W} 
       \\
 =  \intd\Bigl( \frac{1}{2} \xi B_a \tI_{ab} B_b + B_a \tI_{ab} \partial V_b 
 \, + i\zeta M_W (B_+ \phi_- - B_- \phi_+ ) - \zeta M_Z B_A \chi
\Bigr) . \nonumber
}
 The gauge fixing \eqref{gaugesp}
 is indeed a special gauge choice and has to be
replaced by the most general one,
compatible with rigid symmetry, in higher orders of perturbation theory:
\begin{equation}
\label{Gagfgen}
\G_{g.f.} = \intd\left(
 \frac{1}{2} \xi B_a \tI_{ab} B_b + B_a \tI_{ab} {\cal F}_b (\hat G, \zeta)
+ \frac 12 \hat \xi (\sin \theta_W B_Z +  \cos \theta_W B_A) ^2 \right).
\end{equation}
Here the four parameters $\xi, \hat \xi, G$ and $\zeta$ 
are independent parameters of the gauge fixing. 
$\G_{g.f.} $  \eqref{Gagfgen}
is characterized by being linear in the
propagating fields, by CP invariance and by rigid invariance,
\equ{
{\cal W}_\a \G_{g.f.} = 0.
}
The operator of rigid $SU(2)$ transformations is now given by the sum of all
field transformations introduced by now (cf.\  
(\ref{eq: w^vector}), (\ref{eq: w^fermion}), (\ref{eq: w^scalar}),
(\ref{eq: w^B}) and (\ref{gtextscal}))
\equ{
\label{wfields}
{\cal W}_\a
={\cal W}_\a^{fermion}  + {\cal W}_\a^{scalar} +{\cal W}_\a^{vector}    
  + {\cal W}_\a^{B}+  {\cal W}_\a^{\hat \Phi}  
}

\newsubsection{BRS symmetry and Faddeev-Popov ghosts}

In the previous subsection we have shown that the gauge fixing breaks local
gauge symmetry non-linearly and we have argued that as a consequence of the
broken gauge Ward identity the unphysical part of the vector bosons interacts
and contributes to the physical scattering matrix in the tree approximation
(cf.\ (\ref{emwibr})).
 In order to cancel 
these interactions in the scattering matrix further fields, the Faddeev-Popov
 ghosts, are needed. The conventional way for introducing Faddeev-Popov
fields into
gauge theories does not start from unitarity arguments but from the path
integral formulation of quantum field theory:
To implement the gauge fixing program in path integrals one needs a
compensating determinant. 
This determinant can be rewritten in the form a path integral over a set of
anti-commuting scalar fields. 
Since these scalar fields have the wrong statistics (they should have
been bosons instead of fermions) they are not physical and 
therefore called ghosts \cite{FAD67}.
\par
A third way of introducing Faddeev-Popov ghosts in the theory is provided by 
the algebraic method of BRS quantization.
Since this method is close to the  algebraic characterization of renormalized
perturbation theory, we want to discuss it in the following:
In a  first step one considers BRS transformations
  as an alternative way to 
characterize the Lie algebra of the gauge group and replaces the infinitesimal
parameters of gauge transformation
$\epsilon_\a (x) $ by anti-commuting scalars $c_\a (x) $.
With this substitution the infinitesimal transformations on the 
fermion, scalars and vectors are become BRS transformations denoted
by $\brs \varphi$:
\begin{eqnarray}
\label{brs}
{\mathrm s} V_{\mu a}&=&\partial _{\mu}c_a + 
\frac e{\sin \theta _W} \tilde I _{aa'} f_{a'bc}V_{\mu
  b}c_c, \nonumber{}\\
{\mathrm s} \Phi &=& i \frac e{\sin{\theta_W}}
\frac{{\tau}_a(G_s)}{2} (\Phi+ {\mathrm v}) c_{
 a},  \\
{\mathrm s} F^L_{\delta} & =& i
\frac e{\sin{\theta_W}}
 \frac{{\tau}_a(G_\delta)}{2} F^L_{\delta_i} c_{ a} 
\quad \mbox{with} \quad
 \delta = l,q \nonumber\\
{\mathrm s}  f_i^R&= & - i e Q_f \frac {\sin{\theta_W}}{\cos{\theta _W}}  
 f_i^R   c_Z-            ie Q_f f_i^R c_A \nonumber 
\end{eqnarray}
Here we have transformed the fields $c_\a, \a = +,-,3,4$ into physical
fields by the orthogonal transformation matrix $O_{\a a}(\theta_W)$
\eqref{eq: thetaW rot},
\equ{
c_\a = O_{\a a} (\theta_W) c_a \/,
}
and have given the transformations
 in the physical  on-shell fields and in the QED-like
parameterization (\ref{QEDonshellpar}). The structure constants are defined by
\begin{equation}
f_{abc} = \epsilon_{\a \b \ga} O _{\a a} (\theta_W)
O _{\b b} (\theta_W) O _{\ga c} (\theta_W).
\end{equation}
 The matrices $\tau_a (G)$ are given in \eqref{tauphys} and satisfy the
algebra:
\begin{equation}
\label{taualg}
\bigl[ \tau_a ( G) , \tau_b (G) \bigr] = f_{abc} \tilde I_{cc'} \tau_{c'} (G).
\end{equation}
 The parameters $G_k$
are related to the weak hypercharge of the respective $SU(2)$-doublets.
\equ{
G_k = -\frac{\sw}{\cw} Y_W^k \qquad
\mbox{with}\qquad
Y_W^k = 
\begin{cases}
1 & \mbox{for the scalar doublet} \quad (k = s), \\
-1& \mbox{for the lepton doublet} \quad (k = l), \\
\frac 13 & \mbox{for the quark doublet} \quad (k = q).
\end{cases}
}

Since the action $\G_{GSW}$ is gauge invariant, it is 
is invariant under BRS transformations by construction
\equ{
\brs \G_{GSW} = 0.
}
The Lie algebra of functional generators is  
 translated into the
nilpotency of the BRS operator $\brs$
\begin{equation}
\brs^2 = 0.
\end{equation}
It includes the commutation relations as well as the Jacobi identities.
For illustration we calculate the BRS transformations of ghosts by
requiring nilpotency of the BRS operator when acting on the fermion doublets:
\begin{eqnarray}
\label{nilill}
0 = \brs^2 F^L& =& \brs (c_a \delta_a F^L) \\
              & = & \brs c_a \delta_a F^L + c_b c_a \delta_b \delta_a F^L 
\nonumber \\
              & = &  (\brs c_a + \frac 12 f_{bda'} c_b c_d \tilde I_{aa'})
                \delta_a F^L \nonumber
\end{eqnarray}
From the last line one derives:
\equ{
\brs c_a = - \frac 12 \tI_{aa'} f_{a'bc} c_b c_c.
}
In \eqref{nilill} $\delta _a $ denote the infinitesimal $SU(2) \times 
U(1) $ transformations transformed to physical field indices by applying
the orthogonal rotation matrix $O_{\a a} (\theta_W)$. The obey the algebra
\equ{
[\delta_a , \delta_b ] = f_{a b c} \tilde I_{cc'} \delta _{c'}
}

The second and crucial step of the construction
is the observation that one is able to
complete the gauge-fixing action in such a way that it is BRS invariant.
We have shown in the previous subsection that on the gauge-fixing action 
 gauge invariance is  broken by non-linear field polynomials
(\ref{emwibr}). But introducing  the anti-ghosts $\bar c_a $
and their BRS transformations
\equ{
\brs \bar{c}_a = B_a \qquad \mbox{and} \qquad \brs B_a = 0 \quad
\mbox{with} \quad \brs^2 = 0,
}
 it is possible to enlarge the gauge fixing action by a ghost action in such a
 way that the sum is invariant under BRS transformations:
\equ{
\brs \Bigl( \G_{g.f.} + \G_{ghost} \Bigr) = 0.
\label{eq: nil g.f. ghost}
}
If one writes for the ghost action $\G_{ghost} = \intd \bar{c}_a X_a$
one finds according to \eqref{eq: nil g.f. ghost}:
\equ{
0 = \intd \Bigl(
 B_a \tI_{aa'} \brs {\cal F}_{a'} + B_a X_a - \bar{c}_a \brs
X_a \Bigr).
}
By identifying $X_a = - \tI_{aa'} \brs {\cal F}_{a'}$ and noting that then
$\brs X_a = 0$ because of nilpotency, 
 the ghost action is uniquely determined to:
\equ{
\G_{ghost} = - \intd \bar{c}_a \tI_{aa'} \brs {\cal F}_{a'}.
}
One has to note that in the $B_a$-gauges the gauge-fixing and
ghost action is a BRS variation,
\equ{ \G_{g.f.}+ \G_{ghost} = \brs \intd \bar c_a (\frac \xi 2  B_a + 
{\cal F}_a),
}  
and BRS invariant because of nilpotency of the
BRS operator $\brs$.
Finally we have to assign a BRS transformation to the external scalar
doublet 
$\hat \Phi$. Since it couples to a BRS variation it is possible to transform
the external scalar doublet into a scalar external ghost doublet $\bf q$
with Faddeev-Popov charge $1$:
\begin{equation}
\brs \hat \Phi = {\bf q} \qquad \brs {\bf q} = 0.
\end{equation}

The ghost action contains kinetic and mass terms for the fields.
With the gauge choice \eqref{Gspecial} they read
\equ{
\G_{ghost}^{bil.} = -\intd \left( \bar{c}_a \Box \tI_{ab} c_b + \zeta M_W^2
(\bar{c}_+ c_- + \bar{c}_- c_+) + \zeta M_Z^2 \bar{c}_Z c_Z \right).
}
 For this
reason they have to be considered as dynamical fields
with the following free field propagators:
\equa{
\langle T{c}_+ (x) \bar c_- (0) \rangle & =  
\int \frac{d^4 k} {(2\pi)^4} e ^{-ikx}
\frac{i}{k^2 - \zeta M_W^2},
\nonumber \\
\langle T {c}_Z (x) \bar c_Z (0) \rangle & =
\int \frac{d^4 k} {(2\pi)^4} e ^{-ikx}
  \frac{i}{k^2 - \zeta M_Z^2}, \\
\langle T \bar{c}_A (x) \bar c_A (0) \rangle & =  
\int \frac{d^4 k} {(2\pi)^4} e ^{-ikx}
\frac{i}{k^2}.
\nonumber
}
 We want to note, that for the general gauge fixing there
appear non-diagonal ghost propagators in the bilinear action. To diagonalize
the ghost mass matrix one has to introduce in the BRS transformation of
ghosts an additional ghost matrix which allows the diagonalization of the
ghost mass matrix (for details see \cite{KR98,KRWE98} and Appendix A): 
\begin{equation}
\brs \bar c_a = B_b  \hat g _{b a}.
\end{equation}
For higher order loop calculations this observation is crucial for obtaining
infrared finite  results for off-shell Green's functions.

\newsubsection{The defining symmetry transformations}

The  classical action of the electroweak Standard Model is  given by:
\equ{
\G_{cl} = \G_{GSW} + \G_{g.f.} + \G_{ghost}.
\label{eq: classical action}
}
It is the starting point for the perturbative calculation of Green's functions
and determines via
  the Gell-Mann--Low formula and the free field propagators the 
tree approximation completely. Higher orders are, however, subject of
renormalization and have to be properly defined. For this reason
we want to summarize the symmetry properties of the classical action.
In the course of renormalization we have to show that these symmetry
transformations determine the classical action uniquely, if one poses 
appropriate normalization conditions.

Due to the fact that gauge invariance is non-linearly broken by the gauge
fixing we have to replace gauge invariance by invariance with
respect to the nilpotent BRS transformations:
\equ{
\brs \G_{cl} = 0, \qquad \brs ^2 = 0 \/.
\label{eq: classical brs inv}
}
If we want to write BRS transformations in functional form we face the
problem of non-linear symmetry transformations. These symmetry transformation
become insertions\footnote{For an introduction to insertions and
 normal products see the reviews and books on renormalization
 $[$R2$]$ -- $[$R7$]$; see also section 4.2 of these lectures.}, 
 the classical action as the lowest
order of the generating functional of 1PI Green's functions. To make them
well-defined for ordinary as well as connected Green's functions, nonlinear
symmetry transformations have to be coupled to external fields 
$\rho_\a, \sigma_\a, Y, \Psi^ R_l, \Psi^R_q$ and $\psi^ L_f$
(see Appendix A for notations):
\equa{
 \G_{ext.f.} =  \intd & \Bigl( \rho_3 O_{3a}(\theta_W) \brs W_a + \sigma_3
O_{3a}(\theta_W) \brs c_a + \rho_+  \brs W_- + \rho_- \brs W_+ \\
& + \sigma_+ \brs c_- + \sigma_- \brs c_+ + 
Y^\dag \brs \Phi + (\brs \Phi)^\dag Y \nonumber\\
       & +  \bigl(
       \overline {\Psi ^R _{l}} {\brs} F^L _{l}  + 
       \overline {\Psi ^R _{q}} {\brs} F^L _{q}  + 
       \sum_{f}       \overline {\psi ^L _{f}}
{\brs}  f^R_i + \hbox{h.c.} \bigr)\Bigr) .
\nonumber
}
Adding the external field action to the classical action
\equ{
\G_{cl} \longrightarrow \G_{cl} + \G_{ext.f.},
}
BRS invariance is rewritten into the Slavnov-Taylor identity:
\begin{eqnarray}
\label{ST}
{\cal S}
(\Gacl ) &=& \intd \biggl(
\bigl(\sw \partial _\mu c _Z + \cw \partial_\mu c_A\bigr)
             \Bigl(\sw {\delta \Gacl \over \delta Z_\mu } + \cw
       {\delta \Gacl \over \delta A_\mu} \Bigr) \\  
 & & +     {\delta \Gacl \over \delta \rho^\mu_3 }
              \Bigl(\cw {\delta \Gacl \over \delta Z_\mu } - \sw
       {\delta \Gacl \over \delta A_\mu} \Bigr) 
       + {\delta \Gacl \over \delta \sigma _3 }
              \Bigl(\cw {\delta \Gacl \over \delta c_Z } - \sw
       {\delta \Gacl\over \delta c_A} \Bigr) \nonumber \\ 
& &      + {\delta \Gacl \over \delta  \rho^\mu _+ }
               {\delta \Gacl \over \delta W_{\mu,- } }
      + {\delta \Gacl \over \delta \rho^\mu _- }
               {\delta \over \delta W_{\mu,+ } }
+  {\delta \Gacl \over \delta \sigma _+ }
               {\delta \Gacl \over \delta c_{-} }
+  {\delta \Gacl \over \delta \sigma _- }
               {\delta \Gacl \over \delta c_{+} } +
{\delta \Gacl \over \delta Y^\dagger}{ \delta \Gacl \over \delta \Phi } +
{\delta \Gacl \over \delta \Phi^\dagger}{ \delta \Gacl \over \delta Y } 
 \nonumber \\
&& + \sum_{i=1}^{N_F} \Bigl({\delta \Gacl \over \delta \overline{\psi^L_{f_i}}}
{ \Gacl \delta \over \delta f^R_i }
+ {\delta \Gacl \over \delta \overline{\Psi^R_{\delta_i}}}
{ \Gacl \delta \over \delta F^L_{\delta _i }} + 
  \hbox{h.c.} \Bigr)  \nonumber \\
& & + B_a {\delta \Gacl \over \delta \bar c_a }  
   + \hat {\mathbf q}{\delta \Gacl \over \delta \hat \Phi  }  + 
{\delta \Gacl \over \delta \hat \Phi ^\dagger } \hat
{\mathbf q}^\dagger \biggr)  = 0 .
 \nonumber
\end{eqnarray}
The unitarity of the physical S-matrix, i.e.\ cancellation of unphysical
particles  in physical scattering processes, can be derived 
 from the Slavnov-Taylor identity.
To ensure that the physical interpretation also holds to higher orders  
the Slavnov-Taylor identity has to be established
to higher orders of perturbation theory as defining symmetry identity of
non-abelian and spontaneously broken gauge theories \cite{BRS75,BRS76,KUOJ78}.
(For an introduction to unitarity proofs in gauge theories see $[$R1$]$ and
 $[$Q4$]$.) 

In the Standard Model, due to the abelian factor group 
the Slavnov-Taylor identity does not completely characterize the theory.
As we have already mentioned we have to require an abelian Ward identity
for fixing electromagnetic current coupling and also 
 $SU(2)\times U(1)$ rigid symmetry for being able to single out the
abelian operator.  Assigning to the external fields and to the
Faddeev-Popov fields definite transformation
properties  with respect to rigid $SU(2)\times U(1) $ transformations,
we have for the complete classical action
\equ{
\label{wardrig}
{\cal W}_\a \G_{cl} = 0 \qquad \mbox{and} \qquad {\cal W}_{em} \G_{cl} = 0,
}
 The ${\cal W}_\a , \a = +,-,3 $ satisfy  $SU(2) $ algebra:
\equ{
[{\cal W}_\a, {\cal W}_\b] = \epsilon_{\a\b\g} \tI_{\g\g'} W_{\g'}.
}
The operators $ {\cal W}_\a $ are  the sum of all field operators
(cf.~\eqref{wfields})
\equ{
\label{wcomplete}
{\cal W}_\a
={\cal W}_\a^{fermion}  + {\cal W}_\a^{scalar} +{\cal W}_\a^{vector}    
  + {\cal W}_\a^{B}+  {\cal W}_\a^{\hat \Phi} + 
  + {\cal W}_\a^{ghosts}+  {\cal W}_\a^{ext.f} .
}
The complete operators are given in  Appendix A (\ref{wardnaapp}).
Furthermore  we find that the local abelian symmetry defined by the operator
\begin{equation}
\label{wardab}
{\bf w}_4^Q \equiv {\bf w}_{em} - {\bf w}_3\qquad 
\bigl[{\bf w}_4^Q , {\cal W}_\a \bigr] = 0
\end{equation}
is broken linearly and can be interpreted as an abelian Ward identity
for the generating functional of 1PI Green's functions:
\equ{\label{abwardid}
\Bigl(
\frac e {\cw}{\bf w}_4^Q - \cw \partial{\delta \over \delta Z}-
 \sw \partial {\delta \over \delta A} \Bigr)
\G_{cl}=  \Box (\sin \theta_W B_Z + \cos \theta_W B_a)
}
It allows to distinguish electromagnetic current coupling 
from coupling of lepton and baryon number conserving currents in the
construction of the electroweak Standard Model.

As long as we do not consider family mixing in the fermion sector,
 CP invariance is a discrete symmetry of the Standard Model
and conservation of lepton and baryon family are 
 global abelian symmetries \eqref{wlqGSW}:
\equ{\label{wlqcl}
{\cal W}_l \Ga_{cl} = 0 \qquad \mbox{and} \qquad 
{\cal W}_q \Ga_{cl} = 0
}

In the proof of renormalizability to all orders,
 it has to be shown that the Slavnov-Taylor
identity \eqref{ST}, Ward identities of rigid symmetry
\eqref{wardrig} and the local abelian 
Ward identity  \eqref{abwardid} can be established to all orders of
perturbation theory. Furthermore -- and as important as the first part --
 it has to be shown that these symmetry
transformations together with  CP invariance and the global symmetries
\eqref{wlqcl} uniquely
determine all free parameters order to order in perturbation theory, if
appropriate normalization conditions are imposed.
\newpage

\newsection{Proof of renormalizability to all orders}
\newsubsection{Scheme dependence of counterterms}

 For the purpose of illustrating  general properties  of renormalized
 perturbation theory we consider a simple quantum field theoretic model,
the $ \varphi^4$-theory, with the classical action
\begin{equation}
\Ga_{cl} = \intd  \bigl( \frac 12 \partial_\m\varphi \partial^\m \varphi-
\frac 12
m^2\varphi^2  - \frac \lambda{4!}\varphi^4 \bigr).
\end{equation} 
As discussed in section~3.1 the perturbative expansion is formally
governed by
the Gell-Mann--Low  (GML) formula \eqref{eq: Gell-Mann-Low}
and  can be diagrammatically
expressed in the Feynman diagrams. 
There one assigns to propagators and vertices certain diagrammatic expressions
and writes all topological distinct diagrams. 
If one assigns furthermore to the diagrams symmetry factors, diagrams are 
immediately translated into the mathematical expressions of Green's functions.
The correspondence between diagrams and Green's functions
is  summarized in the Feynman rules. For example
 the connected 2- and 4- point Green's functions of the $\varphi^4$-theory
are expanded diagrammatically
as follows
\begin{center}
\begin{tabular}{c}
$G_c^2 = $ \raisebox{16mm}{\epsfig{file = 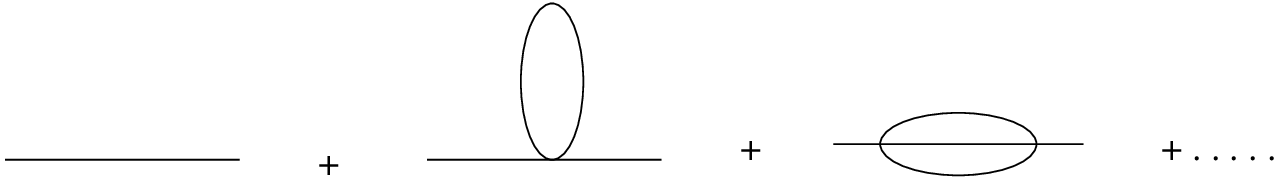, height = 12cm, angle = 270}} \\
\\
$G^4_c = $ \raisebox{13mm}{\epsfig{file = 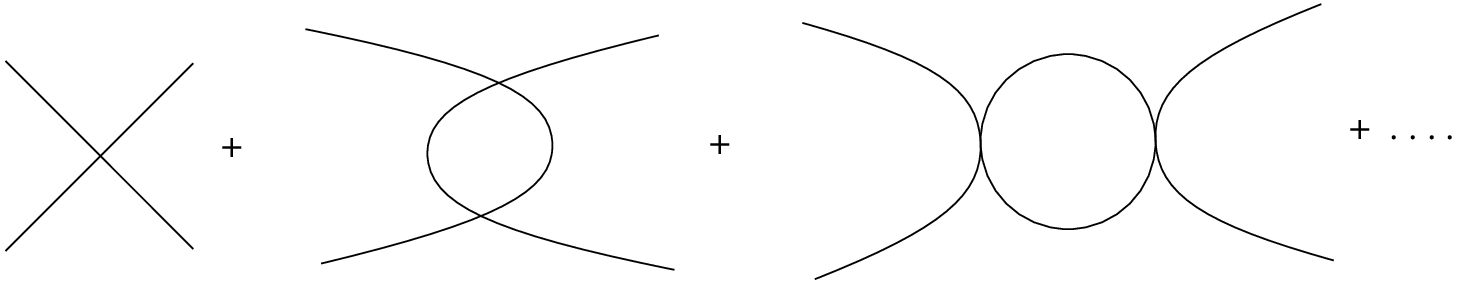, height = 12cm, angle = 270}} \\
\end{tabular}
\end{center}
When writing down the corresponding expressions to the loop diagrams, 
one sees that these integrals 
are not finite and therefore not well defined as they are given
by the GML formula: the integral over the internal loop momenta is unbounded. 
To analyze these divergences it is useful to consider the
one-particle-irreducible (1PI) Green's functions. In the above example they
 are obtained  from the connected ones
by amputating the external legs. For 1PI Green's
functions
the superficial 
degree of the ultraviolet  divergence $d_\G$ of a specific loop  diagram 
$\Ga$
 is given to all orders by the following formula:
\equ{ \label{divdeg}
d_\G = 4 - N_B - \frac 32 N_F + \sum_V (d_V -4) \/.
}
Here $N_B$ and $N_F$ denote the number of external
(amputated) boson  ($B$) and 
 fermion ($F$) legs.
The sum is taken over all vertices $V$ appearing in the respective 1PI diagram
 and $d_V$ denotes
the dimension  of the vertex   $V$. 
For the 2 and 4 point Green's functions in the $\varphi^4$-theory, we
find $d_{\G_2} = 2$ and $d_{\G_4} = 0$ respectively, whereas all the 1PI
Green's functions with more than four external (amputated) legs are finite.
Since all the propagators behave  not worse than
$\frac 1 {p^2} $ when $p^2 \to \infty $  (cf.~section 3.1 and 3.2)
the formula  (\ref{divdeg}) is also valid in the electroweak
Standard Model. Furthermore, since
 all the interaction vertices of the classical
action have dimension less than or equal 4 ($ d_V \leq 4 $ ),
  the divergencies of the Standard Model
 are restricted to 2-, 3-, and 4-point 1PI Green's functions
and are quadratic, linear  and logarithmic depending on the number of
external fermion and boson lines. This property is called naive
renormalizability by power counting (see e.g.~$[$R2$]$ for an introduction
to renormalization).
\par 
Next we consider the explicit expressions of the divergent 1-loop 1PI diagrams
in the $\varphi^4$-model\footnote{The definitions of 1PI Green's functions
  differs by a factor $i$ from other conventions (cf.~(\ref{1PIdef})), since
we want to identify the lowest order contributions to 1PI Green's functions
with the classical action (cf.~\ref{Gao}).}:
\begin{eqnarray}
\label{Gadiv}
\G_2 (p^2 ) &=   &
p^2 - m^2 
- i \lambda {\cal R} 
\int  \frac{\mbox{d}^4k}{(2\pi)^4}\frac{1}{k^2 - m^2} , \\ 
\G_4 (p_1 ,p_2, p_3, p_4) &  = &  -\lambda  - i \frac{\lambda^2}{2} {\cal
R} 
\int  \frac{\mbox{d}^4k}{(2\pi)^4}
 \Bigl(\frac{1}{k^2 - m^2} 
\frac{1}{(k+p_1 + p_2)^2 -
m^2}\nonumber \\
& & \phantom{-\lambda  - i \frac{\lambda^2}{2} {\cal
R} 
\int  \frac{\mbox{d}^4k}{(2\pi)^4}}  +  (p_2 \leftrightarrow p_3 ) + ( p_2
\leftrightarrow p_4)\Bigr).\nonumber
\end{eqnarray}
Here ${\cal R}$ denotes that the integral has to be made meaningful in the
course of renormalization. There are several
 schemes which allow to define Green's functions to all orders consistently.
Here we want to mention two of them: For practical calculations the most 
commonly used scheme is the  scheme of dimensional regularization together
with a subtraction prescription for removing the  poles in the limit of
4 dimensions \cite{HOVE72a,BEMA77}.
 In the abstract approach one refers to the momentum subtraction
scheme in the version of BPHZ and, if one includes massless particles, 
to its generalization,
the BPHZL scheme. (The scheme is called according to 
 Bogoliubov, Parasiuk \cite{BOG57}, Hepp \cite{HEP66} and Zimmermann 
\cite{ZIM69}
and in its massless version in addition to Lowenstein \cite{LOW76}.) 
\newpage

\begin{enumerate}
\item {\em Dimensional regularization} \\
In this scheme the dimension of space-time is analytically continued
to $D$-dimensions in the complex plane.
The  integrals (\ref{Gadiv})  become in $D$-dimensions:
\equa{\label{DIMreg2}
\G_2^{DIM} & = 
p^2 - m^2 
- i \lambda \mu^{4-D}
\int  \frac{\mbox{d}^Dk}{(2\pi)^D}\frac{1}{k^2 - m^2} , \\ 
& \stackrel{\epsilon \to 0}{=} p^2 - m^2 + m^2 \lambda \frac{1}{(4\pi)^2}
\Bigl( \frac{2}{\epsilon} - \g_E + \ln 4\pi +1 +
\ln {\mu^2 \over m^2} \Bigr) \nonumber\\
\G_4^{DIM} &=  -\lambda - i
\frac{\lambda^2}{2}
\m^{4-D}
\int \frac{\mbox{d}^Dk} {(2\pi)^D}\frac{1}{k^2 - m^2}
\Bigl( \frac{1}{(k+p_1 + p_2)^2-
m^2} + \mbox{ 2 Perm.} \Bigr) \label{DIMreg4}
\\
&\stackrel{\epsilon \to O}{=}  -\lambda + \frac{1}{(4\pi)^2}
\frac{3 \lambda^2}{2} 
\Bigl( \frac{2}{\epsilon} - \g_E + \ln 4\pi\Bigr) \nonumber \\
&\phantom{ \stackrel{\epsilon \to o}{=}  -\lambda } -
 \frac{1}{(4\pi)^2} \frac{ \lambda^2}{2} \Bigl(
\int_0^1  \mbox{d}z \ln \frac{m^2 - z(1-z)p^2 -i \varepsilon }{\m^2} +
\mbox{2 Perm.} \Bigr),
\nonumber 
}
where $\epsilon = D - 4 $. 
The auxiliary mass $\m$ is introduced for having dimensionless couplings
also in $D$ dimensions.
 $\Ga^{DIM}$ denotes the dimensionally regularized
integral. From there the finite renormalized Green's
functions in 4 dimensions
 are defined by  an additional subtraction prescription for removing
the poles in the limit of 4 dimensions, i.e.\ $\epsilon \to 0$.
This procedure is well-defined only up to constants: 
In 
the minimal subtraction scheme (MS) \cite{HO73} only the poles $\frac 2 \epsilon$
are subtracted, whereas in
the modified minimal subtraction scheme ($\overline{\mbox{MS}}$)
\cite{BABU78} the 
poles $\frac 2 \epsilon$
 and  the constants $ - \g_E + \ln 4\pi$ are removed from the $D$-dimensional
expression.
In the $\overline{\mbox{MS}}$ scheme   we find for the renormalized
integrals:
\equa{
\G_2^{\overline{\mathrm MS}} & = 
p^2 - m^2 + m^2 \lambda \Bigl(  1 +
\ln {\mu^2 \over m^2} \Bigr) \nonumber\\
\G_4^{\overline {\mathrm MS}} &=  -\lambda 
+ 
 \frac{1}{(4\pi)^2} \frac{ \lambda^2}{2} \Bigl(
\int_0^1  \mbox{d}z \ln \frac{m^2 - z(1-z)(p_1 + p_2)^2
 -i \varepsilon }{\m^2} +
\mbox{2 Perm.} \Bigr) \/ .
}

\item {\em Momentum subtraction scheme of BPHZ}\\
The renormalized Green's functions in the BPHZ scheme are defined without
a regularization procedure. The finite Green's functions in 1-loop order are
readily  obtained by subtracting the first powers of the 
 Taylor expansion in the external momenta $p_i$
 from the integrand
at  $p_i = 0$.
 The subtraction operator is denoted by $
t^d_{p_1 ... p_n}.$ (Divergent  subdiagrams of higher orders
 are subtracted according to  the
forest formula.) The order $d$ of Taylor subtractions is called the
subtraction degree and coincides in the case considered here with
the  degree of divergency:
\equa{\label{BPHZren2}
\G_2 ^{BPHZ} & =  p^2 - m^2 + 
i \lambda (1- t_p^2 ) 
\int  \frac{\mbox{d}^4k}{(2\pi)^4}\frac{1}{k^2 - m^2}  
  =  p^2 - m^2  \/ ; \\ 
\G_4^{BPHZ} &=  -\lambda -i \frac{ \lambda^2}{2} (1 - t_p^0) 
\int  \frac{\mbox{d}^4k} {(2\pi)^4}  \Bigl( \frac{1}{k^2 - m^2}
 \frac{1}{(k+p_1 + p_2)^2-
m^2} + \mbox{ 2 Perm.} \Bigr)
\nonumber \\
&=   -\lambda -i \left(\frac{\lambda^2}{2} \int\!\!
  \frac{\mbox{d}^4k} {(2\pi)^4} 
\Bigl( 
\frac{1}{k^2 - m^2} \frac{1}{(k+p_1 + p_2 )^2-m^2} -
\frac{1}{(k^2 - m^2)^2}\Bigr)\! +\! \mbox{2 P.} \right)
\nonumber \\
&=  - \lambda + \frac 1{(4\pi)^2}\frac{\lambda^2}{2} 
\int_0^1  \mbox{d}z 
\Bigl(\ln \frac{m^2 - z(1-z)(p_1 + p_2)^2}{m^2} + \mbox{2 Perm.} \Bigr)\/.
\label{BPHZren4}
}
\end{enumerate}

Comparing the finite Green's functions
  of  the MS and  $\overline{ \hbox{MS}}$ scheme with  the ones of the
 BPHZL scheme it is seen, that
the renormalized expressions differ  by constants, but that
 the non-local (logarithmic) contributions
coincide as they stand and illustrate the equivalence of different schemes.
In particular we have
\equa{
\G_2^{BPHZ}& = \G_2^{\overline{ \mathrm{MS}}} - m^2
\lambda (1 + \ln {\mu^2 \over m^2})
\nonumber \\
\G_4^{BPHZ}& = \G_4^{\overline{ \mathrm{MS}}} + \frac{3 \lambda^2}{32\pi^2} \ln
\frac{\mu^2}{m^2}.
}
These constants can be related to counterterms, which are
 added order by order to 
the classical action and appear in the GML formula in higher orders of
perturbation theory. Of course these
counterterms are restricted to have dimension less than or equal to 4
in order not to  violate the properties of naive
renormalizability. 
In the $\varphi ^4$-theory the most general counterterms 
are  given by
\equ{
\label{Gact}
\G_{ct} = \intd 
\sum_{n=1}^\infty\bigl( a^{(n)} \lambda^n \varphi \Box \varphi + b^{(n)}
\lambda^n m^2 \varphi^2 + c^{(n)}\lambda^{n+1} \varphi^4 + \sqrt{\l} \l^ {n}
f^{(n)} \varphi^3
\bigr). 
}
In fact the above calculation has demonstrated, that these counterterms
 are fixed arbitrarily in different schemes
 and have
to be defined uniquely by normalization conditions and symmetries.
Let us consider first the $\varphi^3 $-interaction,
 which can be added from pure power counting arguments. Since the classical
action is invariant under the discrete transformation
\begin{equation}
\label{varphicp}
\varphi \to -\varphi \qquad \Gacl(\varphi) \to \Gacl(- \varphi ) = \Gacl
(\varphi)
\end{equation}
and since the discrete symmetries are not violated in the course of 
renormalization,
this term can be omitted  from the counterterm action.
All the other terms have to be fixed by normalization conditions 
and are interpreted as the wave function renormalization, coupling and
mass renormalization. The classical action and the counterterms 
are summarized in a 
$\Ga_{eff} $:
\begin{eqnarray}
\label{Gaeff}
\Ga_{eff} & = & \intd \bigl(
 \frac 12  z_\varphi^2 \partial_\m  \varphi \partial^\m 
 \varphi- \frac 12
(m + \delta m )^2 z_\varphi^2 \varphi^2  - \frac {z_\lambda \lambda }{4!}
z_\varphi ^4\varphi^4 \bigr) \\
 & \equiv & \intd \bigl( \frac 12  \partial_\m\varphi \partial^\m \varphi-
\frac 12
m^2\varphi^2 \bigr)  + \Ga_{int}  \nonumber
\end{eqnarray} 
with
\begin{equation}
z_\varphi = 1  + O (\hbar)\/ , \quad
z_\lambda = 1  + O ( \hbar)\/, \quad \d m = O (\hbar).
\end{equation}
These coefficients
 are uniquely related to the coefficients $a, b$ and $c $ (\ref{Gact})
order by
order in perturbation theory.
The  arbitrary  coefficients of counterterms
 have to be fixed by three normalization conditions,
namely $z_\l$ on the interaction vertex
\equ{\label{norm4}
\Bigl. \G_4(p_1,p_2,p_3,p_4) \Bigr|_{p_i^2 = \kappa^2
                                          \atop p_i \cdot p_j = - \frac 13
\kappa^2 } = - \lambda,
}
and $z_\varphi$ and $\delta m$ on the 2-point function:
\equ{\label{norm2}
\Bigl. \G_2 \Bigr|_{p^2 = m^2} = 0 
\qquad \mbox{and} \qquad
\Bigl. \partial_{p^2} \G_2 \Bigr|_{p^2 = \kappa^2} = 1.
}
The condition on the 4-point function fixes the coupling
at the symmetric Euclidean  momentum  $p^2 = \kappa^2 $ to its
tree value.
The first condition on the 2-point function means that the  mass 
parameter appearing
in the Green's functions is the physical mass since it is the pole of the
propagator, 
the second condition fixes the residue of the pole to  unity at the
 normalization 
momentum $p^2 = \kappa^2$. Applying 
the
normalization conditions to the renormalized  Green's functions,
scheme dependence of counterterms is removed and
the result is unique and does not depend on the scheme, which  one has used
for  making the Green's functions finite:
\begin{eqnarray}
\Ga_2 (p^2)& = & p^2 - m^2   \\
\Ga_4 (p^2)& = & - \l + 
\frac 1{(4\pi)^2}\frac{\lambda^2}{2} 
\int_0^1  \mbox{d}z 
\Bigl(\ln \frac{m^2 - z(1-z)(p_1 + p_2)^2- i \varepsilon}
{m^2 - z(1-z)\frac 43 \kappa^2}
 + \mbox{2 Perm.} \Bigr)\nonumber
\end{eqnarray}
The $\Ga_{eff}$ (\ref{Gaeff}), however, which governs the evaluation of Green's functions
in the GML formula,  depends on the scheme
which one has used for making finite
 the infinite integrals. Therefore, in a scheme independent proof of
renormalizability one never refers to the properties of a $\Ga_{eff}$,
but only to properties of the finite renormalized Green's functions.
\par
In the Standard Model there are a lot of
 counterterms, which can be added from pure
power counting arguments to the action. As it was in the $\varphi^4$-model
(cf.~(\ref{varphicp})),
discrete and global symmetries as electric charge conservation and
lepton and baryon number conservation
are conserved in the procedure of
 renormalization and we are  able to restrict the counterterms
according to these symmetries.
 For example the general renormalizable 
action with bilinear terms in the vector-scalar fields is given by
\equa{
\label{Gabilgen}
\Ga^{bil}_{gen} &  = 
\intd
 \Bigl( - \frac 14 Z^V_{ab} (\partial^\mu V_a^\nu - \partial^\nu V_a ^\mu) 
(\partial_\mu V_{b\nu} - \partial_\nu V_{b \mu})- 
           \frac 12 \tilde Z^V_{ab} \bigl( \partial^\mu V_{a\mu}\bigr)^2
          + \frac 12 {\cal M}^ V_{ab} V^\mu_a V_{\mu b}  \nonumber \\
& \phantom{ = \int} +
\frac 12 Z^S_a \tilde I_{ab}
          \partial^\mu
 \phi_a \partial_\mu \phi_b
          - \frac 12 m^S_{a} \phi_a \tilde I_{ab} \phi_b
              + D_{a,b} V^\mu_a \partial_\mu \varphi_b \Bigr).
}

Applying CP invariance and charge neutrality it is seen 
 that $Z^S$ as well as $m^S$ are diagonal in the neutral sector 
and off-diagonal and real
in the charged sector, whereas $Z^V$ and $M^V$ are
 real but non-diagonal
in the neutral sector. Of course some of these constants are fixed
by normalization conditions, as it is for the mass matrix of vectors
and the mass of the Higgs.  Other counterterms in (\ref{Gabilgen}) 
 as $D_{a,b}$ are determined
by the symmetries as it is seen
from the classical action. 

 In the (complete) on-shell scheme
the mass matrix
of vectors is fixed by the following normalization conditions 
\cite{FLJE81, AOHI82, Denner}: 
\equa{
\label{onshell}
\RE \Bigl. \G_{W^+W^-} \Bigr|_{p^2 = M_W^2} = 0, 
\qquad
\RE \Bigl. \G_{ZZ} \Bigr|_{p^2 = M_Z^2} &= 0, 
\qquad 
\Bigl. \G_{AA} \Bigr|_{p^2 = 0} = 0, 
\nonumber\\
\RE \Bigl. \G_{ZA} \Bigr|_{p^2 = M_Z^2} = 0, 
\qquad &
 \Bigl. \G_{ZA} \Bigr|_{p^2 = 0} = 0.
}
With these conditions  the mass matrix of vectors is diagonalized  on-shell. 
Since $Z$ and $W^\pm$ are unstable particles, their self energies are not
real.  The on-shell conditions do not seem  to be  the appropriate conditions
for describing unstable particle in higher orders, but we want to indicate
here that there are free counterterms available for fixing the 
masses of particles and for diagonalizing the mass matrix. Then 
 on-shell conditions can be replaced
immediately by the appropriate normalization conditions
as for example pole conditions in higher orders.

In the course of algebraic renormalization counterterms have to be
characterized algebraically by the symmetries of the model.
In particular one has to 
 distinguish
the invariant
  counterterms that are fixed by normalization conditions  from non-invariant
counterterms
which are fixed by the symmetries. This classification is carried out
when  one solves the defining symmetries, 
 the Ward identities and the
ST identity, for the most general local field polynomial compatible with
power counting renormalizability. 
The proof of renormalizability is finished by proving that the defining
symmetries
of the model can be established in higher orders by adjusting non-invariant
counterterms appropriately. (For an introduction to algebraic renormalization
see $[$R5,R6$]$.)
The basis for this proof is the quantum action principle for off-shell Green's
functions,
whose content and consequences for renormalization  
we outline in the following subsection.

\newsubsection{The quantum action principle}

The classical action of the Standard Model satisfies
the ST identity (\ref{ST})
\begin{equation}
{\cal S}(\Gacl) = 0,
\end{equation}
Ward identities of rigid $SU(2)$ symmetry and global electromagnetic charge
conservation (\ref{wardrig})
\begin{equation}
{\cal W}_\a \Gacl = 0, \qquad {\cal W} _{em} \Gacl = 0.
\end{equation}
Starting from the classical action one can immediately calculate the
Green's functions of 1-loop order, by using the Gell-Mann--Low formula and
Wick's theorem, or equivalently using Feynman diagrams and Feynman rules
as described in the last subsection. The divergent Green's functions are
renormalized by a well-defined subtraction scheme as we have presented
in the example of the $\varphi^4 $-theory. (Feynman rules of the Standard
Model
and standard 1-loop 
diagrams  evaluated in dimensional regularization are 
given in several publications. See e.g.~\cite{AOHI82,BOHO86,Denner}). 

The finite 1PI Green's functions are summarized in the generating functional
of 1PI Green's functions.
\begin{equation}
\Ga [\varphi_k] = \sum_{n=0}^\infty \frac 1{n!} \int \mbox{d}^4 x_1 \ldots
\mbox{d} ^4 x_n
\sum_{i_1, \ldots i_n} \varphi_{i_1} (x_1)\varphi_{i_2} (x_2)  \cdots 
\varphi_{i_n} (x_n) \G_{\varphi_{i_1}\ldots\varphi_{i_n}}
( x_1, \ldots x_n)
\end{equation}
Here $\varphi_k$ denotes the  different fields of the Standard Model
and $\G_{\varphi_{i_1}\ldots\varphi_{i_n}} $ the 1PI Green's functions with
external amputated legs ${\varphi_{i_1},\ldots \varphi_{i_n}}$
\begin{equation}
\label{1PIdef}
\G_{\varphi_{i_1}\ldots\varphi_{i_n}}(x_1,\ldots x_n)  =
\frac 1i < T {\varphi_{i_1} (x_1)
\ldots\varphi_{i_n} (x_n)} > \Big|_{\mathrm{ 1PI\ diagrams} \atop
                                        \mathrm{ amputated\ legs}}  
\end{equation}
In perturbation theory the generating functional of 1PI Green's functions is expanded in orders of $\hbar $, which agrees
with the loop order and the expansion in the coupling constant. The lowest
order is the classical action:
\begin{equation}
\label{Gao}
\Ga = \sum_{k= 0}^\infty \Ga^{(k)} \qquad \Ga^{(0)} = \Ga_{cl} 
\end{equation}

The proof of renormalizability is an induction proof;
therefore  we have in a first
step to prove that the symmetries of the tree approximation
 can be established also in 1-loop order
\begin{eqnarray}
\label{ST1loop}
{\cal S} (\Gacl) = 0 & \Longrightarrow &
\Bigl( {\cal S} (\Ga ) \Bigr) ^{(\leq 1)}
 = 0  \\
{\cal W}_\a \Gacl = 0 & \Longrightarrow & 
\Bigl( {\cal W}_\a \Ga  \Bigr) ^{(\leq 1)} = 0 \nonumber
\end{eqnarray}
Finally we have also to establish the local Ward identity 
(\ref{abwardid}) in 1-loop order.
The global symmetries as electric charge conservation, lepton and baryon
number conservation as well as discrete CP symmetry are trivially established.
Having carried out the step from the classical approximation to 1-loop order
the step from order $n$ to $n+1$ can be done in analogy if none of the
initial conditions as power counting renormalizability and infrared
existence have changed.

In the following we denote the finite 
scheme-dependent renormalized 1-loop Green's
functions by $\Ga^{(1)}_{ren}$. 
As in the example of the $\varphi^4 $-theory 
 we are able to add arbitrary counterterms
 in 1-loop order. The Green's functions of the Standard Model
are finally determined as a sum of the renormalized scheme-dependent
contributions and local counterterms (see (\ref{Gabilgen})):
\begin{equation}
\Ga^{(\leq 1)} = \Gacl + ( \Ga_{ren}^{(1)} + \Ga^{(1)}_{ct} )
\end{equation}
Applying the ST operator and the Ward operators of the tree 
approximation (\ref{ST}) and (\ref{wardrig}) to this expression we
obtain:
\begin{eqnarray}
\label{STcount}
{\cal S} (\Ga^{(\leq 1)} ) &= & {\cal S}
\bigl( \Gacl + ( \Ga_{ren}^{(1)} + \Ga^{(1)}_{ct} ) 
\bigr)  \\
& = & \brs _{\Gacl}  \Ga_{ren}^{(1)} + \brs _{\Gacl} \Ga^{(1)}_{ct}  
+ O (\hbar^2) \nonumber\\
{\cal W}_\a \Ga^{(\leq 1)}  &= & {\cal W}_\a
\bigl( \Gacl + ( \Ga_{ren}^{(1)} + \Ga^{(1)}_{ct} ) 
\bigr) \nonumber \\
& = & {\cal W}_\a \Ga_{ren}^{(1)} + {\cal W}_\a \Ga^{(1)}_{ct}   \nonumber
\end{eqnarray}
The operator $\brs_\Ga $ is  the linearized version of the
ST operator:
\begin{eqnarray}
\label{brsop}
\brs _ \Gamma &=& \intd  \biggl(  
\bigl(\sw \partial _\mu c _Z + \cw \partial_\mu c_A\bigr)
             \Bigl(\sw {\delta  \over \delta Z_\mu } + \cw
       {\delta  \over \delta A_\mu} \Bigr)  \\
& & \phantom{\int \biggl(} +  B_a {\delta  \over \delta \bar c_a }  
   + \hat {\mathbf q}{\delta  \over \delta \hat \Phi  }  + 
{\delta  \over \delta \hat \Phi ^\dagger } \hat {\mathbf q}^\dagger
   \phantom{\int \biggl(} + \sum_{\varphi_k, \Upsilon_k} u_k
\Bigl({\delta \Ga\over \delta \Upsilon_k} {\delta\over \delta \varphi_k}
+ {\delta\Ga\over \delta \varphi_k} {\delta \over \delta \Upsilon_k}
\Bigr)
\biggr) \nonumber
\end{eqnarray}
 We have generically written $\varphi_k$ for the fields and
$\Upsilon_k$ for the corresponding external sources in the theory. 

If we want to prove that symmetries can be established in 1-loop 
order, we have to show that all scheme dependent breakings in 1-loop order
 can be cancelled by adding appropriate counterterms, i.e.:
\begin{equation}
\label{renct}
 {\cal W}_\a \Ga_{ren}^{(1)} \stackrel != -
 {\cal W}_\a \Ga^{(1)}_{ct}  \qquad \quad
\brs _{\Gacl}  \Ga_{ren}^{(1)} \stackrel !=  - \brs _{\Gacl} \Ga^{(1)}_{ct}  
\end{equation}
(Here $ \stackrel != $ denotes that the equality of both sides of these
equations has to be proven.) For proving these equalities
the most important input comes from the quantum action principle 
\cite{LOW71,LAM73}.
It relates differentiations with respect to parameters and with respect to
fields to insertions (see $[$R3$]$ -- $[$R6$]$
and  below,  in particular (\ref{AP})).
 In its most general form it has been formulated
even independent of a specific renormalization scheme \cite{LAM73}.
Applying the quantum action principle to the symmetry operators involved here
we find that the symmetries of the tree approximation can be at most
broken by {\it local} field polynomials
 with UV-dimension less or equal than 4 in 1-loop order: 
\begin{eqnarray}
\label{Delta1loopa}
{\cal S} (\Gacl + \Ga_{ren} ^{(1)} ) &= & \Delta_{brs}^{(1)} + O(\hbar^2), \\
\label{Delta1loopb}
{\cal W}_\a (\Gacl + \Ga_{ren} ^{(1)} ) &= & \Delta_\a^{(1)} + O(\hbar^2) 
\end{eqnarray}
with
\begin{equation}
\label{DeltaUV}
\dim^{UV} \Delta_{brs}^{(1)} \leq 4 \qquad \mbox{and}\quad
\dim^{UV} \Delta_{\a}^{(1)} \leq 4  .
\end{equation}
In particular the proof of renormalizability is completely traced
back to an (algebraic) analysis of local field polynomials and (\ref{renct})
is simplified to the expressions
\begin{eqnarray}
\label{breakct}
\Delta_{brs}^{(1)} + \brs_{\Gacl} \Ga_{ct}^{(1)} & \stackrel{!}{=} &  0 \\
\Delta_{\a}^{(1)} + {\cal W}_{\a} \Ga_{ct}^{(1)} & \stackrel{!}{=} & 0 .
\nonumber
\end{eqnarray}
A characterization of all possible breakings is obtained by the algebraic
method, which  will be presented in the following section. 
Before we turn to the
algebraic method we want
to make a few remarks on the quantum action principle. 

In its general form the action principle relates field and parameter
differentiations acting  on the generating functional of Green's functions
 to 
insertions into the respective Green's functions. According to the dimension
of fields appearing in the  differential operators 
the field polynomials of the insertions
 have a definite upper
UV dimension in all power counting renormalizable theories.
 In the BPHZL scheme the quantum 
action principle takes a simple form and relates the differential operators
to Zimmermann's normal products \cite{LOW71, CLLO76}.
 Furthermore  the insertions can be 
expressed in terms of the
(scheme-dependent) $\Ga_{eff}$. Here we will restrict ourselves to
the most important properties of insertions.
First we want to give the
definition of an insertion.  Green's functions with insertions are  quite
analogously
determined  as 
 ordinary Green's functions: Factors and Feynman rules are given by the 
formal expansion of the generalized Gell-Mann--Low formula, which
defines  Green's functions with insertions in a formal way:
\equ{ 
\langle T O (x) \varphi_{i_1}(x_1) \cdots \varphi_{i_n}(x_n) \rangle = {\cal R}
\langle T 
: O^{(0)}(x): \varphi^{(0)}_{i_1}(x_1) \cdots \varphi^{(0)}_{i_n}(x_n) e^{i 
\G_{int}(\varphi^{(0)}_k)} \rangle.
\label{eq: GMLins}
} 
With $O(x)$ we denote an arbitrary field polynomial composed of propagating
fields of the model. Examples  for such polynomials
are $\varphi^2 (x) $ and $\varphi^4 (x)$ in the $ \varphi^4$-theory.
 Integrated insertions 
usually denoted by $\Delta $ are defined
by carrying out the $x$ integration  in  the above formula
$\Delta = \int \mbox{d}^4 x O (x)$.
The Green's functions with a certain
insertion are again summarized in the generating functional
of Green's functions with insertion.
From here one is able to define  connected and finally 
1PI Green's functions with insertion
 by  Legendre transformation. The generating functional of 
1PI Green's functions with  the non-integrated insertion $O(x)$ and the
integrated insertion $\Delta $
are  denoted by
\begin{equation}
[O (x)] \cdot \Ga \qquad \mbox{and} \qquad [ \Delta ] \cdot \Ga
\end{equation}
and 
\begin{equation}
{\delta^n \over \delta \varphi_{i_1} (x_1) \ldots 
\varphi_{i_n} (x_n)}  [O (x)] \cdot \Ga 
=  {\cal R } < T :O (x): {\varphi_{i_1} (x_1)
\ldots\varphi_{i_n} (x_n)} > \Big|_{\mathrm{1PI\ diagrams} \atop
                                         \mathrm{amputated\ legs} } 
\nonumber
\end{equation}
1PI Green's functions  have the same obvious
diagrammatic interpretation  as ordinary Green's functions.
It is important to note that the lowest order in the perturbative
expansion is a local expression and given by the field polynomial $O(x)$:
\begin{equation}
[O(x)] \cdot \Ga = O(x) + O(\hbar) .
\end{equation}
(This is analogous to the observation, that  $\Gacl$ is the lowest
order of the generating functional of 1PI Green's functions (see (\ref{Gao}).)

Of course insertions of field polynomials into loop diagrams are in general
divergent and have also to be made meaningful by renormalization.
Similarly as for ordinary 1PI Green's functions we find the following
 degree  of divergency $d_{\Ga_O}$ of a 1PI Green's functions
with one insertion $O(x)$: 
\begin{equation}
d_{\Ga_{O}} = 4- N_B - \frac 32 N_F + \sum_V (d_V -4) + (d_O -4). 
\end{equation}
Here the notation is the same as in (\ref{divdeg}), and $d_O$ denotes
the dimension of the field polynomial $ O (x)$.
(For  example in the $\varphi^4$-theory we have $d_{\varphi^4}= 4$
and $ d_{\varphi ^2} = 2$.) In the BPHZ scheme the renormalized Green's
functions with insertions are defined by Taylor subtraction. The number
of Taylor subtractions are given by the subtraction degree, which
is in a renormalizable theory given by
\begin{equation}
\delta_{\Ga_{O}} = 4-N_B - \frac 32 N_F +  (\delta_O -4), 
\end{equation}
and $\delta _O \geq d_O $ defines the subtraction degree. 
(For  example in the $\varphi^4$-theory one often has to
consider ${\varphi ^2} $-insertions with  $\d_{\varphi^2}= 4$.)
 In the BPHZ scheme the  Green's functions with insertions are therefore given
together with their subtraction degree $\delta$: With the notation
$$
[O(x)]_\delta \cdot \Ga
$$
 the Green's functions with insertion are completely defined.

The quantum action principle relates field differentiations to insertions with
a well defined UV-degree $\delta$. For the purpose of the present lectures
we need the following forms of the action principle:
 variations of propagating fields as they appear
in the Ward operators of rigid symmetry  (\ref{wardrig})
and products of field variations
 with respect to
a propagating and an external fields as they appear in the ST operator
(\ref{ST}): 
\begin{eqnarray}
\label{AP}
\varphi_k(x) {\delta\Ga \over \delta  \varphi_l (x)} & = & 
[ O (x) ]_{\delta_O} \cdot \Ga \quad\mbox{with} \quad 
\delta _O = 4- \dim^{UV} \varphi_l + \dim^{UV} \varphi_k \\
\int \mbox{d}^4 x {\delta\Ga \over \delta  \Upsilon_k (x)}
  {\delta\Ga \over \delta  \varphi_l(x)} & = & [\Delta ]_{\delta_\Delta}
 \cdot \Ga
\quad\mbox{with} \quad 
\delta _\Delta  = 4- \dim^{UV} \varphi_l + 4 - \dim^{UV} \Upsilon _k 
\nonumber
\end{eqnarray}
The lowest order of $\Delta$ and $ O(x)$ is given  in expressions of
the classical action:
\begin{eqnarray}
O(x) & = &  \varphi_k {\delta\Gacl \over \delta  \varphi_l}  + O(\hbar) \\
\Delta & = & \int \mbox{d}^4x {\delta\Gacl \over \delta  \Upsilon_k (x)}
  {\delta\Gacl \over \delta  \varphi_l (x)}  + O(\hbar) .
\nonumber
\end{eqnarray}
Field polynomials appearing in higher orders are scheme dependent but
restricted
by the UV-degree $\delta_O$ and $\delta_\Delta$:
\begin{equation}
\dim^{UV} O(x) \leq  \delta_O \qquad \quad \dim^{UV} \Delta \leq 
\delta_\Delta .
\end{equation}

Applying the quantum action principle as given in the above formula to the
Standard Model we find
\begin{eqnarray}
{\cal S}(\Ga) & = & [ \Delta_{brs}]_4 \cdot \Ga \quad \mbox{with} \quad
\Delta_{brs} = {\cal S} (\Gacl) + O (\hbar) = O (\hbar), \\
{\cal W}_\a\Ga  & = & [ \Delta_{\a}]_4 \cdot \Ga \quad \mbox{with} \quad
\Delta_{\a} = {\cal W}_\a \Gacl + O (\hbar) = O (\hbar). 
\end{eqnarray}
Using that the lowest order of the insertion is a local field polynomial
we  arrive immediately at (\ref{Delta1loopa}) and (\ref{Delta1loopb}), 
 where the upper
UV dimension of field polynomials
is given by the subtraction degree of the insertion (\ref{DeltaUV}).

In the Standard Model and quite generally in  gauge theories with unbroken
gauge groups
there are massless particles. For this reason, one has to assign to
every field also an infrared (IR) dimension \cite{LOW76}.
 Insertions are  defined
by giving an subtraction degree not only with respect to their
UV but also with respect to their IR dimension \cite{CLLO76}.
Then the local field polynomials $\Delta_{brs}$ and
$\Delta_\a $ are in addition restricted with respect
to their infrared degree \cite{LOW76,CLLO76}.
In the Standard Model we obtain from the pure power counting analysis
\begin{equation}
\dim^{IR} \Delta_ {brs} \geq  3, \qquad \quad
\dim^{IR} \Delta_\a \geq 2.
\end{equation}
A complete list of the UV- and IR-dimension of the fields
appearing the Standard Model is given in ref.~\cite{KR98}.

\newsubsection{The algebraic method}

With the algebraic method  one has to characterize the 
counterterms and the breakings by the defining symmetries of the model.
 In the algebraic characterization of counterterms 
the free parameters of the model are determined and the normalization
conditions and symmetries are identified. Then
the Green's functions   can  be uniquely defined 
independently of a specific (invariant) scheme. 
In the second step the possible breakings of the symmetry operators
are restricted by algebraic consistency, and in this way it is possible
to find out, if  eq.~(\ref{breakct}) can be solved by adjusting appropriate
counterterms. 

The first step is called in the literature the general classical
solution, since one solves the defining symmetry identities for
all integrated 
 field polynomials allowed by the power counting renormalizability. Neglecting
in a first step the local Ward identity (\ref{abwardid}),
 the defining symmetries 
are the ST identity (\ref{ST})
 and the Ward identities of rigid symmetry (\ref{wardrig}):
\begin{equation}
{\cal S} (\Ga ) = 0\/, \qquad
{\cal W}_\a \G = 0 \qquad \mbox{and} \qquad {\cal W}_{em} \Ga = 0.
\end{equation}
In usual gauge theories with simple gauge groups these symmetry operators
are defined by their tree approximation. Since the gauge group of the
Standard Model is non-semisimple and since the unbroken gauge group does
not correspond to the $U(1)$-group, such a procedure is not satisfactory
for renormalizing the Standard Model. 
In particular, when we try to proceed as usually, it is seen that there
are 
 not  available enough free parameters to establish the normalization
 conditions
of the  on-shell scheme for the vector and ghost fields
(cf.~(\ref{onshell})). Due to the presence of the massless
photon such normalization conditions are crucial for obtaining 
off-shell finite Green's functions
 in higher orders. 
Therefore the symmetry operators have to be themselves subject of
renormalization, especially the weak mixing angle expressed in the on-shell
scheme by the mass ratio of vector bosons, does get higher order corrections
and cannot be fixed to its tree value in the symmetry operators. 

For this reason we have to generalized the notion of invariant counterterms:
Instead of taking the ST identity and Ward identities
of the tree approximation, we take the
most general operators compatible with the algebra (\ref{brsnil}) --
(\ref{cons})
and call counterterms
invariant if they satisfy these generalized identities (\ref{Gagencl}).
For the ST operator we require the
following properties of nilpotency:
\begin{eqnarray}
\label{brsnil}
\brs _\Gamma \,{\cal S}
 (\Gamma) &=& 0 \quad \hbox{for any functional}\, \Gamma, \\
\brs _\Gamma \, \brs_\Gamma &=& 0 \quad \hbox{if} \quad {\cal S}(\Gamma)
=0. 
\nonumber
\end{eqnarray}
The 
Ward operators  ${\cal W}_\a$ are required to fulfil the $SU(2)$
algebra
\begin{equation}
\label{rigalg}
[{\cal W}_\a, {\cal W}_\b] = \epsilon_{\a\b\g} \tI_{\g\g'} {\cal W}_{\g'}.
\end{equation}
Finally ST operator and the Ward operators have to fulfil the
consistency equation:
\equ{
\label{cons}
\brs_\G {\cal W}_\a \Ga - {\cal W}_\a {\cal S}(\G) = 0.
}
These properties are valid for the  operators of the tree approximation
(\ref{ST}) and (\ref{brsop}) and (\ref{wardrig}).

For determining the general classical solution of general symmetry operators,
i.e.~the invariant counterterms,
one has to solve the algebra as well as the defining symmetry identities
for the most general power counting renormalizable action:
\begin{equation}
\label{Gagencl}
{\cal S}^{gen} (\Gacl^{gen}) = 0 \qquad
{\cal W}_\a ^{gen} (\Gacl^{gen}) =0
\end{equation}
and
${\cal W}_\a^{gen} $ and ${\cal S} ^{gen}$ fulfil  equations (\ref{brsnil}),
(\ref{rigalg}) and (\ref{cons}) and 
\begin{equation}
\dim ^{UV} \Gacl^{gen} \leq 4 
\end{equation}
$\Gacl^{gen}$ as well as the symmetry operators are restricted according to
the global and discrete symmetries (CP invariance !) of the model
(cf.~the discussion after eq.~(\ref{Gabilgen})).
An outline of the main steps  of the solution can be found in
\cite{KR98}. Here we give the most important results:

The most general solution is gained from the special solution of the
classical approximation $\Gacl$ by  redefining all fields with the
most general matrix allowed by discrete and global symmetries.
Of course these field redefinitions  have to be carried out in
the ST operator and in the Ward operators of rigid symmetry. It is seen
that such field redefinitions renormalize the operators in accordance with 
the algebra.
For the vectors one is able to introduce a non-diagonal wave function
redefinition matrix $z^V_{ab}$ in the neutral sector, whereas the
redefinition matrix of scalars is diagonal due to CP invariance.
\begin{equation}
\label{fred}
z^V_{ a b} = \left(\begin{array}{cccc}
            \hat  z_W& 0& 0 \\
               0       &    z_W & 0& 0\\
               0 & 0 &   z_Z \cos \theta_Z & - 
                            z_A \sin \theta_A    \\
              0 & 0 &    z_Z \sin \theta_Z &     z_A \cos \theta_A 
                   \end{array} \right)  
\qquad
z^S_{ab}  = \left(\begin{array}{cccc}
               z_+ &0& 0 & 0 \\
               0 & z_+& 0 & 0 \\
               0 & 0  &  z_H & 0\\
               0 & 0 & 0 &  z_\chi 
        \end{array} \right)
\end{equation}
Similar general field redefinitions can be carried out
 for the Faddeev-Popov ghosts and the  
fermions. Further free parameters are the parameters listed in
(\ref{onshellpar}) and the gauge parameters 
$\xi, \hat \xi , \zeta$ and $  G $ of 
the general gauge fixing (\ref{Gagfgen}).
In this way one is able to carry out mass diagonalization on-shell and
to give 
  normalization  conditions
for   all the residua   in accordance with the symmetry operators.
Due to the fact, that the general field redefinitions enter the
Ward operators, 
invariant counterterms in 1-loop order
are  characterized by the
equations
\begin{equation}
\label{invgen}
\brs_{\Gacl} \Ga^{(1)}_{inv} +  \delta {\cal S}^{(1)} \Gacl = 0\/ ,
\qquad
{\cal W}_\a  \Ga^{(1)}_{inv} +  \delta {\cal W}_\a^{(1)} \Gacl = 0 .
\end{equation}
Consequently non-invariant counterterms are called such counterterms
which cannot be arranged to fulfil equations (\ref{invgen})
by an adjustment of parameters in the 1-loop operators.
By solving the general classical approximation we have now
splitted uniquely the counterterms into invariant and non-invariant
counterterms and have specified at the same time all the possible
normalization conditions. In the fermion sector of course not all
the abelian couplings are specified by the solution the ST identity and
Ward identities of rigid symmetry, but
we find the couplings of the abelian field combination to lepton and
baryon number conserving currents order to order as free parameters of the
model. For this reason one has finally to establish the Ward identity of
local abelian gauge symmetry (\ref{abwardid}) also in higher orders.

According to eq.~(\ref{breakct}) we have finally to prove that
all breakings can be written as variations of the counterterm action.
Again scheme invariance of global and discrete symmetries immediately
restricts
breakings according to their electric and Faddeev-Popov charge and according to
their behaviour under CP transformations.  
Then we  apply the classical ST operator 
$\brs_{\Gacl}$ and Ward operators ${\cal W }_\a$ on eqs.~(\ref{Delta1loopa})
and (\ref{Delta1loopb}). Using the algebraic properties of the operators
(\ref{brsnil}) -- (\ref{cons}) 
we get:
\equa{
\brs_{\Gacl} \Delta_{brs}^{(1)} &= 0,  \nonumber \\
\brs_{\Gacl} \Delta_{\a}^{(1)} - {\cal W}_\a
\Delta_{brs}^{(1)} &= 0, \\
{\cal W}_\a \Delta_\b^{(1)} - {\cal W}_\b \Delta_\a^{(1)}
&= 0. \nonumber
}
These equations restrict strongly the  possible breakings.
It is seen immediately that all breakings, which are variations
\begin{equation}
\Delta^{var}_{brs} = \brs_{\Gacl} P_{ct}  
 \qquad \Delta^{var}_\a = {\cal W}_\a P_{ct}
\end{equation}
and
\begin{equation}
\dim^{UV} P_{ct}\leq   4
\end{equation}
 satisfy
the above consistency equations. Further solutions of the equations, which
cannot be written in the form of a variation are the 
Adler-Bardeen anomalies \cite{ADL69,BEJA69,BAR69}.
 For their explicit form in the Standard Model
we refer to \cite{KR98}. They are seen to cancel in 1-loop order according
to the appearance of lepton and quark pairs and vanish to all orders
according to the non-renormalization theorems proven in \cite{BABE80}.
Therefore all breakings can be written as variations of  4-dimensional
field polynomials.

Finally we have to show, that we are able to add the field polynomials
$P_{ct} $ to the counterterm action without being in conflict with
infrared existence and on-shell normalization conditions conditions.
Indeed it turns out that on-shell schemes and a complete normalization
of residua fix uniquely all field polynomials appearing in $\Ga_{inv}$.
Establishing the normalization conditions by adding such counterterms we find
\begin{eqnarray}
{\cal S} (\Gacl + \Ga_{ren}^{(1)} + \Ga_{inv}^{(1)}) 
 & = & \brs_{\Gacl} P_{ct} ^{(1)} + \brs_{\Gacl} 
\Ga_{inv}^{(1)} 
  + O(\hbar^2)  \\
{\cal W}_\a (\Gacl + \Ga_{ren}^{(1)} + \Ga_{inv}^{(1)}  )
 & = & {\cal W}_\a P^{(1)} _{ct} + {\cal W}_\a 
\Ga_{inv}^{(1)} 
  + O(\hbar^2)  
\end{eqnarray}
From the definition of invariant counterterms (\ref{invgen}) it is
obvious that some invariants are naive invariants of the tree operators
and other invariants break the symmetry of the tree operators:
\begin{eqnarray}
& &\Ga_{inv} = \Ga^{o}_{inv} + \Ga_{break}^{o}  \qquad \mbox{with}\nonumber \\
& & \brs_{\Gacl} \Ga^{o}_{break} = - \delta {\cal S}^{(1)} \Gacl 
\quad \hbox{and} \quad \brs_{\Gacl} \Ga^{o}_{inv} = 0
\end{eqnarray}
(The superscript $o$ indicates
that we have splitted the generalized invariants (\ref{invgen}) 
into invariants and breakings of
tree operators.)
In the same way $P_{ct}$ can be splitted into non-invariant counterterms
and such counterterms which are invariant in the generalized sense
of (\ref{invgen}) but break the symmetry of the tree operators.
Having already disposed of invariant counterterms for establishing
the normalization conditions we are not able to dispose of
the invariant counterterms for establishing the symmetry. 
But according to their definitions
these breakings can just be absorbed into a redefinition of
the ST operator  and Ward operators. These redefinitions become unique
if we take into account the algebraic properties of the
symmetry operators. Finally we obtain the following equations:
\begin{eqnarray}
{\cal S} (\Gacl + \Ga_{ren}^{(1)} + \Ga_{inv}^{(1)}  ) & = & 
\Delta_{brs} ^{(1)} + \brs_{\Gacl} \bigl(
\Ga^{o}_{break} \bigr) ^{(1)}   + O(\hbar^2)  \\
& = & \brs_{\Gacl} ( P^{o}_{break} + \Ga^{o}_{break}) ^{(1)} + \brs_{\Gacl} 
P_{noninv}^{(1)}  + O(\hbar^2)  \nonumber \\
& = & - \delta {\cal S }^{(1)} \Gacl -
\brs _{\Gacl} \Ga_{noninv}^{(1)}   + O(\hbar^2 ) \nonumber \\
{\cal W}_\a (\Gacl + \Ga_{ren}^{(1)} + \Ga_{inv}^{(1)}  )
 & = & \Delta_{\a} ^{(1)} + {\cal W}_\a \bigl(
{\Ga^{o}_{break}}\bigr)^{(1)}   + O(\hbar^2)   \\
& = & {\cal W}_\a (P^{o}_{break} + \Ga^{o}_{break})^{(1)}
+ {\cal W}_\a P_{noninv}^{(1)}   + O(\hbar^2 ) \nonumber  \\
&=& - \delta {\cal W}^{(1)}_\a \Gacl - {\cal W}_\a \Ga_{noninv}^{(1)}  
 + O(\hbar^2 )  \nonumber 
\end{eqnarray}
Therefore we are able to establish all normalization conditions and
to remove all the breakings by adjusting non-invariant counterterms
and symmetry operators:
\begin{eqnarray}
 {\cal S}^{gen} (\Ga) \equiv
{\cal S} (\Gacl + \Ga_{ren}^{(1)} + \Ga_{inv}^{(1)} +
\Ga_{noninv}^{(1)} ) + \delta {\cal S}^{(1)}  \Gacl + O(\hbar^2) & = &
O(\hbar^2)  \nonumber  \\
 {\cal W}^{gen}_\a \Ga \equiv
{\cal W}_\a (\Gacl + \Ga_{ren}^{(1)} + \Ga_{inv}^{(1)} +
\Ga_{noninv}^{(1)} ) + \delta {\cal W}_a^{(1)}  \Gacl + O(\hbar^2) & = &  
O(\hbar^2) .
\end{eqnarray}

The proof to all orders can be immediately finished by induction, i.e.\ 
one has to  go  through all the  steps above
from order $n$ to order  $n +1 $
and one has to  realize that none of the initial conditions as power counting
renormalizability, infrared existence and global symmetries have changed 
by the adjustment of 1-loop counterterms. Then
 the quantum action principle can be applied in the same way
as in 1-loop order.  The point where one has to be careful in proving
renormalizability of the Standard Model is infrared existence of
Green's functions.
  Due to the fact,  that the mass matrix of vector
bosons (and the one of Faddeev Popov ghosts) 
can be diagonalized in accordance with the 
symmetries on-shell,
 we are indeed able to proceed to higher orders as it was
from the classical approximation to the 1-loop order and the renormalizability
as well as infrared existence  is proven to all orders.

 \newpage
\newsection{Summary}
In these lecture notes we have discussed the 
renormalization of the electroweak Standard Model by using the method 
of algebraic renormalization.
According to the fact, that the renormalization of the electroweak
Standard Model cannot be based on an invariant scheme, we have to characterize
the model completely by its symmetries. Due to the non-semisimple gauge group
and the specific form of the spontaneous symmetry breaking the characterization
by symmetries requires quite a few generalizations compared to theories
with simple groups. 
 For clarity we review the main steps of our lectures  here briefly again. 
\par
We started from the free massless Dirac action of fermions and constructed 
the symmetry operators which produce the currents of weak and
electromagnetic interactions. In this way we found quite naturally to
the $SU(2) \times U(1) $ gauge structure of electroweak interactions.
When we coupled the currents to vector fields,
we required  a local gauge symmetry
to hold for the enlarged theory. Then the interactions as well as
the transformation of vectors are fixed. 
\par
So far we have worked with the massless gauge theory.
Mass terms for fermions were not allowed 
since they break $SU(2) \times U(1)$ symmetry of the theory.
We noted however that the mass terms transform  covariantly
under $SU(2) \times U(1)$. Therefore we are able to couple them to scalars
and require again that the transformations satisfy the $SU(2) \times U(1) $
algebra. Then the transformation of scalars is fixed. The action of
the Glashow-Salam-Weinberg model is then constructed by giving the most
general 4-dimensional action invariant under the spontaneously broken
symmetry transformations.
Apart from the $SU(2)\times U(1)$ gauge symmetry with the unbroken
electromagnetic  gauge symmetry, we identified two further global symmetries:
 the conservation of lepton and baryon family number. 
In these lectures we did not consider mixing of different fermion families,
especially we have been able to require CP invariance in the construction
of higher
orders.
\par
In order to have renormalizability by power counting we added to the
Glashow-Salam-Weinberg action the
  gauge-fixing functions in the so called $R_\xi$-gauges.
For having nilpotency of the BRS transformations the gauge fixing functions
have been coupled to the auxiliary fields $B_a$. 
Furthermore it was noticed that the $R_\xi$-gauges break not only
local but also rigid symmetry. For maintaining rigid $SU(2) \times
U(1)$ invariance external scalars have been introduced. 
In this way one is able
to construct even a local abelian Ward identity in the tree approximation.
This Ward identity proven to all orders ensures electromagnetic current
coupling in the model and is the functional form of the Gell-Mann--Nishijima
relation.
The gauge fixing breaks the gauge symmetry non-linearly. Therefore one
had to replace gauge invariance by BRS invariance, introducing the
Faddeev-Popov ghost fields. BRS transformations act on the matter fields and
vectors as gauge transformations, but allow to complete  the gauge fixing
 to a BRS symmetric action by adding  the ghost action.
The algebra of $SU(2)\times U(1)$ transformations
is then translated to nilpotency of the BRS transformations.
Having determined the gauge fixing and ghost part, the construction
of the classical action has been finished by giving all the symmetry
transformations in their functional form. BRS invariance is replaced by
the Slavnov-Taylor identity and invariance 
under rigid and local gauge transformations
by the Ward identities. In the proof of renormalizability it has
to be proven, that these symmetries can be established to all orders
of perturbation theory and define the Green's functions of the 
Standard Model uniquely 
to all orders. 
\par
In the last section we first illustrated in the $\varphi^4$-theory
some special properties of renormalized perturbation theory.
By comparing
two renormalization prescriptions,  dimensional regularization
with (modified) minimal subtraction and the BPHZ momentum subtraction scheme, 
we have shown, that in
the procedure of 
 renormalization  Green's functions  are only defined up to local counterterms.
To remove this scheme dependence one has to introduce normalization conditions
for  the free parameters of the model.
For  the Standard Model we have chosen
a normalization scheme, which allows to fix all mass parameters of the
theory and  all the residua independently. In particular
we required 
 the photon and $Z$ boson  mass matrix to be diagonal
at the $Z$-mass and at $p^2 = 0$.
The latter normalization condition is crucial for ensuring infrared existence
for off-shell Green's functions.
\par
Finally the most important ingredient 
for the algebraic proof of renormalizability,
the quantum action principle, has been given. In particular we have discussed
consequences of the quantum action principle for the symmetries of the 
Standard Model to higher orders. 
The
 notes ended with an outline of the algebraic method. We have shown, that
by the algebraic characterization of all possible counterterms and
all possible breakings renormalizability can be proven in a scheme
independent way. Indeed the symmetries, the Slavnov-Taylor identity,
the rigid $SU(2)$ and the local abelian Ward identity, which
we have derived in the classical approximation, 
completely characterize the model 
and can be established to all orders of perturbation theory since the anomalies
are
cancelled by the lepton and quark loops.
\par

{\it Acknowledgements} It is a pleasure 
  to thank the organizers of the Saalburg
Summer School 1997, 
Olaf Lechtenfeld (Hannover University),
Jan Louis (Halle University),
Stefan Theisen (Munich University) and  
Andreas Wipf (Jena University) for the invitation  to lecture 
at this interesting 
 school.
We would also like to thank Torsten Tok (Leipzig University) for taking notes
during the lectures.
\newpage

\begin{appendix}
\appsection{List of important formulae}

In this appendix we summarize  the important formulae of the electroweak 
Standard Model, the  action  and the defining
symmetry operators, in the tree approximation. All expressions are given  
 in the QED-like on-shell parameterization 
(\ref{QEDonshellpar}), in particular we use the on-shell definition of the weak
mixing angle throughout (\ref{wma}):
$\cos \theta_W \equiv \frac {M_W} {M_Z} $.

\begin{flushleft}
{\bf Fields of the Standard Model}
\end{flushleft}

Left-handed fermion doublets: 
\begin{equation}
F^{L}_{l_i} = \left( \begin{array}{c} \nu^L _i \\
                                    e^L _i \end{array} \right),
\qquad 
F^{L}_{q_i}  = \left( \begin{array}{c} u^L _i \\
                                    d^L _i \end{array} \right), 
\qquad i= 1 ... N_F,
\end{equation}
right-handed-fermion singlets:
\begin{equation}
f^R_i = e_i^R ,\/ u_i^R, \/  d_i^R , \qquad i = 1 ...N_F.
\end{equation}
With three generations of fermions  $(N_F = 3 )$ one has explicitly:
\begin{equation}
 \begin{array}{lcccc}
\nu_{e_i}& =& \nu_{e} ,\/& \nu_{\mu} , \/ &\nu_{\tau} \\
e_i& =& e ,\/& \mu , \/ &\tau \\
u_i& =& u ,\/&  c, \/ & t \\
d_i& =& d ,\/ & s,  \/ & b ;
\end{array}
\end{equation}
quarks are colour vectors, $q = (q_r, q_b, q_g), q = u_i, d_i$. \\
Vector fields:
\begin{equation}
V^\mu_a = (W^  \mu _+, W^\mu_-, Z^\mu, A^\mu),
\end{equation}
auxiliary $B$-fields:
\begin{equation}
B_a = (B _+, B_-, B_Z,B_ A) ,
\end{equation}
Faddeev-Popov ghosts with Faddeev-Popov charge $1(c_a)$ and $-1 (\bar c_a)$:
\begin{equation}
c_a = (c _+, c_-, c_Z,c_ A) 
\qquad  \bar c_a = (\bar c _+, \bar c_-, \bar c_Z,\bar c_ A) .
\end{equation}
The scalar doublet and its hermitian conjugate:
\begin{equation}
 \Phi \equiv\left(
    \begin{array}{c}
      \phi^+(x)\\
    \frac  1{\sqrt 2}(H(x) + i\chi(x))
    \end{array}
  \right) 
\qquad \tilde \Phi \equiv i\tau_2 \Phi^* = \left(
    \begin{array}{c}
            \frac 1{\sqrt 2}(H(x) - i\chi(x)) \\
            -\phi^-(x) 
    \end{array}
  \right) \/,
\end{equation}
the external scalar doublet and its hermitian conjugate:
\begin{equation}
\hat \Phi \equiv\left(
    \begin{array}{c}
    \hat  \phi^+(x)\\
    \frac  1{\sqrt 2}(\hat H(x) + i\hat \chi(x))
    \end{array}
  \right) 
\qquad \tilde {\hat\Phi} \equiv i\tau_2 {\hat \Phi}^* = \left(
    \begin{array}{c}
            \frac 1{\sqrt 2}(\hat H(x) - i\hat \chi(x)) \\
            -\hat \phi^-(x) 
    \end{array}
  \right) \/.
\end{equation}
External fields with  Faddeev-Popov charge $-1 (\rho^\mu_\a)$ and $ -2 
(\sigma_\a)$:
\begin{equation}
\rho_\a = (\rho_+ ,\rho_-, \rho_3)
\qquad 
\sigma_\a = (\sigma_+ ,\sigma_-, \sigma_3),
\end{equation}
scalar fields with Faddeev-Popov charge $-1$:
\begin{equation}
Y \equiv\left(
    \begin{array}{c}
      Y^+\\
    \frac  1{\sqrt 2}( Y _H + iY_ \chi)
    \end{array}
  \right) 
\qquad Y^* = \left(
    \begin{array}{c} Y^- \\
            \frac 1{\sqrt 2}(Y_H - iY_ \chi) 
     \end{array}
  \right),
\end{equation}
right-handed fermion doublets with Faddeev-Popov charge $-1$:
\begin{equation}
\Psi^{R}_{l_i} = \left( \begin{array}{c} \psi^R_{\nu _i} \\
                                  \psi^L_{  e _i} \end{array} \right)
\qquad \Psi^{R}_{q_i}  = \left( \begin{array}{c} \psi_{u_i}^R \\
                                  \psi_{d _i}^R \end{array} \right) \/,
\end{equation}
left-handed fermion singlets with Faddeev-Popov charge $-1$
\begin{equation}
\psi^{L}_{f_i} 
= \psi^{L}_{e_i} ,\psi^{L}_{u_i} ,\psi^{L}_{d_i}\/.
\end{equation} 
The BRS variation of the external scalar doublet $\hat \Phi$
with Faddeev-Popov charge $1$:
\begin{equation}
{\bf q} \equiv\left(
    \begin{array}{c}
      q^+\\
    \frac  1{\sqrt 2}( q _H + iq_ \chi)
    \end{array}
  \right) 
\qquad {\bf q}^* = \left(
    \begin{array}{c} q^- \\
            \frac 1{\sqrt 2}(q_H - iq_ \chi) 
     \end{array}
  \right) \/ .
\end{equation}

\begin{flushleft}
{\bf The classical action}
\end{flushleft}

The classical action of the Standard Model  can be decomposed in
the gauge invariant Glashow-Salam-Weinberg action and the gauge-fixing
and ghost action:
 \begin{equation}
\Gamma_{cl} = \Ga_{GSW} + \Ga_{g.f.} + \Ga_{ghost}
\end{equation}
The Glashow-Salam-Weinberg action is given by
\begin{equation}
\Ga_{GSW} = \Gamma_{YM} + \Gamma_{scalar} + \Gamma_{matter} + \Gamma_{Yuk},
\end{equation}
with
\equa{
  \Gamma_{YM} & = 
  -\frac 1 4 \intd \/G_a^{\mu\nu}\tilde{I}_{aa'}G_{\mu\nu a'}\\
  \Gamma_{scalar}&=\intd\Bigl( (D^\mu(\Phi +   {\hbox{v}}
 ))^\dagger D_{\mu}(\Phi
+   {\hbox{v}}) 
  -\frac 1 8\frac{m_H^2}{M^2_W} {e^2 \over \sws}
(\Phi^\dagger\Phi +   {\hbox{v}} 
 ^\dagger \Phi
+ \Phi^\dagger   {\hbox{v}})  ^2 \Bigr) \\  
\Gamma_{matter} &= \sum_{i=1}^{N_F} \intd \Bigl( \overline
 {F^L_{l_i}} i \Ds  F^L_{l_i}
       + \overline {F^L_{q_i}} i \Ds F^L_{q_i} + \overline
 {f^R_i} i \Ds f^R_{i} \Bigr)\\ 
\Gamma_{Yuk}& = 
\sum_{i=1}^{N_F}\intd   {-  e  \over M_W \sqrt 2 \sin \theta_W
} \biggl( m_{e_i} 
\overline
 {F^L_{l_i}} (\Phi +   {\hbox{v}}) e^R_i
                 +  m_{u_i}  \overline
 {F^L_{q_i}} (\Phi +   {\hbox{v}}) u^R_i \nonumber \\
                &  
\phantom{= \sum\intd}  \qquad + m_{d_i}  \overline
 {F^L_{q_i}} (\tilde \Phi + \tilde{  {\hbox{v}}}) d^R_i + \hbox{h.c.}
 \biggr)  ,
}
The gauge-fixing in the most general linear gauge compatible with
rigid symmetry is given by
\begin{eqnarray}
\label{fixext}
\Gamma_{{g.f.}}&\hspace{-3mm}=&\hspace{-3mm}
\intd 
\biggl( \frac 12\xi B_a \tilde I _{ab} B_b +
 \frac 12 \hat \xi (\sw B_Z + \cw B_A )^2 +
B_a \tilde I_{ab} \partial V_b \\
&\hspace{-3mm} &\hspace{-3mm}-\frac {ie}{\sin \theta _W }
\bigl( (\hat \Phi  +  \zeta {\mathrm v}
)^\dagger \frac { \tau _a (\hat G) } 2  B_a ( \Phi +{\mathrm v})
- (\Phi + {\mathrm v})^\dagger \frac {\tau _a (\hat G) } 
2  B_a  (\hat \Phi + \zeta 
{\mathrm v})\bigr) \biggr) . \nonumber
\end{eqnarray} 
The Faddeev-Popov ghost action for arbitrary $\hat G$ is derived from
the BRS trans\-for\-mations (\ref{BRS}) by postulating $\brs \Ga_{ghost} +
\brs\Ga_{g.f.} = 0$. (The matrix $\hat g$ depends on $\hat G$ and $\theta_W$;
it is defined in (\ref{hatg})) : 
\begin{eqnarray}
\label{Gaghostapp}
\Gamma_{ghost} & = &   \intd\biggl(- \bar c _a \Box \tilde I_{ab} c_b
 - \frac e \sw \bar c_a \hat g^{-1}_{aa'}f_{a'bc'} \partial( V_b \hat g_{c'c}
 c_c) \\
   & & \phantom{\int} + i \frac e {2\sw} \bar c _a \hat g ^{-1}_{aa'}
\bigl(\hat{\mathbf q}
^\dagger \tau_{a'} (\hat G) (\Phi + {\mathrm v}) -
(\Phi + {\mathrm v})^\dagger \tau_{a'} (G_s) \hat{\mathbf q} \bigr)
 \nonumber \\
& & \phantom{\int}
- \frac {e^2} {4\sin^2 \theta_W} \Bigl( 
\bar c_a \hat g^{-1}_{aa'}( \hat \Phi + \zeta {\mathrm v} ) ^\dagger 
\tau_{a'}(\hat G) \tau_{b'}(G_s) ( \Phi +  {\mathrm v} )  \nonumber \\
 & & \phantom{\int - \frac {e^2} {4\sin^2 \theta _W}\bar c_a \hat g^{-1}_{aa'}
 } +
( \Phi +  {\mathrm v} ) ^\dagger \tau_{b'}(G_s) \tau_{a'}(\hat G) (
\hat  \Phi +  \zeta {\mathrm v} ) \Bigr)\hat g_{b' b} c_b \biggr)\nonumber 
\end{eqnarray}
 We want to note that the bilinear part of the ghost action is diagonal
with arbitrary ghost masses $\zeta_W M_W^2$ and $\zeta_Z M_Z^2 $
\begin{equation}
\label{Gaghostbilapp}
\Ga_{ghost}^{bil}
 =  
\intd \Bigl(- \bar c _a \Box \tilde I_{ab} c_b - 
  \zeta _W M_W^2 (\bar c_+ c_- + \bar c_- c_+)  -\zeta _Z M_Z^2 \bar c_Z c_Z 
\Bigr) + \Ga_{ghost}^{int}
\end{equation}
with
$ \zeta_W \equiv \zeta $ and 
$ \zeta_Z = \zeta \cos\theta_W(\cos\theta_W - \hat G \sin \theta_W). $ 

In the above formulae we have used the following conventions and
abbreviations:\\
$ { \hbox{v}} $ denotes the shift of the scalar field doublet:
\begin{equation}
\label{shift}
{ \hbox{v}} = \left( \begin{array}{c} 0 \\
                                 \hbox{$\frac 1{\sqrt 2}$}
   v \end{array} \right)
\quad \hbox{with} \quad v = \frac 2e M_Z \cos\theta_W \sin \theta_W \; .
\end{equation}
The field strength tensor and the covariant derivatives have the form
\begin{eqnarray}
  G^{\mu\nu}_a & = & \partial ^\mu V_a^{\nu} - \partial^{\nu} V_a^\mu
  + \frac e{\sin{\theta_W}}
 \tilde{I}_{aa'} 
  {f}_{a'bc} V^{\mu}_b V^{\nu}_c\\
  D_{\mu}\Phi &=& \partial_{\mu}\Phi - i \frac e{\sin{\theta_W}}
 \frac{{\tau}_a(G_s)}{2} \Phi V_{\mu a}
\\
  D_{\mu} F^L_{\delta_i} &=& \partial_\mu F^L_{\delta_i} - i 
\frac e{\sin{\theta_W}}
 \frac{{\tau}_a(G_\delta)}{2} F^L_{\delta,i} V_{\mu a} \qquad \delta= l,q\\
  D_{\mu} f_i^R &=& \partial_\mu f_i^R  +
 i e Q_f \frac {\sin{\theta_W}}{\cos{\theta _W}}   f_i^R Z_\mu +
            ie Q_f f_i^R A_\mu \; .
\end{eqnarray}
The structure constants are defined by the antisymmetric tensor
\begin{equation}
   {f}_{abc}  =  \left\{
    \begin{array} {ccc}
       {f}_{+-Z}&=&- i \cos\theta_W\\
       {f}_{+-A}& =&i \sin\theta_W 
    \end{array}\right.
\end{equation}
 The matrices $ \tau_a\ \ (a=
+,-,Z,A)$ form a representation of $SU(2) \times U(1)$:
\begin{equation}
  \left[\frac{ {\tau}_a}2, \frac{ {\tau}_b}2\right]= 
  i  {f}_{abc} 
  \tilde{I}_{cc'} \frac{ \tau_{c'}}2 \; .
\end{equation}
They are explicitly given by 
\begin{eqnarray}
\label{taudef}
  {\tau}_+  
= \hbox{$\frac 1{\sqrt{2}}$} (\tau_1 + i\tau_2) &  &
  {\tau}_Z(G)  =\tau_3 \cos\theta_W + G {\mathbf 1} \sin\theta_W
    \nonumber\\
  {\tau}_-\!
 =\hbox{$\frac 1{\sqrt{2}}$} (\tau_1 - i\tau_2 ) &  &
  {\tau}_A (G) = - \tau_3 \sin\theta_W \! + G{\mathbf 1}\cos\theta_W \; .
\end{eqnarray}
 $\tau_i $, $i = 1,2,3$, are the Pauli matrices and 
\begin{equation}
\tau_+ = \left( \begin{array} {cc} 0& \sqrt 2 \\ 
                                  0 & 0 \end{array}\right) \quad 
\tau_- = \left( \begin{array} {cc} 0& 0 \\ 
                                  \sqrt 2 & 0 \end{array}\right) \quad 
\tau_3 = \left( \begin{array} {cc} 1& 0 \\ 
                                  0 & -1 \end{array}\right) .
\end{equation}
The matrices $\tau_Z $ and $\tau_A$
depend on the abelian coupling  $G$, which is related to
the weak hypercharge $Y_W$ of the different $SU(2)$-doublets:
\begin{equation}\label{hyper}
  G_k = - Y^k_W \frac{\sin\theta_W}{\cos\theta_W} \qquad
Y_W^{k}  =  \left\{
    \begin{array} {ccc}
       &1 & \hbox{for the scalar ($k=s$)} \\
       &$-1$& \hbox{for the lepton doublets ($k=l$)} \\
       & \hbox{$\frac 13$} & \hbox{for the quark doublets ($k=q$) . } 
    \end{array}\right. 
\end{equation}
The matrix $\tilde{I}_{aa'}$ guarantees the charge neutrality of the
classical action
\begin{eqnarray}
\label{tildeI}
\tilde{I}_{+-}&=&\tilde{I}_{-+}=\tilde{I}_{ZZ}=\tilde{I}_{AA}=1\\
\tilde{I}_{ab}& =&0 \mathrm{\ else} . \nonumber
\end{eqnarray}

The parameter $\hat G$ appearing in the gauge-fixing  and ghost action
is arbitrary and not restricted by the symmetries of the Standard Model.
Defining
\begin{equation}
\label{hatG}
\hat G = - \frac {\sin \theta_G} {\cos \theta_G}
\end{equation}
the matrix $\hat g_{ab}$ depends on $\hat G$ in the following way:
\begin{eqnarray}
\label{hatgab}
\hat g _{+-}  = 1 &\quad & \hat g _{-+} = 1 \nonumber\\
\hat g _{ZZ}  =  \cwg &\quad &\hat g_{AZ} = -\swg \\
\hat g_{ZA}  =  0 &\quad& \hat g_{AA} = 1 . \nonumber
\end{eqnarray}
In matrix notation it reads:
\begin{equation}
\label{hatg}
\hat g _{ab} = \left(
    \begin{array}{cccc}
      1& 0& 0& 0 \\
      0& 1& 0& 0 \\
      0& 0& \cwg & 0\\
      0& 0& - \swg& 1 
    \end{array}
  \right) .
\end{equation}
A natural choice in the tree approximation is given by
\begin{equation}
\hat G \equiv - \frac {\sin \theta_W} {\cos \theta_W};
\end{equation}
then one has
\begin{equation}
\hat g_{ab} =  \delta_{ab} \qquad \mbox{and} \qquad \zeta_W = \zeta_Z .
\end{equation}
 It turns out, that this choice is not stable under renormalization.

\begin{flushleft}
{\bf BRS transformations}
\end{flushleft}

The classical action $\Ga_{cl}$, explicitly 
$\Ga_{GSW}$ and $\Ga_{g.f.} + \Ga_{ghost}$, are invariant under 
BRS transformations:
\begin{equation}
\brs \Ga_{cl} = 0 \qquad \mbox{and} \qquad \brs \Ga_{GSW} = 0,
\qquad \brs \bigl(\Gamma_{g.f.} + \Gamma_{ghost} \bigr) =0. 
\end{equation}
BRS transformations are given by:
\equa{
\label{BRS}
{\mathrm s} V_{\mu a}&=\partial _{\mu}c_a + 
\frac e{\sin \theta _W} \tilde I _{aa'} f_{a'bc}V_{\mu
  b}\hat g_{cc'}c_{c'}\\
{\mathrm s} \Phi &= i \frac e{\sin{\theta_W}}
\frac{{\tau}_a(G_s)}{2} (\Phi+ {\mathrm v}) \hat g_{aa'} c_{
 a'} \nonumber  \\
{\mathrm s} F^L_{\delta_i} & = i
\frac e{\sin{\theta_W}}
 \frac{{\tau}_a(G_\delta)}{2} F^L_{\delta_i} \hat g_{aa'}
c_{ a'} \qquad \delta = l,q\nonumber \\
{\mathrm s}  f_i^R&=  - i e Q_f \frac {\sin{\theta_G}}{\cos{\theta _W}}  
 f_i^R   c_Z-            ie Q_f f_i^R c_A \nonumber \\
{\mathrm s} c_{a}&=- \frac e{2 \sin \theta_W } \tilde I _{aa'} 
 f_{a'bc}\hat g_{bb'} c_{b'} \hat g _{cc'} c_{c'} \nonumber \\
{\mathrm s} \bar c_a & =  \hat B_{a'} \hat g_{a'a} \mbox{ (i.e. }
\brs \bar c_Z = \cwg B_Z - \swg B_A,\mbox{ }\brs \bar c _A  = B_A) \nonumber \\
{\mathrm s} B_a & = 0 \nonumber \\
{\mathrm s} \hat \Phi &=   \hat {\mathbf q} \nonumber \\
{\mathrm s} \hat {\mathbf q} & = 0 \nonumber
}
The BRS transformations are nilpotent on all fields:
\begin{equation}
\brs^2 \varphi_k = 0
\end{equation}
We have given the 
BRS transformations  for arbitrary $\hat G $ (cf~(\ref{hatG})). 

\begin{flushleft}
{\bf Slavnov-Taylor identity}
\end{flushleft}

For renormalization the  BRS transformations are encoded in
the Slavnov-Taylor identity.
For this reason one adds to the classical action the external field part:
\begin{equation}
\label{Gaclcomp}
\Gacl \longrightarrow \Ga_{cl} = \Ga_{GSW} +  \Ga_{g.f.} + \Ga_{ghost}
                               + \Ga_{ext.f.}    
\end{equation}
with
\begin{eqnarray}
\label{gaext}
\Gamma_{ext.f.} & = & \intd \Bigl( \rho ^\mu _+ {\brs} W_{\mu,-} +
                           \rho ^\mu _- {\brs} W_{\mu,+ } +
                           \rho ^\mu _3  (\cw {\brs} Z_{\mu} - \sw 
{\brs} A_\mu) \\
               &   & \! \phantom{\intd} 
+ \sigma _+ {\brs} c_{-} +
                                        \sigma _- {\brs} c_{+} +
                               \sigma _3 (\cg {\brs}  c_Z - \sw {\brs} c_A) 
\nonumber \\
                              &   & \! \phantom{\intd} 
       + Y^\dagger {\brs} \Phi  + ({\brs} \Phi) ^\dagger Y 
       + \sum _{i=1}^{N_F} \bigl(
       \overline {\Psi ^R _{l_i}} {\brs} F^L _{l_i}  + 
       \overline {\Psi ^R _{q_i}} {\brs} F^L _{q_i}  + 
       \sum_{f}       \overline {\psi ^L _{f_i}}
{\brs}  f^R_i + \hbox{h.c.} \bigr)\Bigr) .
\nonumber
\end{eqnarray}
The Slavnov-Taylor identity of the 
tree approximation  reads (again for arbitrary
$\hat G$ (\ref{hatG}) -- (\ref{hatg})):
\begin{eqnarray}
\label{STG}
{\cal S}
(\Gacl )&\hspace{-3mm} = & \hspace{-2mm}\intd \biggl(
\bigl(\sg \partial _\mu c _Z + \cw \partial_\mu c_A\bigr)
             \Bigl(\sw {\delta \Gamma \over \delta Z_\mu } + \cw
       {\delta \Gamma \over \delta A_\mu} \Bigr) \\  
 &\hspace{-3mm} & \hspace{-2mm}+     {\delta \Gamma \over \delta \rho^\mu_3 }
              \Bigl(\cw {\delta \Gamma \over \delta Z_\mu } - \sw
       {\delta \Gamma \over \delta A_\mu} \Bigr) 
       + {\delta \Gamma \over \delta \sigma _3 } 
              \Bigl(\cw {\delta \Gamma \over \delta c_Z } - \sg
       {\delta \Gamma\over \delta c_A} \Bigr) 
  \frac 1{\cwg}\nonumber \\ 
&\hspace{-3mm} & \hspace{-2mm}
+ \Bigl(\cwg B_Z -\swg B_A\Bigr)
{\delta \Gamma \over \delta \bar c_Z } +
B_A{\delta \Gamma \over \delta \bar c_A }   \nonumber \\
&\hspace{-3mm} &  \hspace{-2mm}    + {\delta \Gamma \over \delta  \rho^\mu _+ }
               {\delta \Gamma \over \delta W_{\mu,- } }
      + {\delta \Gamma \over \delta \rho^\mu _- }
               {\delta \Gamma\over \delta W_{\mu,+ } }
+  {\delta \Gamma \over \delta \sigma _+ }
               {\delta \Gamma \over \delta c_{-} }
+  {\delta \Gamma \over \delta \sigma _- }
               {\delta \Gamma \over \delta c_{+} } 
+ B_+{\delta \Gamma \over \delta \bar c_+ } +  
B_-{\delta \Gamma \over \delta \bar c_- }    
\nonumber \\
&\hspace{-3mm}& \hspace{-2mm}+ \sum_{i=1}^{N_F} \Bigl(\sum_{f}
{\delta \Gamma \over \delta \overline{\psi^L_{f_i}}
}{ \delta \Gamma \over \delta f^R_i }
+ \sum _{\delta= l,q}{\delta \Gamma \over \delta \overline{\Psi^R_{\delta_i}}}
{ \delta \Gamma \over \delta F^L_{\delta_i} } + 
  \hbox{h.c.} \Bigr) +
\Bigl({\delta \Gamma \over \delta Y^\dagger}
{ \delta \Gamma \over \delta \Phi } 
+ {\bf q}
{\delta \Gamma \over \delta \hat \Phi  }  + \hbox{h.c.} \Bigr)\biggr) 
   \nonumber \\
& \hspace{-3mm}= & \hspace{-2mm} 0 .\nonumber
   \end{eqnarray}

\begin{flushleft}
{\bf Ward identities of rigid symmetry}
\end{flushleft}

The classical action including the gauge-fixing and ghost action and external 
field action (\ref{Gaclcomp}) is  constructed in a way, 
that it is invariant under rigid
$SU(2) \times U(1)$ transformations. 
The Ward operators of rigid SU(2) transformations satisfy the algebra
\begin{eqnarray}
\label{wardalg}
\bigl[ {\cal W }_\alpha , {\cal W } _\beta \bigr] & =& 
\varepsilon 
_{\alpha\beta\gamma} 
\tilde I_{\gamma \gamma'} {\cal W} _{\gamma'} \qquad
\a = +,-,3
\end{eqnarray}
The structure constants are defined by the completely antisymmetric tensor
$\varepsilon _{\alpha\beta\gamma} $ with $\varepsilon _{+-3}  = -i $.
The Ward identities of rigid $SU(2)$ transformations are given in the
tree approximation by
\begin{eqnarray}
\label{wardnaapp} 
{\cal W}_\alpha \Ga_{cl} =  \tilde I_{\alpha\alpha'} 
\intd & \hspace{-3mm} \Biggl( & \hspace{-3mm} V^\mu _b 
      \hat\varepsilon _{b c, \alpha'} 
      \tilde I _{cc'}
\frac{\delta}{\delta V^\mu _{c'}} +
B _b 
      \hat\varepsilon _{b c, \alpha'} 
      \tilde I _{cc'}
\frac{\delta}{\delta B _{c'}}  \\ 
& + &  c _b  \hat g^T_{bb'}
      \hat\varepsilon _{b' c', \alpha'} \hat g^{-1T}_{c'c} 
      \tilde I _{cd}
\frac{\delta}{\delta c _{d}} 
+ \bar c _b  \hat g^{-1}_{bb'}
      \hat\varepsilon _{b' c', \alpha'} \hat g_{c'c} 
      \tilde I _{cd}
\frac{\delta}{\delta \bar c _{d}} \nonumber \\
&+ &
\rho^\mu _\beta \varepsilon _{\beta\gamma \alpha'}  \tilde I _{\gamma \gamma'}
\frac{\delta}{\delta \rho^\mu _{\gamma'}} 
+\sigma _\beta \varepsilon _{\beta\gamma \alpha'}  \tilde I _{\gamma \gamma'}
\frac{\delta}{\delta \sigma _{\gamma'}}  \nonumber \\
& + & 
  i (\Phi + {\mathrm v}) ^\dagger
\frac { \tau _{\alpha'} } 2  \frac{\overrightarrow 
\delta}{\delta \Phi^\dagger}
 -  i \frac{\overleftarrow
\delta}{\delta \Phi} \frac { \tau _{\alpha'} } 2  (\Phi +{\mathrm v})   \nonumber \\
& + & 
  i (\hat \Phi + \zeta {\mathrm v}) ^\dagger
\frac { \tau _{\alpha'} } 2  \frac{\overrightarrow 
\delta}{\delta \hat \Phi^\dagger}
 -  i \frac{\overleftarrow
\delta}{\delta \hat \Phi} \frac { \tau _{\alpha'} } 2  (\hat \Phi +
\zeta {\mathrm v}) \nonumber \\  
& + & 
  i Y ^\dagger
\frac { \tau _{\alpha'} } 2  \frac{\overrightarrow 
\delta}{\delta Y^\dagger}
 -  i \frac{\overleftarrow
\delta}{\delta Y} \frac { \tau _{\alpha'} } 2  Y
+  i {\bf q} ^\dagger
\frac { \tau _{\alpha'} } 2  \frac{\overrightarrow 
\delta}{\delta {\bf q}^\dagger}
 -  i \frac{\overleftarrow
\delta}{\bf q} \frac { \tau _{\alpha'} } 2  {\bf q} \nonumber \\
&+ &  
\sum_{i = 1}^{N_F} \sum_{\delta = l,q} \Bigl(
 i \overline{F ^L_{\delta_i}} \frac {\tau _{\alpha'} } 2  
\frac{\overrightarrow 
\delta}{\delta \overline{ F^L_{\delta_i}} }
 -  i \frac{\overleftarrow
\delta}{\delta F^L_{\delta_i}} \frac { \tau _{\alpha'} } 2  F^L_{\delta_i}   \nonumber \\
& &   \phantom{ \sum_{i = 1}^{N_F} \sum_{\delta = l,q}}+
 i \overline{\Psi ^R_{\delta_i}} \frac {\tau _{\alpha'} } 2  
\frac{\overrightarrow 
\delta}{\delta \overline{ \Psi^R_{\delta_i}} }
 -  i \frac{\overleftarrow
\delta}{\delta \Psi^R_{\delta_i}} \frac { \tau _{\alpha'} } 2  \Psi
^R_{\delta_i}  
 \Bigr)\Biggr) \Ga_{cl} = 0  \nonumber
\end{eqnarray}
The matrix $\hat g _{ab}$ is defined in (\ref{hatg}), the tensor 
$\hat \varepsilon_{bc,\a } , b,c = +,-,Z,A $ and $ \a = +,-,3$ 
is given by
\begin{equation}
\label{hateabc}
O^T_{b\beta}(\theta_W) \ve _{\b \ga \a}
O_{\ga c} (\theta_W) \equiv \hat \varepsilon_{bc,\alpha} = \left\{
    \begin{array} {ccc}
       {\hat \ve}_{Z+,-} &=& -i \cos\theta_W\\
       {\hat \ve}_{A+,-} &=& i \sin\theta_W \\
        \hat \ve_{+-,3}  &=& -i 
    \end{array}\right.
\end{equation}
The matrix $O_{\a a} (\theta_W) $ 
(\ref{eq: thetaW rot}) transforms the $SU(2) \times U(1)$ fields
$W^\mu_+, W^\mu_-, W^\mu_3, W^\mu_4$
to physical on-shell fields $W^\mu_+, W^\mu_-, Z^\mu, A^\mu$.

In the Standard Model there are three types of abelian rigid symmetries:
the abelian operator connected with electromagnetic charge conservation
${\cal W}_4^Q = {\cal W}_{em} - {\cal W}_3 $, and the abelian
operators of lepton and baryon
conservation $ {\cal W} _{l_i}$ and $ {\cal W}_{q_i} $: 
\begin{equation}
\bigl[ {\cal W }_\alpha , {\cal W } _4^Q \bigr]  = 0 ,
\quad 
\bigl[ {\cal W }_\alpha , {\cal W } _{l_i} \bigr]  = 0,
\quad
\bigl[ {\cal W }_\alpha , {\cal W } _{q_i} \bigr]  = 0 
\end{equation}
These operators are given by
\begin{eqnarray}
\label{chargecons}
{\cal W}_{em} = - i \intd &\hspace{-3mm}\biggl( & \hspace{-3mm}
W^\mu_+{\delta \over \delta W^\mu_+} -W^\mu_-{\delta \over \delta W^\mu_-} +
B_+{\delta \over \delta B_+} -B_-{\delta \over \delta B_-}\\
&\hspace{-3mm}+& \hspace{-3mm}
c_+{\delta \over \delta c_+} -c_-{\delta \over \delta c_-} +
\bar c_+
{\delta \over \delta \bar c_+} -\bar c_-{\delta \over \delta \bar c_-} 
\nonumber\\
&\hspace{-3mm}+& \hspace{-3mm}
\rho_+{\delta \over \delta \rho_+} -\rho_-{\delta \over \delta \rho_-} + 
\sigma_+{\delta \over \delta \sigma_+} -\sigma_-{\delta \over \delta 
\sigma_-} \nonumber \\
&\hspace{-3mm}+& \hspace{-3mm}
\phi^+{\delta \over \delta \phi^+} -\phi^-{\delta \over \delta \phi^-} +
Y^+{\delta \over \delta Y^+} -Y^-{\delta \over \delta Y^-} \nonumber \\
&\hspace{-3mm}+& \hspace{-3mm}
\hat \phi^+{\delta \over \delta \hat \phi^+} -\hat \phi^-{\delta \over 
\delta \hat \phi^-} +
q^+{\delta \over \delta q^+} -q^-{\delta \over \delta q^-} \nonumber \\
&\hspace{-3mm}-& \hspace{-3mm}
 \sum _{i=1}^{N_F} 
\Bigl( Q_e
(\bar e_i {\delta \over \delta \bar e_i} - {\delta \over \delta e_i} e_i  +
\bar \psi_{e_i} {\delta \over \delta \bar \psi_{e_i}} -
 {\delta \over \delta \psi_{e_i}} \psi _{e_i} ) \nonumber\\
&\hspace{-3mm}& 
+ Q_u 
(\bar u_i {\delta \over \delta \bar u_i} - {\delta \over \delta u_i} u_i 
+ \bar \psi_{u_i} {\delta \over \delta \bar \psi_{u_i}}
 - {\delta \over \delta \psi_{u_i}} \psi_{u_i} ) \nonumber \\
&\hspace{-3mm}& 
+ Q_d 
(\bar d_i {\delta 
\over \delta \bar d_i} - {\delta \over \delta d_i} d_i +
\bar \psi_{d_i} {\delta 
\over \delta \bar \psi_{d_i}} - {\delta \over \delta \psi_{d_i}} \psi_{d_i}
 ) \Bigr) \biggr)  \nonumber
\end{eqnarray}
and
\begin{eqnarray} 
\label{qlfcons}
{\cal W}_{l_i} = i \intd &\hspace{-3mm}\biggl( & \hspace{-3mm}
\bar e_i {\delta \over \delta \bar e_i} - {\delta \over \delta e_i} e_i  +
\bar \psi_{e_i} {\delta \over \delta \bar \psi_{e_i}} -
 {\delta \over \delta \psi_{e_i}} \psi _{e_i} \\ 
&\hspace{-3mm}+& \hspace{-3mm}
\overline {\nu^L_i} {\delta \over \delta \overline {\nu^L_i}} 
- {\delta \over \delta \nu^L_i} \nu^L_i
  +\overline {\psi^R_{\nu_i}} {\delta \over \delta \overline
 {\psi^R_{\nu_i}}} -
 {\delta \over \delta \psi^R_{\nu_i}} \psi^R _{\nu_i} \biggr)\nonumber\\
{\cal W}_{q_i} = i \intd &\hspace{-3mm}\biggl( & \hspace{-3mm}
\bar d_i {\delta \over \delta \bar d_i} - {\delta \over \delta d_i} d_i  +
\bar \psi_{d_i} {\delta \over \delta \bar \psi_{d_i}} -
 {\delta \over \delta \psi_{d_i}} \psi _{d_i} \\ 
&\hspace{-3mm}+& \hspace{-3mm}
\bar u_i {\delta \over \delta \bar u_i} 
- {\delta \over \delta u_i} u_i
  + \bar \psi_{u_i} {\delta \over \delta \bar \psi_{u_i}} -
 {\delta \over \delta \psi_{u_i}} \psi _{u_i} \biggr)\/ .\nonumber
\end{eqnarray}
The classical action  is invariant under these global symmetries: 
\begin{equation}
{\cal W}_{em} \Ga_{cl} =  0  \qquad
{\cal W}_{l_i} \Ga_{cl} =  0  \qquad
{\cal W}_{q_i} \Ga_{cl}  =   0.
\end{equation}
Since these global symmetries are not broken by renormalization the generating
functional
of 1PI Green's functions $\Ga$ is invariant  by
construction
to all orders.

\begin{flushleft}
{\bf The local U(1) Ward identity}
\end{flushleft}

The  local $U(1)$ Ward operator, which is  defined by the relation
\begin{equation}
{\cal W}_4^Q = \intd \,{\bf w}_4^Q ,
\end{equation}
is (to all orders of perturbation theory) given by the following expression:
\begin{eqnarray}
{\bf w}_4 ^Q& = &  
\frac i2 (\Phi +{\mathrm v} ) ^\dagger   \frac{\overrightarrow 
\delta}{\delta \Phi^\dagger}
 -  \frac i2 \frac{\overleftarrow
\delta}{\delta \Phi}  ( \Phi + {\mathrm v} )  
+  \{ Y , \hat \Phi + \zeta {\mathrm v} , {\bf q} \}  \\
& & \phantom{ \biggl(} 
+ \sum_{i =1}^{N_F} \biggl( \sum_{\delta = l,q} 
Y_W^\delta \Bigl( \frac i2 
 \overline{F _{\delta_i}^L}   \frac{\overrightarrow 
\delta}{\delta \overline{ F_{\delta_i}^L} }
 -  \frac i2 \frac{\overleftarrow 
\delta}{\delta F_{\delta_i}^L}    F_{\delta_i}^L  
+  \{ \Psi^R _{\delta_i} \} \Bigr)    \nonumber \\
& & \phantom{ \biggl( + \sum_{i =1}^{N_F} }
 + \sum_f  Q_f \Bigl( i \overline{f_i ^R}   \frac{\overrightarrow 
\delta}{\delta \overline{ f_i^R} }
 -  i \frac{\overleftarrow
\delta}{\delta f_i^R}   f_i^R  
+  \{ \psi^L _{f_i} \} \Bigr) \biggr) .    \nonumber 
\end{eqnarray}
This operator is continued to a local $U(1)$ Ward identity which is the
functional form of the Gell-Mann--Nishijima relation:
\begin{equation}
\label{wardloc}
\biggl(\frac e {\cw} {\mathbf w}^Q_4 - 
 \Bigl(\sw \partial {\delta \over \delta Z} + \cw \partial 
{\delta \over \delta
A}\Bigr) \biggr) \Ga_{cl} = 
 (\sw \Box B_Z + \cw \Box B_A )
\end{equation}

\appsection{Exercises}

\begin{enumerate}
\item
  Parity transformation $P$ is defined by $(x^0,\vec x) \overset{P}{\longrightarrow} 
(x^0, - \vec x)$. \\
Show that under the parity transformation $P$
\begin{enumerate}
\item 
\(
\psi(x^\mu) \overset{P}{\longrightarrow}  \eta_P \g^0 \psi (x^o,-\vec x),
\) 
where $\eta_P$ is a phase factor,
\item
\(
 \bar \psi \g^\m \psi \overset{P}{\longrightarrow}(\bar \psi \g^0 \psi, - \bar \psi \vec \g \psi) 
\),
\item
\(
 \bar \psi \g^\m \g^5\psi \overset{P}{\longrightarrow} (-\bar \psi \g^0 \g^5 \psi, - \bar \psi \vec \g \g^5 \psi) 
\),
\item 
\(
\int \bar \psi \g_\m \psi V^\m \overset{P}{\longrightarrow} \int \bar \psi \g_\m \psi V^\m 
\),
if $V^\m$ is a vector,
\item 
\(
\int \bar \psi \g_\m \g^5\psi A^\m \overset{P}{\longrightarrow} \int \bar \psi \g_\m \g^5 \psi A^\m 
\),
if $A^\m$ is an axial vector.
\item Show that one cannot assign a well-defined parity to
\(
\intd W^\m_- \bar e \g^\m \frac 12 (\Identity-\g^5) \n 
\).
\end{enumerate}

\item Left- and right-handed projectors $P^L = \frac12 (\Identity - \g^5)$
 and  $P^R = \frac 12(\Identity + \g^5)$.
\begin{enumerate}
\item Verify the projector properties 
\[
P^L + P^R = \Identity, \qquad P^i P^i = P^i,\, i= L, R, 
\qquad P^L P^R = 0.
\]
\item Show that 
\(
\overline{f^L} = \bar f P^R
\).
\item Express the action $\G^{bil}_{Dirac}$ (\ref{eq: Free Dirac})
in terms of left- and right-handed fermion fields.
\end{enumerate}
\item Fields are representations of the $SU(2)\times U(1)$ symmetry.
\begin{enumerate}
\item  Transformations have been given in terms of left- and
 right-handed
fields.  Calculate on the
Dirac spinors the transformations
$\d_+ f$, $\d_- f$ and $\d_3 f$ for $f = u, d, e, \n$
  explicitly. Use these results to find an expression for the
local functional operators ${\bf w}_\a$ in expressions of Dirac spinors.
Show that the transformations depend on $\ga_5$.
\item Verify that $\d_+$ raises and $\d_-$ lowers the electric charge.
\item Check  the commutator relation of local $SU(2)$-operators 
\equ{
[{\bf w}_\a(x), {\bf w}_\b(y)] = \ve_{\a\b\g} {\bf w}^\dagger_\g (x) \d^4(x-y).
\nonumber }
\end{enumerate}
\item The $SU(2)$ invariance is broken by mass terms. 
\begin{enumerate}
\item Calculate the currents $J^\m_\pm, J^\m_3$ and $j_{em}^\m$ from the identities
\equa{
{\bf w}_\a &\G^{bil}_ {Dirac} \Big |_{m_f = 0}  = - \pa_\m J^\m_\a, \nonumber
\\
{\bf w}_{em} &\G^{bil}_{Dirac}   = -\pa_\m j^\m_{em}. \nonumber
} 
\item  Take a non-vanishing electron mass $m_e \neq 0$ and 
show that 
\[
{\bf w}_\a \G^{Dirac}_{leptons} (m_e \neq 0)  = \pa_\m J^\m_{\a\; leptons} + 
 i Q_\a^{leptons} 
\]
with
\equa{
Q_+^{leptons} & = - m_e\frac 1{\sqrt{2}} \bar e^R \n^L, \nonumber \\
Q_-^{leptons} & = m_e\frac 1{\sqrt{2}} \bar \n^L e^R, \nonumber \\
Q_3^{leptons} & = m_e \frac 12 (\bar e^R e^L- \bar e^L e^R). \nonumber 
}
\end{enumerate}
\item This exercise shows that $\phi^\pm$ and $\chi$ are not physical
fields in the Standard Model.
\begin{enumerate}
\item Give the bilinear part of the Glashow-Salam-Weinberg action $\G_{GSW}$.
\item Eliminate the fields $\phi^\pm$ and $\chi$ from the free field
action by  redefining 
the $W$ and $Z$ bosons:
\equa{
{W'}^\m_\pm & = W^\m_\pm \pm \frac i {M_W} \partial^\m \phi_\pm, \nonumber \\
{Z'}^\m & = Z^\m +\frac1 {M_Z}   \partial^\m \chi. \nonumber 
}
\item Give the respective
free  field equations  for these redefined   fields.
\end{enumerate}
\item Calculate $\G_{matter}$ in terms of the physical fields
$W^\mu_+, W^\mu_-, Z^\mu,A^\mu$.
\item Lepton and quark numbers.
\begin{enumerate}
\item Show that the operators of lepton and quark number
conservation  commute with rigid $SU(2)$-operators (see
(\ref{wardnaapp}) and (\ref{qlfcons})):
\[
[{\cal W}_l, {\cal W}_\a] = 0, \qquad [{\cal W}_q, {\cal W}_\a] = 0.
\]
\item Determine  the corresponding currents $j_\m^l$ and $j_\m^q$ in the
classical approximation.
\item Construct the $SU(2) \times U(1)$
 gauge theory, in which  apart from $J^\pm_\m$ and $J_\m^3$ 
the currents
\(
q_l j^l_\m + q_q j^q_\m
\)
are gauged, but not the electromagnetic current. Discuss the result!
\end{enumerate}
\item Consider the renormalization of the $\varphi^4$-theory 
as discussed in the text (section 4.1).
 Two renormalization prescriptions were given: dimensional regularization 
with MS-subtraction and BPHZ renormalization. Take the one-loop expressions
we have given in (\ref{DIMreg2}) and (\ref{DIMreg4}), (\ref{BPHZren2}) and
(\ref{BPHZren4}). Compare the $\Ga_{eff}$, i.e.~the  1-loop counterterms,
in different schemes. The vertex functions are normalized according to
the conditions (\ref{norm4}) and (\ref{norm2}).
\begin{enumerate}
\item Calculate the counterterms 
to the $\Ga_{eff}$ in dimensional regularization with  MS-subtraction.
\item Calculate the counterterms  to the
$\Ga_{eff}$ in BPHZL subtraction.
\item Discuss the result!
\end{enumerate}
\item The 't Hooft gauges versus unitary gauge.
\begin{enumerate}
\item Calculate the propagators of the scalar and vector fields of the
  Standard  Model in the 't Hooft gauges.
\item Compare these results with  the unitary gauge.
\end{enumerate}
\item BRS transformations.
\begin{enumerate}
\item Check the nilpotency of the BRS operator s
 on the vector and Higgs fields explicitly (see (\ref{BRS})).
\item Determine the bilinear part of the Faddeev-Popov ghost action 
(see (\ref{Gaghostapp}) and (\ref{Gaghostbilapp})).
\end{enumerate}
\end{enumerate}

\end{appendix}
\clearpage

\addcontentsline{toc}{section}{Books and reviews on quantum field theory
       and renormalization}
\begin{flushleft}
{\large\bf  Books and reviews on quantum field theory
       and renormalization}
\end{flushleft}
\medskip
\begin{flushleft}
{\bf Quantum field theory}
\end{flushleft}

\newcounter{bookcounter}
\begin{list}{$[$Q\arabic{bookcounter}$]$ \hfil}
{\usecounter{bookcounter}
}
\item N.N. Bogoliubov, D.V. Shirkov, {\sl Introduction
 to the Theory of Quantized
Fields}, Wiley-Intersciences, NewYork, 1959.
\item S. Gasiorowicz, {\sl Elementary Particle Physics}, J.~Wiley \& Sons,
  Inc., New York, 1966.
\item C. Itzykson, J.-B. Zuber, {\sl Quantum field theory}, McGraw-Hill, 
New York, 1985.
\item T. Kugo, {\sl Eichtheorien}, Springer, Berlin, Heidelberg, 1997.
\item S. Weinberg, {\sl The Quantum Theory of Fields,
Volume I Foundations}, Cambridge University
  Press, Cambridge,  1995.
\end{list}

\begin{flushleft}
{\bf Renormalization}
\end{flushleft}

\begin{list}{$[$R\arabic{bookcounter}$]$ \hfil}
{\usecounter{bookcounter}
}
\item C. Becchi, {\sl Lectures on the Renormalization of Gauge Theories}
, in `Relativity, groups and topology II', Les Houches 1983 
(eds.~B.S. DeWitt and R. Stora)
North Holland, Amsterdam, 1984.
\item J.C. Collins, {\sl Renormalization}, 
Cambridge University Press, Cambridge, 1984.
\item J.H. Lowenstein, {\sl Seminars on Renormalization Theory} vol.II,
University of  Maryland Techn.~Rep.~No 73-068 (1972).
\item J.H. Lowenstein, {\sl BPHZ Renormalization}, in
`Renormalization Theory' (eds.~G. Velo and A.S. Wightman) D. Reidel,
    Dordrecht, 1976.
\item  O. Piguet, {\sl Renormalisation en th\'eorie quantique des champs} and
                  {\sl Renormalisation des th\'eories de jauge}, Lectures of
the `Troisieme cycle de la physique en suisse romande', 1982-83.
\item O. Piguet and S. Sorella, {\sl Algebraic Renormalization},
                 Lecture Notes in Physics 28, Springer, Berlin, Heidelberg,
                     1995. 
\item W. Zimmermann, {\sl Local Operator Products and Renormalization}
in  `Lectures on Elementary Particles and Quantum Field Theory', Vol.~I,
1970 Brandeis Lectures (eds.~S. Deser, M. Grisaru, H. Pendleton) 
M.I.T. Press, Cambridge, 1970.
\end{list}

\newpage

\end{document}